\newif\ifacm
\newif\ifusenix
\newif\ifneurips

\def\conftemplate{neurips}

\expandafter\def\csname confswitch@acm\endcsname{\acmtrue}
\expandafter\def\csname confswitch@usenix\endcsname{\usenixtrue}
\expandafter\def\csname confswitch@neurips\endcsname{\neuripstrue}
\expandafter\ifx\csname confswitch@\conftemplate\endcsname\relax
  \PackageError{confswitch}{Unknown conference template `\conftemplate'}%
    {Set \string\conftemplate\space to acm, usenix, or neurips.}
\else
  \csname confswitch@\conftemplate\endcsname
\fi

\makeatletter
\def\input@path{{templates/acmart/}{templates/usenix/}{templates/neurips/}}
\makeatother

\PassOptionsToPackage{table,xcdraw,x11names}{xcolor}

\ifneurips
  \PassOptionsToPackage{numbers,compress}{natbib}
  \documentclass{article}
  \usepackage[eandd, final]{neurips_2026}
\else\ifusenix
  \documentclass[letterpaper,twocolumn,10pt]{article}
  \usepackage{usenix}
\else
  \documentclass[sigplan,twocolumn,anonymous]{acmart}
\fi\fi

\usepackage{confswitch}

\usepackage{tikz}
\usepackage{amsmath}

\ifneurips
  \usepackage{amssymb,amsmath,amsfonts,amsthm}
  \usepackage[T1]{fontenc}
  \usepackage[utf8]{inputenc}
  \usepackage{microtype}
  \UseMicrotypeSet[protrusion]{basicmath}
  \usepackage{hyperref}
  \usepackage{graphicx}
  \usepackage{xcolor}
  \usepackage{booktabs}
  \usepackage{etoolbox}
  \usepackage{comment}
  \usepackage{caption}
\else\ifusenix
  \usepackage{amssymb,amsmath,amsfonts,amsthm}
  \usepackage[T1]{fontenc}
  \usepackage[utf8]{inputenc}
  \usepackage{microtype}
  \UseMicrotypeSet[protrusion]{basicmath}
  \usepackage{hyperref}
  \usepackage{graphicx}
  \usepackage{xcolor}
  \usepackage{booktabs}
  \usepackage{etoolbox}
  \usepackage{comment}
  \usepackage{caption}
\else
\fi\fi

\usepackage{listings}
\usepackage{upquote}
\usepackage{grffile}
\usepackage[normalem]{ulem}
\usepackage{multirow}
\usepackage{xspace}
\usepackage{tabularx}
\usepackage{ragged2e}
\usepackage{paralist}

\ifneurips
\else\ifusenix
  \usepackage[american]{babel}
\fi\fi

\usepackage{wrapfig}
\usepackage{balance}
\usepackage{enumitem}
\usepackage{url}
\usepackage{pifont}
\usepackage{tikz}
\usepackage{mathtools}

\ifneurips
\else\ifusenix
  \usepackage{xkeyval}
\fi\fi

\usepackage{threeparttable}
\usepackage{makecell}

\usepackage[tworuled, vlined]{algorithm2e}
\usepackage{sepfootnotes}
\ifneurips
\else\ifusenix
  \usepackage[compact]{titlesec}  
\fi\fi
\usepackage[capitalize]{cleveref}
\crefformat{section}{§#2#1#3}
\usepackage{textcomp}
\usepackage{fvextra}
\usepackage[most]{tcolorbox}
\tcbuselibrary{listings,breakable}
\usepackage{subcaption}

\PassOptionsToPackage{usenames,dvipsnames}{color}
\hypersetup{unicode=true,
  colorlinks=true,
  linkcolor=blue,
  citecolor=blue,
  anchorcolor=blue,
  urlcolor=blue,
  breaklinks=true}
\setlist{topsep=1pt, partopsep=0pt, itemsep=1pt, parsep=1pt, leftmargin=10pt}

\makeatletter
\def\maxwidth{\ifdim\Gin@nat@width>\linewidth\linewidth\else\Gin@nat@width\fi}
\def\maxheight{\ifdim\Gin@nat@height>\textheight\textheight\else\Gin@nat@height\fi}
\makeatother
\setkeys{Gin}{width=\maxwidth,height=\maxheight,keepaspectratio}
\makeatletter
\g@addto@macro{\UrlBreaks}{\UrlOrds}
\makeatother

\newcommand\paraspace{\vspace*{0.4ex}}
\providecommand\parab[1]{\paraspace\noindent\textbf{#1}}

\newtoggle{reviewmode}
\newtoggle{anony}

\newcommand{\projurl}[1]{%
  \iftoggle{anony}{URL is hidden for review purpose}{\url{#1}}%
}

\newcommand{\ie}{\emph{i.e.,}\xspace}
\newcommand{\eg}{\emph{e.g.,}\xspace}

\newcommand{\secref}[1]{\S\ref{#1}}
\newcommand{\figref}[1]{Figure~\ref{#1}}
\newcommand{\tabref}[1]{Table~\ref{#1}}

\definecolor{teal}{rgb}{0.0, 0.5, 0.5}
\definecolor{olive}{rgb}{0.5, 0.5, 0.0}
\definecolor{pink}{rgb}{1.0, 0.75, 0.8}
\definecolor{lightgray}{gray}{0.75}
\definecolor{mediumgray}{gray}{0.5}
\definecolor{darkgray}{gray}{0.25}
\definecolor{charcoal}{gray}{0.2}
\definecolor{turquoise}{rgb}{0.25, 0.88, 0.82}
\definecolor{coral}{rgb}{1.0, 0.5, 0.31}
\definecolor{navyblue}{rgb}{0.0, 0.0, 0.5}
\definecolor{lime}{rgb}{0.75, 1.0, 0.0}
\definecolor{darkgreen}{rgb}{0.0, 0.5, 0.0}
\definecolor{violet}{rgb}{0.56, 0.0, 1.0}
\definecolor{lightgreen}{rgb}{0.85, 1.0, 0.85}
\definecolor{burgundy}{cmyk}{0.5, 1.0, 0.7, 0.4}
\definecolor{olivegreen}{cmyk}{0.64, 0, 0.95, 0.4}
\definecolor{peach}{cmyk}{0, 0.5, 0.7, 0}
\definecolor{mustard}{cmyk}{0, 0.3, 1, 0}



\newcolumntype{N}{>{\raggedleft\arraybackslash}p{1.4cm}}
\newcolumntype{U}{>{\raggedright\arraybackslash}p{0.8cm}}

\ifneurips
    \usepackage{titlesec}
    \titlespacing*{\section}{0pt}{7pt plus 3pt minus 3pt}{3pt plus 3pt minus 2pt}
    \titlespacing*{\subsection}{0pt}{4pt plus 3pt minus 2pt}{2pt plus 3pt minus 1pt}
    \titlespacing*{\subsubsection}{0pt}{4pt plus 3pt minus 2pt}{1pt plus 3pt minus 1pt}
\else\ifusenix
    \usepackage{titlesec}
    \titlespacing*{\section}{0pt}{7pt plus 3pt minus 3pt}{3pt plus 3pt minus 2pt}
    \titlespacing*{\subsection}{0pt}{4pt plus 3pt minus 2pt}{1pt plus 3pt minus 1pt}
    \titlespacing*{\subsubsection}{0pt}{4pt plus 3pt minus 2pt}{0pt plus 3pt minus 1pt}
    \titleformat{\section}{\large\bfseries}{\thesection}{1em}{}
    \titleformat{\subsection}{\normalsize\bfseries}{\thesubsection}{1em}{}
  \fi\fi

\toggletrue{anony}  

\newtoggle{showmarks}
\toggletrue{showmarks} 

\iftoggle{showmarks}{
  \newcommand\red[1]{\textcolor{red}{#1}}
  \newcommand\redstrike[1]{\red{\sout{#1}}}
  \newcommand\green[1]{\textcolor{\green}{#1}}
  \newcommand\greenstrike[1]{\green{\sout{#1}}}
  \newcommand\orange[1]{\textcolor{orange}{#1}}
  \newcommand\orangestrike[1]{\orange{\sout{#1}}}
  \newcommand\blue[1]{\textcolor{blue}{#1}}
  \newcommand\bluestrike[1]{\blue{\sout{#1}}}
  \newcommand\purple[1]{\textcolor{purple}{#1}}
  \newcommand\purplestrike[1]{\purple{\sout{#1}}}
  \newcommand\teal[1]{\textcolor{teal}{#1}}
  \newcommand\tealstrike[1]{\teal{\sout{#1}}}
  \newcommand\turquoise[1]{\textcolor{turquoise}{#1}}
  \newcommand\turquoisestrike[1]{\turquoise{\sout{#1}}}
  \newcommand\darkgreen[1]{\textcolor{darkgreen}{#1}}
  \newcommand\darkgreenstrike[1]{\darkgreen{\sout{#1}}}
  \newcommand\lime[1]{\textcolor{lime}{#1}}
  \newcommand\limestrike[1]{\lime{\sout{#1}}}
  \newcommand\olivegreen[1]{\textcolor{olivegreen}{#1}}
  \newcommand\olivegreenstrike[1]{\olivegreen{\sout{#1}}}

  \newcommand{\yibo}[1]{[\purple{\sf\textit{#1 - Yibo}}]}
  \newcommand{\draft}[1]{\textcolor{turquoise}{\sf\textit{#1}}}
  \newcommand{\todo}[1]{[\textcolor{turquoise}{\sf\textbf{TODO: }\textit{#1}}]}
}{
  \newcommand\red[1]{#1}
  \newcommand\redstrike[1]{\unskip}
  \newcommand\green[1]{#1}
  \newcommand\greenstrike[1]{\unskip}
  \newcommand\orange[1]{#1}
  \newcommand\orangestrike[1]{\unskip}
  \newcommand\blue[1]{#1}
  \newcommand\bluestrike[1]{\unskip}
  \newcommand\purple[1]{\unskip}
  \newcommand\purplestrike[1]{\unskip}
  \newcommand\teal[1]{\unskip}
  \newcommand\tealstrike[1]{\unskip}
  \newcommand\turquoise[1]{\unskip}
  \newcommand\turquoisestrike[1]{\unskip}
  \newcommand\darkgreen[1]{\unskip}
  \newcommand\darkgreenstrike[1]{\unskip}
  \newcommand\lime[1]{\unskip}
  \newcommand\limestrike[1]{\unskip}
  \newcommand\olivegreen[1]{\unskip}
  \newcommand\olivegreenstrike[1]{\unskip}

  \newcommand{\yibo}[1]{}
  \newcommand{\draft}[1]{}
  \newcommand{\todo}[1]{}
}

\newcommand{\sysname}{\textsc{DDBench}\xspace}

\newcommand{\cmark}{\ding{51}}
\newcommand{\xmark}{\ding{55}}

\newcommand{\tierone}{{tier-1}\xspace}
\newcommand{\tiertwo}{{tier-2}\xspace}
\newcommand{\tierthree}{{tier-3}\xspace}

\newcommand{\ddbenchNumModels}{10\xspace}
\newcommand{\ddbenchTierOneSize}{31\xspace}
\newcommand{\ddbenchTierTwoSize}{15\xspace}
\newcommand{\ddbenchTierThreeSize}{14\xspace}
\newcommand{\ddbenchReleaseSize}{60\xspace}
\newcommand{\ddbenchReleaseSysSize}{13\xspace}
\newcommand{\ddbenchNumSystems}{13\xspace}

\begin{document}

\title{DDBench: A Benchmark for Agentic Debugging on Distributed Systems}

\ifneurips
  \author{%
    Yibo Yan \quad Huijuan Wang \quad Junzhou He \quad Yizhuo Liang \\
    \textbf{Shaoyu Wang} \quad \textbf{Huanchen Sun} \quad \textbf{Seo Jin Park} \\
    University of Southern California \\
    \texttt{\{yiboyan, huijuanw, junzhouh, yizhuoli,} \\
    \texttt{wangshao, huanchen, seojinpa\}@usc.edu}
  }
\else\ifusenix
    \author{
      {\rm Author Name}\\
      University or Institution
    }
  \else
    \author{Author Name}
    \affiliation{
      \institution{University or Institution}
    }
    \email{author@example.com}
  \fi\fi

\ifacm
  \begin{abstract}
    LLM-based coding agents have advanced rapidly on single-process SWE tasks, with frontier models now clustering in the high-70s on SWE-bench Verified.
    Distributed-system debugging, however, remains an under-explored regime: bugs span processes, nodes, and protocol interactions, with root causes rarely recoverable from source alone and brute-force exploration intractable across non-deterministic interleavings.
    This leaves two gaps in LLM and agent evaluation: 
    no code-repair benchmark targets distributed-system bugs, and no controlled study isolates how much externally provided debugging context changes agent success on them.
    We introduce \sysname, a code-repair benchmark of \ddbenchReleaseSize{} historical bugs mined from \ddbenchReleaseSysSize{} open-source distributed systems, partitioned into three reasoning-scope tiers.
    \sysname evaluates every case under two matched conditions: a \emph{symptom-only} condition where the agent receives only the bug symptom and repository, and a \emph{context-augmented} condition where it additionally receives bounded debugging context (logs, traces, runtime state, and targeted code-investigation notes), isolating the effect of debugging context from model capability.
    The evaluation of \ddbenchNumModels{} LLMs on \sysname reveals several findings.
    First, distributed debugging exercises a reasoning dimension that single-process benchmarks do not surface: models' pass rates span 45~pp, and pairwise bootstrap separates 6 of 15 top-tier model pairs at $p<0.05$ on \sysname's hardest case-set.
    Second, bounded debugging context lifts aggregate pass rate by $+19.0$~pp, and the lift is asymmetric: weaker models gain more in pass rate, while stronger models gain more in efficiency.
    Third, debugging context requires careful curation, as even faithful debugging context can anchor a strong model on the wrong layer of the system when its observation sits far from the fault.
    \looseness=-1
\end{abstract}

\fi

\ifusenix\date{}\fi
\confmaketitle
\ifacm\else
  
\fi
\ifneurips\else
  \pagestyle{plain}
\fi


\section{Introduction}
\label{sec:intro}

Large Language Models (LLMs) and LLM-based coding agents have advanced rapidly on software engineering tasks~\cite{jin2025llms, yang2025swesmith, wang2025openhands, wei2025swerl, thai2026sweevo, deng2025swebench}.
Frontier models now routinely resolve multi-file bugs even through simple agent harnesses~\citep{xia2024agentless, yang2024sweagent}, and SWE-bench Verified~\citep{jimenez2023swebench}
provides an established test set and evaluation protocol for tracking continued progress in this regime.

Debugging distributed systems, however, remains an under-explored regime for LLM-based agents, and is substantially harder than single-process code repair.
A distributed-system bug rarely surfaces in a single stack trace as its root cause spans processes, nodes, and network paths.
Diagnosing such a bug requires reasoning about cross-process causality, non-deterministic interleavings, data races, and protocol-level invariant violations.
Consequently, the critical debugging context (information that helps the debugging process) is often hard to surface from static reasoning on the source code alone
making logs, traces, and runtime state snapshot valuable assets.
Nor can an agent reconstruct critical debugging context by brute-force enumeration and exploration, since a distributed bug often depends on the precise sequence of events across processes (\eg whether a timeout fires before or after a reply arrives), and the number of such sequences is intractable.
Therefore, existing LLM and agent benchmarks that target single-process repair tasks can poorly capture distributed debugging capability.
\looseness=-1

This leaves two concrete gaps in LLM and agent evaluation.
First, no code-repair benchmark targets distributed-system bugs.
Existing code-repair benchmarks (\eg SWE-bench~\citep{jimenez2023swebench}, DebugBench~\citep{tian2024debugbench}, and Multi-SWE-Bench~\citep{zan2025multiswebench}) focus primarily on single-process workloads.
IT-operations benchmarks such as AIOpsLab~\citep{chen2025aiopslab} and CloudOpsBench~\citep{wang2026cloudopsbench} reach across distributed services but evaluate operational diagnosis from telemetry rather than source-level code repair.
There is therefore no principled way to compare models on debugging distributed-system bugs at the source-level.
Second, no existing evaluation quantifies how much controlled debugging context (\eg logs, traces, diagnostic code exploration notes) changes an agent's success rate and behavior in solving these distributed-system bugs, relative to working from the bug symptom and codebase alone.
\looseness=-1

We introduce \sysname, a code-repair benchmark that contains \ddbenchReleaseSize{} historical bugs mined from \ddbenchReleaseSysSize{} open-source distributed systems. 
It is partitioned into three disjoint reasoning-scope tiers: \tierone{} (\ddbenchTierOneSize{} cases, the most challenging set), \tiertwo{} (\ddbenchTierTwoSize{} cases), and \tierthree{} (\ddbenchTierThreeSize{} cases).
\sysname therefore enables principled measurement of model capability on distributed-system debugging, which addresses the first gap.

Since brute-force exploration is prohibitively expensive in this regime, the value of externally-provided debugging context becomes measurable, making the context channel itself a natural experimental knob.
Therefore, \sysname is designed to run every case under two conditions:
\begin{inparaenum}[(i)]
    \item a \emph{symptom-only} condition, where the agent receives only the bug symptom and codebase and must surface any additional information through its own exploration;
    and
    \item a \emph{context-augmented} condition, where the agent additionally receives a curated debug-context bundle along with the bug symptom and codebase.
\end{inparaenum}
This design allows us to isolate the effect of additional debugging context from model capability, addressing the second gap by providing a controlled experimental setup.

We evaluated \ddbenchNumModels{} proprietary and open-weight LLMs with \texttt{mini-swe-agent}~\citep{yang2024sweagent} through \sysname's two-condition harness. 
Two major findings emerge:

\begin{wrapfigure}[21]{r}{0.48\linewidth}
    \vspace{-1.1em}
    \centering
    \includegraphics[width=\linewidth]{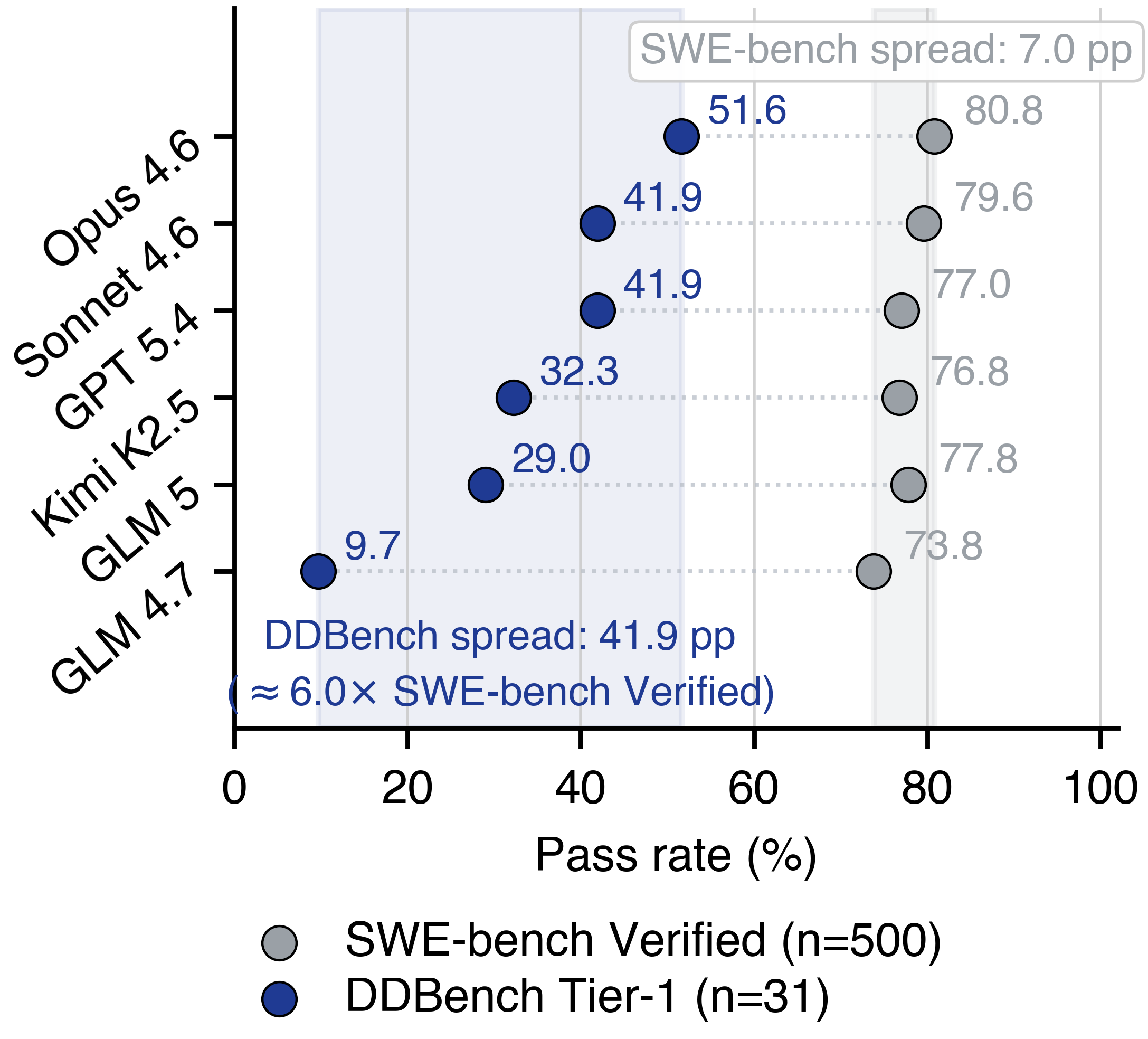}
    \caption{\sysname sharply differentiates models on distributed debugging. Under the \emph{symptom-only} condition, six models span 42~pp on \sysname\xspace \tierone{} ($n{=}31$), compared with only 7~pp on SWE-bench Verified ($n{=}500$). 
    }
    \label{fig:swe-vs-dd-dumbbell}
\end{wrapfigure}

\noindent\textbf{(1) Cross-model separation on distributed debugging.}
Under the \emph{symptom-only} condition on \tierone{} case set, \ddbenchNumModels{} models spanned 45~pp in pass rate.
Pairwise bootstrap tests ($n=\ddbenchTierOneSize{}$ cases) distinguished 6 of 15 top-tier model pairs at $p<0.05$.
For the six models we sampled with publicly reported SWE-bench Verified scores\footnote{We sampled Claude Opus~4.6, Claude Sonnet~4.6, GPT~5.4, GLM~4.7, GLM~5, and Kimi~K2.5. We obtained the score for GPT-5.4 (high thinking) from Epoch AI as OpenAI did not officially release the number; other scores come from model vendors' official reports.}, the same models span only 7~pp on SWE-bench but 42~pp on \sysname\xspace \tierone{} (\figref{fig:swe-vs-dd-dumbbell}).
The contrast is consistent with the fact that distributed debugging exercises a reasoning dimension that is not directly measurable by single-process code-repair tasks.

\noindent\textbf{(2) Additional debugging context lifts pass-rate unevenly and cuts token cost consistently.}
Adding curated debugging context lifted aggregate pass rate from 31.0\% to 50.0\% (+19.0~pp).
The gains break down differently across models. 
Weaker models gained primarily in pass rate (GLM~4.7: 9.7\% $\rightarrow$ 48.4\%). 
In contrast, stronger models saw smaller pass-rate gains but larger efficiency gains (Opus~4.6: +12.9~pp in pass rate, with 61\% lower token consumption and 39\% fewer steps-to-completion).
The gain of adding debugging context was also large enough to reorder model ranking. Specifically, context-augmented GPT~5.4 exceeded Opus~4.6's symptom-only pass-rate (67.7\% vs.\ 51.6\%) while consuming 70\% fewer tokens.
Furthermore, we observed that the effect of debugging context is not uniform across cases: a faithful debugging context (\eg the failing output of a test that reproduces the bug) can anchor even a strong model on the wrong layer when its observation sits far from the fault (\secref{sec:eval:case-studies}), revealing the hidden risk of adding more information without careful curation.

The two findings reveal that the same regime that makes curated debugging context valuable also spreads models apart, because the agents that suffer most from missing critical debugging information that cannot be easily compensated through expensive exploration and reasoning.
\sysname therefore measures not only whether an agent can fix a distributed bug, but also how much adding debugging context helps.
One prior work shows that the context curated via lightweight pre-exploration of the target codebase can substantially improve the resolution rate~\cite{suri2026codescout}; following this same vein, \sysname concretely anchors this insight in the distributed debugging regime.
Thereby, we summarize our contributions as follows:
\begin{itemize}[leftmargin=1.5em]
    \item \textbf{\sysname}: a benchmark for agentic distributed-system debugging, comprising \ddbenchReleaseSize{} cases from \ddbenchReleaseSysSize{} open-source systems, each independently reproduced, sanitized against leakage, and reviewed by five distributed-systems researchers. The size reflects this quality bar, not a scale ceiling.
    \item \textbf{Model differentiation on distributed debugging}: an evaluation of \ddbenchNumModels{} models that yields a spread of 45~pp on \sysname's hardest case set and separates 6 of 15 model pairs, indicating that distributed debugging exercises a reasoning dimension different from single-process code repair.\looseness=-1
    \item \textbf{The effect of debugging context on agent performance}: a symptom-only vs.\ context-augmented study that isolates debugging context from model capability. Bounded context disproportionately lifts models' pass rate while uniformly reducing cost across all models.
    \item \textbf{An extensible testbed for agent-tool research}: \sysname is agent-agnostic, and its debugging context channel is a drop-in slot for alternative debugging tools (\eg tracers, static analyzers, code-search tools), enabling head-to-head comparison of future agentic tool proposals on the same cases and the same capability baseline.
\end{itemize}

The dataset and benchmark harness is available at: \href{https://github.com/USC-NSL/DDBench}{https://github.com/USC-NSL/DDBench}.

\section{\sysname}
\label{sec:benchmark}

\sysname is a code-repair benchmark containing \ddbenchReleaseSize{} historical bugs from \ddbenchNumSystems{} open-source distributed systems, partitioned into three disjoint reasoning-scope tiers (\tierone{}/\tiertwo{}/\tierthree{}) ordered from the widest to the narrowest scope of reasoning that diagnosis requires.
For each case, we localize the faulty commit from the upstream issues and PRs.
Each case ships with a \emph{git repository} pinned to the faulty commit with no later history, a \emph{symptom description} of the fault (\texttt{SYMPTOM.md}) derived from the original GitHub issue/PR, a hidden \emph{oracle} returning PASS/FAIL on the patched codebase, and a \emph{debug-context} bundle (\secref{sec:debug-context}).
\sysname is agent-agnostic, so that any agent that can read a repository and propose a patch can be evaluated.

\begin{figure}[t]
    \centering
    \includegraphics[width=\linewidth]{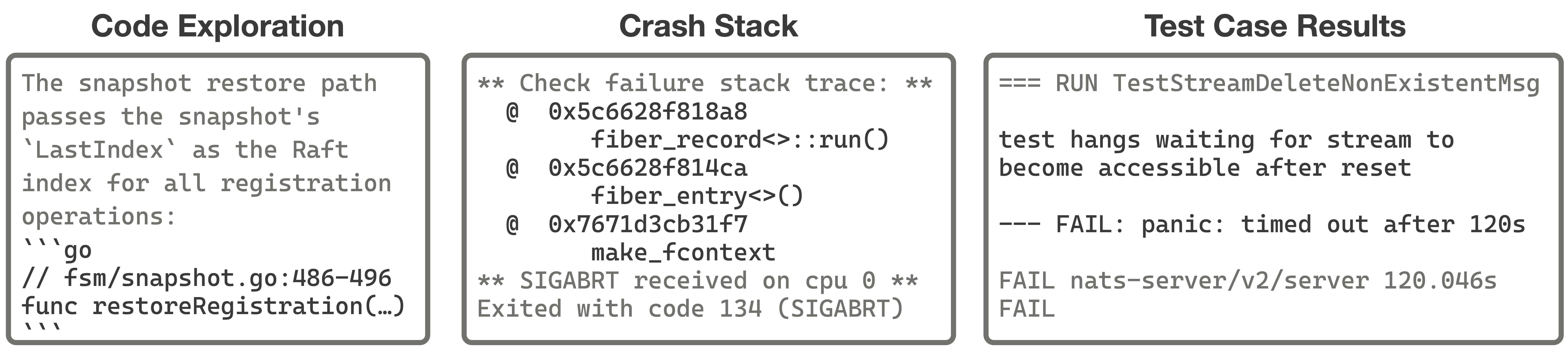}
    \caption{Examples of debug-context files shipped in \sysname. 
    }
    \label{fig:debug-context}
\end{figure}

\subsection{Debug Context: What It Is and How It Is Supplied}
\label{sec:debug-context}

Symptom description merely describes the observation of the fault without providing additional debugging information, such as logs, traces, or targeted code exploration on the faulty code.
We treat such additional debugging information an agent receives, referred as \textit{debug-context}, as a first-class experimental variable.
A {debug-context} is any bounded, per-case artifact that a debugging tool could realistically produce for developers.
This covers both \emph{runtime signals} (\eg process and network logs, stack traces, race-detector reports, partial runtime-state snapshots) and \emph{static-but-targeted code exploration} on the fault (\eg the suspicious call sites and cross-file relationships a static analyzer would surface).
Anything a tracer, debugger, invariant checker, or static code analyzer could generate is admissible; however, the fix itself and any direct pointer to it are excluded.
In this \sysname release, each case's debug context bundle contains one to three short Markdown files, median 321 estimated tokens (max 1{,}403) for models to consume.
Every bundle in this release was generated by Claude Code driving Claude Opus~4.6, following the generation skill in Appendix~\secref{appendix:prompt-context}.
This choice does not appear to favor the Claude family: on \tierone{}, the bundle lifts Opus~4.6 by 12.9~pp and Sonnet~4.6 by 16.1~pp, at or below the cohort median of 16.1~pp, while GPT~5.4 gains 25.8~pp~(\secref{sec:eval:context-effect}).
\figref{fig:debug-context} describes three examples of debug-context files included in this release of \sysname.

\begin{figure}[t]
    \centering
    \includegraphics[width=0.98\linewidth]{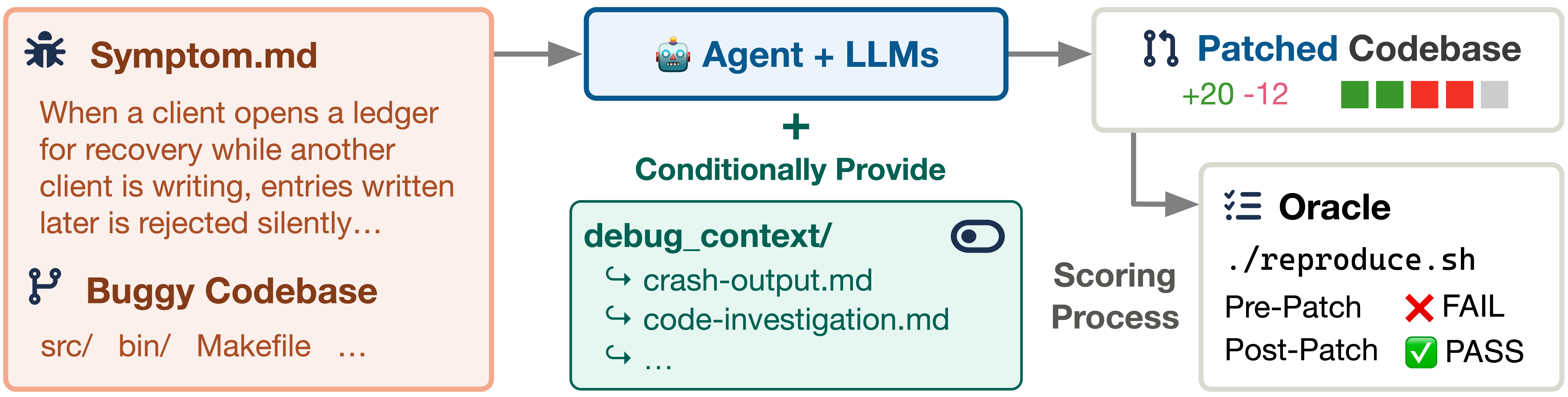}
    \caption{The flow of evaluating a \sysname bug case. The buggy codebase and agent are containerized together, and the agent can access the codebase and a Markdown symptom description. In the context-augmented condition, the agent also has access to a debug-context bundle (green). After the agent concludes its patch, a hidden oracle runs against the codebase and returns PASS/FAIL as the final evaluation.}
    \label{fig:benchmark_overview}
\end{figure}

\parab{Evaluation procedure.}
The evaluation procedure (\figref{fig:benchmark_overview}) is a two-condition test, where the agent diagnoses and repairs the same case with and without the debug-context bundle.
In the symptom-only condition, the agent's container is mounted with the faulty codebase and \texttt{SYMPTOM.md}.
The agent is free to read the repository, run commands, and invoke tools its scaffold provides to gather additional context on its own.
In the context-augmented condition, the same agent on the same case additionally receives the debug-context bundle, mounted in the containerized codebase with additional prompts to use it.
Agents have the same tool call access in both conditions.

\parab{A parameterized extensibility slot.}
The debug-context bundle is designed to be a drop-in slot. 
Namely, a researcher who builds a new structured tracer, debugger, invariant checker, or code-search aid can re-populate debug-context files for cases from their tool's output and evaluate that tool against any \sysname-compatible agent without re-engineering the benchmark harness.
What we ship is therefore just one default bundle; alternative bundles produced by other tools can be dropped into the same slot and compared head-to-head on identical cases and oracles.
The slot establishes \emph{which debug context an agent sees} as a first-class design axis for agent tool augmentation.

\begin{figure}[h]
    \centering
    \includegraphics[width=0.98\linewidth]{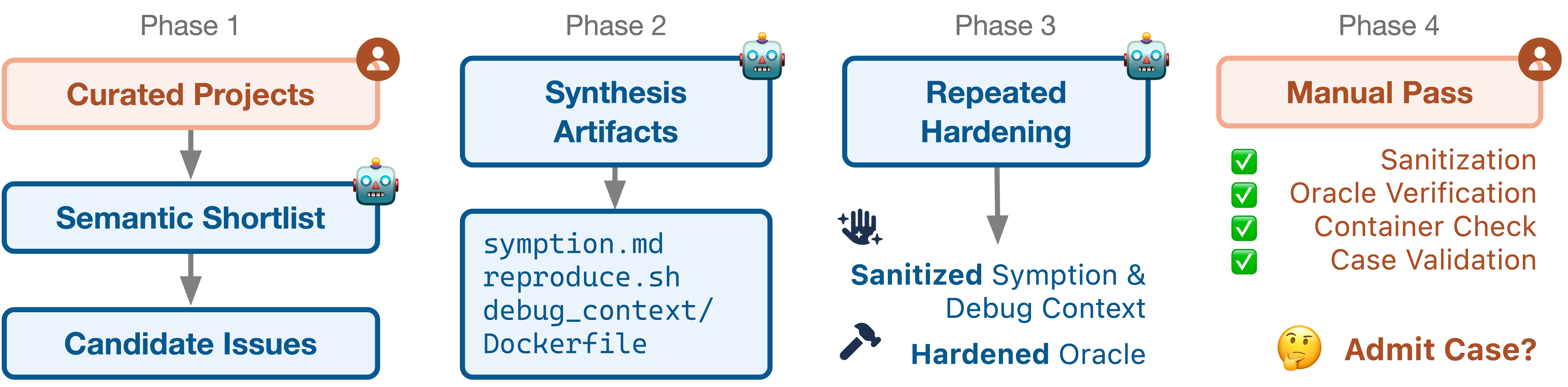}
    \caption{The \sysname curation pipeline, where an LLM agent performs first-pass mining and artifact synthesis under tight specifications, and human researchers perform final sanitization, validation, and acceptance.}
    \label{fig:benchmark_construction}
\end{figure}

\subsection{Dataset Construction}

We construct \sysname through the semi-agentic pipeline (\figref{fig:benchmark_construction}):
an LLM agent performs first-pass mining (phase 1), artifact synthesis (phase 2), and iterative hardening (phase 3) under tightly scoped specifications, and human researchers perform final validation, sanitization, and acceptance (phase 4).
A candidate is admitted only if (a) the bug was confirmed and fixed upstream and (b) its diagnosis demonstrably requires cross-process or interleaving-aware reasoning.
Purely manual curation does not scale to the per-system protocol expertise each case demands, and naive issue scraping cannot discriminate cross-process bugs from single-process ones; the agent is what makes this admission bar enforceable at scale.
The mining protocol, symptom and debug-context synthesis, and answer-leakage sanitization are deferred to Appendix~\secref{appendix:semi-agentic}.

\parab{Oracle reliability and manual review.}
Each oracle is iteratively hardened by an agent with a specialized prompt until it reliably reports \texttt{FAIL} on the faulty commit and \texttt{PASS} on a correctly patched codebase.
Cases that fail to stabilize after five hardening iterations are dropped.
Oracles accept any patch under which the reproducer reports \texttt{PASS}, not just patches that match the upstream fix textually.
A fix-specificity scan additionally flags fix-overfit oracles (\eg ones that pattern-match a specific code substring or use reflection-based assertions) for manual hardening (Appendix~\secref{appendix:prompt-verify-oracle}).
Five distributed-systems researchers then manually review every surviving case for quality and re-check symptom and debug-context sanitization on top of the automated passes before final admission.

\parab{Dataset tiering.}
We divide all cases into three mutually exclusive reasoning-scope tiers (tier-1, tier-2, tier-3), ordered by how far the reasoning needed to diagnose the bug must reach: \tierone{} cases require reasoning over distributed protocol state and concurrent interleavings, \tiertwo{} cases span multi-process state, and \tierthree{} cases localize to a single process and a triggering input.
Tier-1 is the most challenging set, as it concentrates the hardest cross-process, undeterministic timings reasoning scenarios.
The demanded reasoning load decreases from tier-1 to tier-3.
Tier-3 cases are bugs in distributed systems that can be localized to single-process misbehaviors without much complex cross-process reasoning.
Both tier-2 and tier-3 enable us to understand how tier-1 patterns generalize as the reasoning scope narrows.
\looseness=-1

\begin{figure}[t]
    \centering
    \includegraphics[width=\columnwidth]{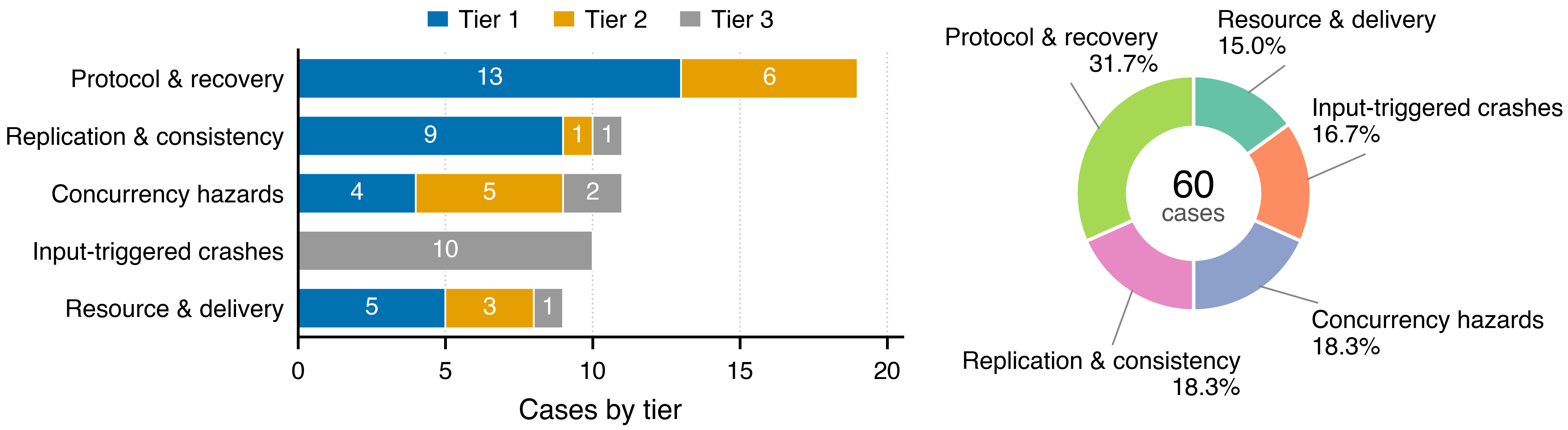}
    \caption{Bug-family composition of \sysname. \textbf{Left}: per-tier breakdown where tier-1 concentrates protocol/recovery and replication/consistency bugs; tier-3 absorbs all input-triggered crashes, which can be localized within a single process; tier-2 bridges the two. \textbf{Right}: each family's overall share of the benchmark.}
    \label{fig:bug-family}
\end{figure}

\subsection{Features}
\label{sec:benchmark-features}

\sysname covers these system categories: key-value stores (22 cases), consensus libraries (21), messaging systems (10), service meshes (3), streaming engines (3), and a storage engine (1).
These cases cover five languages: Go (25), C++ (23), Java (8), Erlang (2), and Rust (2).

\parab{Bug families.}
We group the \ddbenchReleaseSize{} cases into five families that reflect the reasoning each bug demands rather than its surface symptom (\figref{fig:bug-family}).
\emph{Protocol and recovery failures} (31.7\%) are the largest family: bugs in selecting a coordinator (leader election), copying state across the cluster (log replication), re-syncing a recovering node (snapshot transfer), and changing node membership (cluster reconfiguration), where correctness depends on multi-step protocol invariants holding across nodes.
\emph{Concurrency hazards} (18.3\%) cover data races, deadlocks, and livelocks that manifest only under specific runtime interleavings.
\emph{Replication and consistency violations} (18.3\%) produce divergent state across replicas, \eg operations that appear out of order to clients (linearizability violations), and lost updates.
\emph{Input-triggered crashes} (16.7\%) and \emph{resource and delivery anomalies} (15.0\%) account for the remaining cases.
The first three families require reasoning that cannot be discharged from a single process trace.
In other words, the LLM and agent must understand state across nodes or across non-deterministic schedules to reason and localize the fault.
The tier breakdown in \figref{fig:bug-family} (left) shows that tier-1 concentrates the cross-process families (13 of 19 protocol/recovery cases, 9 of 11 replication cases), while every input-triggered crash falls in tier-3.
The tier-1 headline set is therefore not merely a harder version of the same task, but it is structurally different in the reasoning it requires.

See Appendix~\secref{appendix:dataset-details} and \secref{appendix:bench-features} for more dataset details, including the debug-context-bundle-size distribution and a quantitative analysis of \sysname's reasoning difficulty relative to patch size.


\section{Evaluation}
\label{sec:eval}

We evaluate \sysname along three axes: 
\begin{inparaenum}[(1)]
    \item 
    whether distributed-system bugs differentiate frontier models that existing benchmarks compress together, 
    \item 
    whether bounded debug-context improves agent diagnosis success rate and at what cost, and 
    \item
    how these effects vary by bug family, reasoning-scope tier, and individual case.
\end{inparaenum}

\subsection{Setup}
\label{sec:eval:setup}

\parab{Models.}
We evaluated \ddbenchNumModels models, spanning proprietary and open-weight models: Claude Opus~4.6, Claude Sonnet~4.6, GPT~5.4, GPT~5.4~mini, GLM~5.1~(FP8, self-hosted), GLM~5, GLM~4.7, Gemma~4~(31B, self-hosted), Kimi~K2.5, and GPT~OSS~(120B~BF16, self-hosted).
For each case, we use an identical agent scaffold with tool-use access to the repository.
The agent being evaluated runs inside the isolated container and is blind to the test oracle.
We bound the agent's max steps to 450 per case, max token cost to \$5 per case, and max time to 30 minutes per case.

\parab{Agent.}
We used \texttt{mini-swe-agent}~\citep{2026sweagent} (the successor of \texttt{SWE-agent}~\citep{yang2024sweagent}) as the agent scaffold, as it provides a structured approach to collect token usage, action trajectories, and intermediate reasoning steps.
\texttt{mini-swe-agent} has proven effective in solving complex SWE problems~\citep{yang2024sweagent, fan2025sweeffi, deng2025swebench, thai2026sweevo} and has been widely adopted~\citep{sun2026sweworld, yuan2026sweminisandbox, xia2025livesweagent, merrill2025terminalbench}, making it a suitable choice for our evaluation.
\looseness=-1

\parab{Metrics and statistical methodology.}
We report pass rate (fraction of cases where the agent's patch passes the hidden reproduce oracle), agent steps, first-edit step, total tokens consumed, and action-type distributions.
Every case in \tierone{} is evaluated by every model under both conditions, yielding 620 binary outcomes (\ddbenchNumModels{} models $\times$ \ddbenchTierOneSize{} cases $\times$ 2 conditions) with two structural axes: per-case difficulty and per-model capability. Our statistical tests are chosen to respect this structure.

\emph{Pairwise model differentiation.} We performed a pairwise bootstrap test for six models that have both publicly available SWE-bench Verified scores and also evaluated in our study (Opus~4.6, Sonnet~4.6, GPT~5.4, GLM~5, Kimi~K2.5, and GLM~4.7). 
For each of the $\binom{6}{2}=15$ model pairs,
we perform a paired bootstrap over the \ddbenchTierOneSize{} cases (10{,}000 resamples) to estimate the pass-rate difference with 95\% CI.
The SWE-bench Verified panel is computed analogously, but with an unpaired bootstrap over 500 per-case outcomes per model whose pass count equals the vendor-reported rate (\eg 404 of 500 for Opus~4.6), as per-case outcomes are not publicly released for all six models we sampled.


\emph{Aggregate debug-context effect across models.} We fit a mixed-effects logistic regression $\Pr(\text{pass}) = \sigma(\beta_0 + \beta_{\text{ctx}} \cdot \mathbf{1}[\text{debug-context}] + u_{\text{model}} + u_{\text{case}})$ with random intercepts for model and case, so that the debug-context effect is estimated after controlling for baseline capability ($u_{\text{model}}$) and case difficulty ($u_{\text{case}}$).
For the debug-context coefficient we report its estimate $\hat{\beta}_{\text{ctx}}$ with standard error (SE, the estimator's sampling-distribution standard deviation), 95\% CI, odds ratio (OR $= e^{\hat{\beta}_{\text{ctx}}}$, the multiplicative effect of debug-context on the odds of pass), and Wald $z$-statistic ($z = \hat{\beta}/\text{SE}$, asymptotically standard normal under the null).
We prefer this toolkit over resampling-only alternatives (e.g., a bootstrap over the 10-model population) because 10 units are too few to estimate between-model heterogeneity reliably; the mixed model estimates it parametrically from all 620 outcomes.

\subsection{Models Cluster on Single-Process Code-Repair but Separate on \sysname}
\label{sec:eval:differentiation}


We anchor the differentiation analysis on \tierone because it concentrates the cross-process reasoning load \sysname is built to measure: bugs whose diagnosis hinges on causality across processes, concurrency, and protocol invariants.
On the other hand, \tiertwo and \tierthree relax this load and serve as robustness checks (\secref{sec:eval:tiers}) rather than primary differentiation evidence.

\tabref{tab:headline} reports the per-model pass rates on \tierone under both conditions.
Under the symptom-only condition, the cohort spans 45~pp, with Claude Opus~4.6 at the top (51.6\%) and GPT~OSS~(120B) at the bottom (6.5\%).
To check that this spread reflects a reasoning dimension SWE-bench does not exercise (rather than a different point on the same dimension), we contrast the \sysname results with SWE-bench Verified~\citep{jimenez2023swebench}.
Of the \ddbenchNumModels{} models we evaluate, six have published SWE-bench Verified scores: Claude Opus~4.6~(80.8\%), Claude Sonnet~4.6~(79.6\%), GLM~5~(77.8\%), GPT~5.4~(77.0\%), Kimi~K2.5~(76.8\%), and GLM~4.7~(73.8\%).
These six models cluster within a 7-pp band on SWE-bench Verified, but span 42~pp (9.7\%--51.6\%) on \sysname\xspace \tierone case-set.
The wide spread is not an artifact of including weak models. Specifically, Claude Opus~4.6 and GLM~5 are 22.6~pp apart on \sysname ($p=0.001$) despite being only 3.0~pp apart on SWE-bench Verified ($p=0.26$).

We quantify this with a pairwise bootstrap test over the $\binom{6}{2}=15$ model pairs (the full matrix of results can be found in Appendix~\secref{appendix:pairwise-models}).
Significance concentrates on pairings involving GLM~4.7 (5 pairs) or Opus~4.6 (1 additional pair).
The two pairs SWE-bench Verified resolves (Opus~4.6 and Sonnet~4.6, each vs.\ GLM~4.7) are both contained in \sysname's six and involve GLM~4.7, the model with the lowest single-process score; the remaining four \sysname-resolved pairs cluster tightly on SWE-bench Verified yet separate sharply once cross-process reasoning is required, exposing a capability axis the single-process benchmark does not measure.
\looseness=-1

\sysname shows that ``frontier model'' is not a homogeneous category on the cross-process reasoning axis. Concretely, models clustered tightly on single-process code repair benchmarks diverge substantially on distributed debugging tasks that require reasoning about cross-process causality, non-deterministic interleavings, and protocol invariants.

\parab{Contamination check.}
Every \sysname case is a public, fixed issue, so models may have seen the upstream fix during pretraining.
We compared all 204 passing agent patches with the upstream fix commits: among the 155 fully parsed patches that share at least one line with the fix, a Tree-sitter comparison finds only 5 (3.2\%) that contain the upstream fix, while 38 (24.5\%) are logically equivalent but not identical.
A manual audit of the 53 patches closest to the upstream fix traced 47 to six cases whose bug admits few natural fixes.
Under the symptom-only condition, pass rate tracks neither bug recency (Spearman $\rho=-0.109$ against fix-commit date) nor repository popularity ($\rho=+0.014$ against GitHub stars).
We therefore find no strong evidence of memorization~(Appendix~\secref{appendix:contamination}).

\begin{table}[t]
    \caption{Per-model pass rates (\%) on \tierone (\ddbenchTierOneSize cases) under symptom-only and context-augmented conditions. $\Delta$ denotes the percentage-point improvement. All ten models improved; none regressed.}
    \label{tab:headline}
    \centering
    \resizebox{\textwidth}{!}{%
        \begin{tabular}{@{}lll|rrr|rrr@{}}
            \toprule
            \multirow{2}{*}{\textbf{Model}}               &
            \multirow{2}{*}{\textbf{Thinking}}            &
            \multirow{2}{*}{\textbf{Form}}                &
            \multicolumn{3}{c|}{\textbf{Total Cost (\$)}} &
            \multicolumn{3}{c}{\textbf{Pass Rate \%}}                                                                                         \\
                                                          &
                                                          &
                                                          &
            Sym.                                  &
            +Context                                      &
            \multicolumn{1}{l|}{$\Delta$ (\%)}            &
            Sym.                                  &
            +Context                                      &
            $\Delta$                                                                                                                          \\ \midrule
            Claude Opus 4.6                               & Adaptive High & API             & 72.5 & 32.6 & -\,55.1\% & 51.6 & 64.5 & +\,12.9     \\
            Claude Sonnet 4.6                             & Adaptive High & API             & 71.8 & 35.8 & -\,50.1\% & 41.9 & 58.1 & +\,16.1 \\
            GPT 5.4                                       & High          & API             & 25.0 & 21.4 & -\,14.3\% & 41.9 & 67.7 & +\,25.8 \\
            GPT 5.4 mini                                  & High          & API             & 7.9  & 6.1  & -\,23.5\% & 45.2 & 61.3 & +\,16.1 \\
            GLM 5.1 (FP8)                                 & Enabled       & Open, Self-Host & 23.1 & 14.6 & -\,36.7\% & 32.3 & 48.4 & +\,16.1 \\
            GLM 5                                         & Enabled       & Open, API       & 42.3 & 30.3 & -\,28.5\% & 29.0 & 45.2 & +\,16.1 \\
            GLM 4.7                                       & Enabled       & Open, API       & 38.5 & 18.7 & -\,51.5\% & 9.7  & 48.4 & +\,38.7 \\
            Gemma 4 (31B)                                 & Enabled       & Open, Self-Host & 4.9  & 4.5  & -\,8.6\%  & 19.4 & 38.7 & +\,19.4 \\
            Kimi K2.5                                     & N/A           & Open, API       & 11.2 & 7.5  & -\,32.6\% & 32.3 & 38.7 & +\,6.5  \\
            GPT OSS (120B BF16)                           & Enabled       & Open, Self-Host & 3.7  & 3.3  & -\,11.5\% & 6.5  & 29.0 & +\,22.6 \\ \bottomrule
        \end{tabular}%
    }
\end{table}

\subsection{Effect of Debug-Context on Agent Performance}
\label{sec:eval:context-effect}

Across all \ddbenchNumModels{} models on \tierone, the context-augmented condition raised the aggregate pass-rate from 31.0\% to 50.0\% (+19.0~pp).
The lift is broad and uniform in direction: all ten models improved and none regressed.
Across the 310 paired runs, the debug-context flipped 72 cases from \texttt{FAIL} to \texttt{PASS} and only 13 from \texttt{PASS} to \texttt{FAIL}, with the remaining 225 unchanged, a 5.5:1 ratio of gains to regressions (Appendix~\secref{appendix:outcome-matrix}).

\begin{figure}[t]
    \centering
    \begin{subfigure}[b]{0.52\linewidth}
        \centering
        \includegraphics[width=\linewidth]{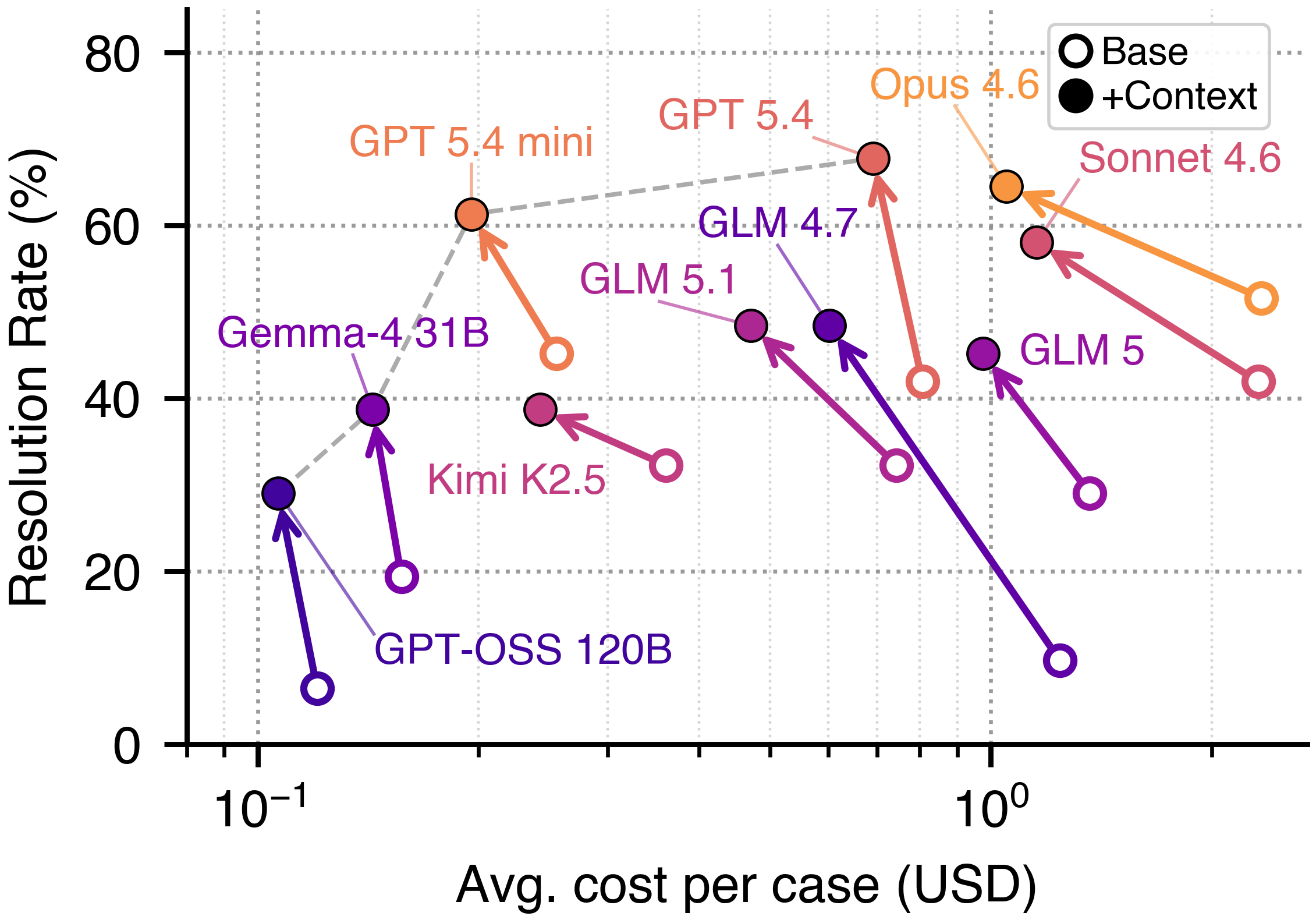}
        \caption{Capability--cost Pareto (log-scale cost).}
        \label{fig:pareto}
    \end{subfigure}\hfill
    \begin{subfigure}[b]{0.46\linewidth}
        \centering
        \includegraphics[width=\linewidth]{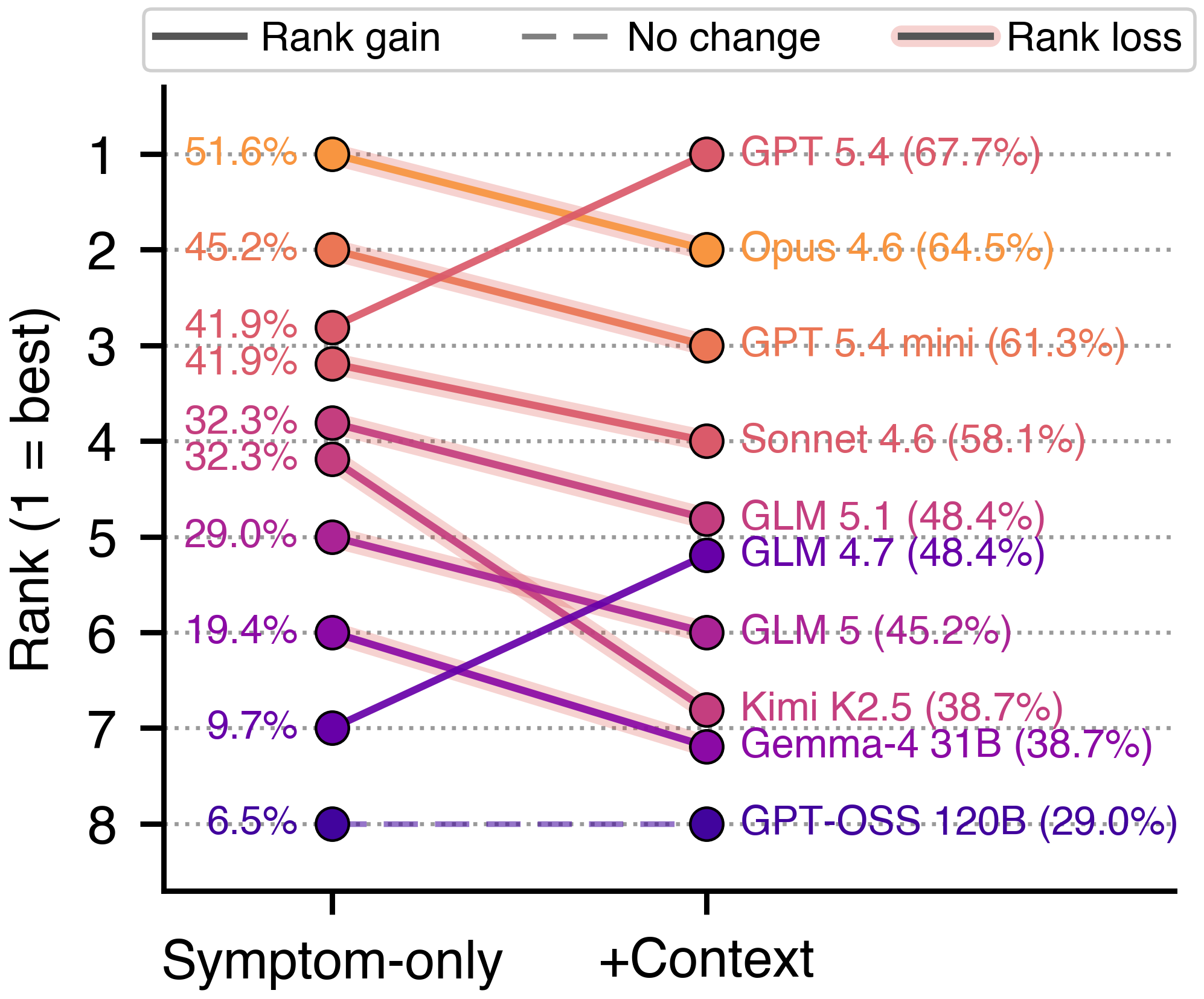}
        \caption{Rank changes on \tierone case-set.}
        \label{fig:bump}
    \end{subfigure}
    \caption{Debug-context reshapes both the capability--cost Pareto and the capability ranking on \tierone. \textbf{(a)} Each arrow traces a model from symptom-only (hollow) to context-augmented (solid). Weaker models' arrows point up (capability gain) and stronger models' arrows point further left (larger cost reduction, smaller pass-rate gain). \textbf{(b)} Solid lines denote rank gains, dashed lines no change, and lines with a soft red halo denote losses; ties share a rank with vertical jitter. The presence of debug-context reshuffles the leaderboard.}
    \label{fig:pareto-bump}
\end{figure}

\parab{Pass-rate gains and efficiency gains divide along model baseline capability.}
Weaker models gain more in pass rate, while stronger models gain more in efficiency: the four strongest models under the symptom-only condition gain +17.7~pp on average and cut per-case tokens by 44.5\%, versus +24.2~pp and 25.2\% for the four weakest.
More specifically,
GLM~4.7 rose from 9.7\% to 48.4\% (+38.7~pp) and GPT~OSS from 6.5\% to 29.0\% (+22.6~pp), the largest absolute pass-rate gains in the cohort.
Claude Opus~4.6, by contrast, showed a smaller, non-significant pass-rate gain (+12.9~pp, $p=0.13$); its token consumption dropped from 3.11M to 1.21M ($-$61\%) and its steps-to-completion dropped from 67.4 to 41.1 ($-$39\%) (\secref{sec:eval:trajectory}).
Mid-range models gained 16--19~pp in pass-rate, with GPT~5.4 the notable exception, gaining +25.8~pp from a 41.9\% baseline to 67.7\%.
Per-model breakdowns appear in \tabref{tab:headline}.
This asymmetry tracks how models cope with missing context.
Strong models have the capacity to compensate by exploring more aggressively and reasoning longer; adding context removes that exploration overhead with a comparatively small pass-rate gain.
Weak models lack that capacity, so a narrowed search space converts directly into solved cases.


\parab{Debug-context can reorder the leaderboard at fixed budget.}
A weaker model with debug-context can outperform a stronger model without it on both axes simultaneously. For instance, GLM~4.7 augmented with context reaches 48.4\% pass rate at \$0.60 per case, beating Sonnet~4.6 in the symptom-only condition (41.9\% at \$2.31) on both pass rate and cost.
The Pareto plot in \figref{fig:pareto} traces this geometry across the cohort: each model's arrow points along its dominant gain, mostly capability for weak models and mostly cost-efficiency for strong models.
The rank diagram in \figref{fig:bump} shows the practical consequence. Concretely, GLM~4.7 climbs from rank 7 to tied 5, and GPT~5.4 overtakes Opus~4.6 for first.
For agent-tool developers, these findings suggest that a well-designed debug-context can lift a weaker, cheaper model close to the operating point of a stronger, more expensive one, making tool design a competitive lever against model upgrades.

\begin{wraptable}[10]{r}{0.53\linewidth}
    \vspace{-1.0em}
    \footnotesize
    \centering
    \caption{Pass-rate lift by bug category on \tierone, pooled across \ddbenchNumModels{} models. ``Input-triggered crashes'' is omitted because its cases reside exclusively in \tierthree.}
    \label{tab:bug-family-lift}
    \begin{tabular}{@{}lrrr@{}}
        \toprule
        \textbf{Bug Category}               & \textbf{Sym.} & \textbf{+Ctx.} & $\boldsymbol{\Delta}$ \textbf{(pp)} \\ \midrule
        Concurrency hazards (n=4)           & 30.0\% & 75.0\% & +45.0 \\
        Replication \& consistency (n=9)    & 36.7\% & 57.8\% & +21.1 \\
        Protocol \& recovery (n=13)         & 27.7\% & 39.2\% & +11.5 \\
        Resource \& delivery (n=5)          & 30.0\% & 44.0\% & +14.0 \\ \bottomrule
    \end{tabular}
\end{wraptable}

\parab{Effect by bug family and the abstraction-distance principle.}
\label{sec:eval:case-studies}
\tabref{tab:bug-family-lift} breaks down the effect by bug category on \tierone, revealing how the lift tracks the size of the diagnostic search space that debug-context helps to narrow.
Concurrency hazards category has the biggest pass-rate lift (+45.0~pp), showing that interleavings observable in stack traces and race-detector reports can substantially reduce the search space. The replication and consistency violation category (state divergence visible across replica logs) gain +21.1~pp. However, the protocol and recovery failure category gain only +11.5~pp, suggesting the protocol-level failures might require a more sophisticated approach to curate a more effective debug-context.

\parab{When can it mislead?} 
Despite the general effectiveness of debug-context, per-case trajectories reveal that it can also steer the agent toward the wrong hypothesis.
We performed a case study on three \tierone cases (\texttt{ZooKeeper-4846}, \texttt{Raft-268}, \texttt{Kvrocks-3356}) in Appendix~\secref{app:case-studies} and observed three regimes governed by the \emph{abstraction distance}, \ie the number of causal steps that separate what the agent observes (the symptom and the debug-context bundle) from the code location where the bug resides:
when this distance is small, debug-context either supplies a runtime-only signal that code reading cannot recover or acts as a search-space reducer;
however, when the gap is large, the additional debug-context anchors the agent on the wrong layer of the system, even for strong models.
For instance, in \texttt{Kvrocks-3356} the debug context faithfully reports the failing test's measurement (disk growth on the master above a 512\,MB threshold), while the fault is a replication thread blocked on a socket write; Claude~Opus~4.6 anchored on the match between that threshold and Kvrocks' default 512\,MB write-ahead-log size limit, and spent 45 of its 205 steps reading the storage engine's source rather than patching the replication thread.
These cases suggest abstraction distance as a key axis along which debug-context for LLM-based debugging agents should be designed.

\parab{Statistical significance.}
Per-model paired bootstraps (10{,}000 resamples) yield positive debug-context effects for all 10 models, with 6 clearing $p < 0.05$ individually at $n = \ddbenchTierOneSize{}$ (Sonnet~4.6, GPT~5.4, GLM~5.1, GLM~4.7, Gemma-4 31B, GPT-OSS 120B).
The remaining four models show positive point estimates whose bootstrap CIs include zero, as expected given limited per-model power.
\begin{wraptable}[12]{r}{0.46\linewidth}
    \vspace{-0.2em}
    \footnotesize
    \centering
    \caption{Behavioral metrics across 310 paired runs on \tierone. Debug-context reduces exploration effort consistently while edit count remains stable.}
    \label{tab:trajectory}
    \begin{tabular}{@{}lrrr@{}}
        \toprule
        \textbf{Metric}      & \textbf{Sym.} & \textbf{+Ctx.} & $\boldsymbol{\Delta}$ \\ \midrule
        Mean steps           & 57.6  & 43.1  & $-$14.4 \\
        Mean first-edit step & 32.3  & 20.8  & $-$11.5 \\
        Mean tokens          & 1.79M & 0.99M & $-$0.80M \\
        Mean read actions    & 24.7  & 17.7  & $-$6.9  \\
        Mean explore actions & 10.9  & 6.8   & $-$4.2  \\
        Mean edit count      & 3.0   & 3.1   & +0.1    \\ \bottomrule
    \end{tabular}
\end{wraptable}
The mixed-effects logistic regression (introduced early in \secref{sec:eval:setup}) on all 620 outcomes gives $\hat{\beta}_{\text{ctx}} = +1.44$ (SE $0.15$, 95\% CI $[+1.14, +1.74]$; OR $4.22$; $z = 9.31$, $p < 10^{-17}$), with an average marginal effect of $\Delta = +18.9$~pp matching the raw $+19.0$~pp lift.
The variance components carry two further messages.
First, case difficulty dominates model identity as a source of outcome variation ($\hat{\sigma}^2_{\text{case}} = 4.62$ vs.\ $\hat{\sigma}^2_{\text{model}} = 0.93$ on the log-odds scale), indicating \sysname{}'s signal comes from genuine per-case difficulty, not from noisy model rankings.
Second, $\hat{\beta}_{\text{ctx}}$ is unchanged when these two random intercepts are included, so debug-context helps across the full difficulty distribution rather than only on easy cases.
\looseness=-1

\subsection{Debug-Context as a Diagnostic Accelerator}
\label{sec:eval:trajectory}

Beyond pass rate, debug-context reshapes how agents debug (\tabref{tab:trajectory}): per-case steps, first-edit time, tokens, and exploratory reads drop by 25--45\%, while edit count is unchanged.
We also observe that the narrowing is over the diagnostic search space rather than over the fix itself, and the savings persist regardless of outcome.
Concretely, our analysis shows 11.4 fewer steps on concordant-PASS pairs (83~cases), 9.0 on concordant-FAIL (142), and 28.7 on FAIL$\to$PASS flips (72).
Claude Opus~4.6, the most expensive model evaluated, dropped from 3.11M to 1.21M tokens per case ($-$61\%) while its pass rate rose by 12.9~pp.
Across the cohort, debug-context roughly halved the mean token consumption without loss in diagnostic quality.
The agent reaches the relevant code region faster and wastes less effort on dead ends. Overall, debug-context acts simultaneously as a capability booster for weaker models and a cost reducer for stronger ones.

We recognize that fixing distributed-system bugs is costly because LLMs can rarely infer the root cause purely by reasoning over the codebase.
The agents oftentimes need to trial and error their way through a vast search space of potential hypotheses.
Therefore, we argue that the future agent-native design can effectively utilize the debug-context slot that \sysname provides to guide their tool design and evaluate what information is most valuable for the agentic debugging process.

\subsection{Debug-Context-Enabled Performance Lift Scales with Diagnostic Search-Space Size}
\label{sec:eval:tiers}

\begin{wrapfigure}[16]{r}{0.5\linewidth}
    \vspace{-0.8em}
    \centering
    \includegraphics[width=\linewidth]{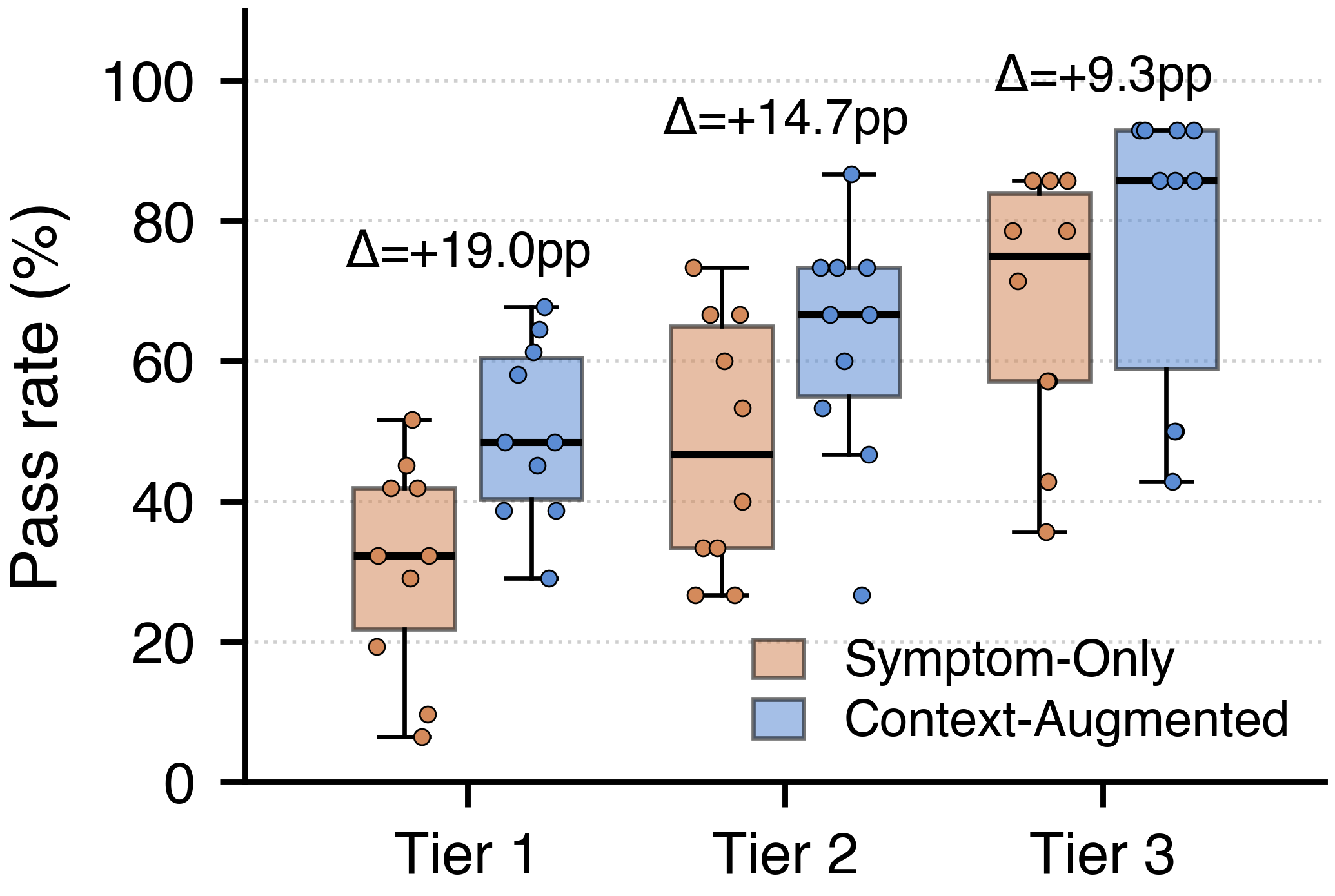}
    \caption{Pass rates across tiers 1--3: debug-context lift on pass-rate attenuates as the diagnostic search space shrinks. Dots show per-model pass rates.}
    \label{fig:pass-rate-boxplot}
\end{wrapfigure}

We also evaluated on \tiertwo (\ddbenchTierTwoSize cases) and \tierthree (\ddbenchTierThreeSize cases) to test whether the \tierone effect is structural to distributed debugging or an artifact of selecting bugs LLMs cannot solve unaided (\figref{fig:pass-rate-boxplot}).

Debug-context lifted aggregated mean pass rates on both tiers: 48.0\%~$\rightarrow$~62.7\% on \tiertwo (+14.7~pp) and 67.9\%~$\rightarrow$~77.1\% on \tierthree (+9.3~pp).
On \tiertwo, 8 of 10 models improved, the remaining 2 tied (Claude Sonnet~4.6 at 73.3\% and GPT~OSS~(120B) at 26.7\%), none regressed, and debug-context flipped 31 cases FAIL~$\rightarrow$~PASS against only 9 in reverse.
On \tierthree, 6 of 10 models im6roved, 2 tied (GLM~5 and GLM~4.7), and 2 showed a minor regression (GLM~5.1 and Gemma~4, $-$7.1~pp each, a net loss of one case, within binomial noise at $n = 14$); however, we note that the symptom-only baseline (67.9\%) is already more than twice that of \tierone (31.0\%), since \tierthree cases are predominantly single-process input-triggered crashes that models can often diagnose from the symptom alone.

\parab{Debug-context value tracks diagnostic search-space size.}
The pass-rate lift narrows monotonically across tiers, yet remains obvious even in the case where the symptom-only baseline approaches 70\%.
A tier's reasoning scope in \sysname{} is, by construction, a proxy for search-space size: \tierthree localizes to a single process and a triggering input; \tiertwo spans multi-process state; \tierone requires reasoning over distributed protocol state and concurrent interleavings.
The \tiertwo and \tierthree results confirm that debug-context produces a measurable lift but the lift shrinks as the search space narrows.
\looseness=-1


\section{Limitations and Future Work}
\label{sec:limitations}

\parab{Single agent scaffold.}
We only used \texttt{mini-swe-agent}~\citep{yang2024sweagent} as the agent scaffold.
Nevertheless, \sysname is designed to be agent-agnostic so that any agent that can read a repository, edit files, and signal completion can be plugged in via the addon mechanism described in Appendix~\secref{appendix:infrastructure}, and the released codebase already ships working addons for Claude Code~\cite{claude} and Codex~\cite{codex}.
We have not yet run a controlled cross-agent-scaffold replication of the symptom-only vs.\ context-augmented contrast because of compute constraints.
A multi-scaffold evaluation isolating the contribution of agent design from model capability is a natural next step.

\parab{No content-level ablation of the debug-context bundle.}
The two-condition contrast (symptom-only vs.\ context-augmented) treats the bundle as an atomic unit and so cannot tell us \emph{which} component of the bundle (logs, traces, goroutine dumps, code-investigation notes) drives the gain, or how the gain scales with the number and type of files supplied.
The debug-context statistics in \secref{sec:benchmark-features} make a content-level ablation tractable on the existing infrastructure, and a follow-up study holding the model fixed while varying the debug-context subset would attribute the headline lift to specific signal types.
\looseness=-1

\parab{Single debug-context generator.}
All bundles were generated by a single agent and model, Claude Code with Claude Opus~4.6.
Claude models gain no more from these bundles than other families do~(\secref{sec:debug-context}), but Opus~4.6 also has the highest symptom-only baseline on \tierone{}, so a small same-family advantage could be masked by the ceiling.
Regenerating the bundles with a different model through the extensibility slot would test this directly.
Such a bundle defines a new context condition, so its pass rates would not be directly comparable to those reported here.

\parab{Single curated release.}
\sysname's \ddbenchReleaseSize{} cases reflect the multi-stage quality bar described in \secref{sec:benchmark} (semi-agentic mining, two-direction oracle verification, three-pass sanitization, five-researcher manual audit) rather than a scale ceiling.
The agent-assisted mining protocol (Appendix~\secref{appendix:semi-agentic}) is released alongside the benchmark so that the case set can grow under the same contract, and we treat scale-up across additional systems and bug families as ongoing work.

\parab{Deployment substrate.}
\sysname runs each case in a single hermetic container.
Where an oracle exercises a cluster, its logical nodes are native server processes inside that container, communicating over loopback sockets, rather than separate containers or Kubernetes pods.
Faults specific to the deployment substrate are therefore absent from this release, including Kubernetes control-plane, container-networking, and storage-provider failures, as is production-scale emergent behavior.
We treat the substrate as orthogonal to \sysname's admission criterion.
Running each node in its own pod would add orchestration-specific failure modes, but it would not change the replication, recovery, or membership semantics a \tierone{} case violates.
Likewise, packaging a single-process parser crash in Kubernetes would not move it out of \tierthree{}.
This separation is consistent with an earlier failure study~\citep{yuan2014simple}, which found that most of the production failures it examined in the systems it studied could be reproduced with a small number of nodes.

\section{Conclusion}
\label{sec:conclusion}

Distributed-system debugging is the regime where two distinct questions about coding agents become cleanly measurable: which models reason better on cross-process interactions and undeterministic timings, and how much additional debugging information (debug-context) changes agent behavior. 
\sysname{} operationalizes this regime as a controlled instrument with \ddbenchReleaseSize{} historical bugs evaluated under matched symptom-only and context-augmented conditions,
surfacing a 45~pp cross-model spread on \tierone{} and an asymmetric context lift that reorders rankings at fixed budget.
\sysname thereby points to two frontiers that still remain underexplored:
\begin{inparaenum}[(1)]
\item how to evaluate and enhance model reasoning capabilities across processes and non-deterministic schedules and timings, and 
\item how can agentic-native tools effectively supply well-curated debugging information and context to lift diagnostic performance and reduce cost simultaneously.
\end{inparaenum}

\newpage
\ifneurips
  \begin{ack}
We thank the anonymous reviewers for their thoughtful discussion and constructive feedback, which substantially improved this paper.
Research supported by the NVIDIA Academic Grant Program.
This work was also supported by a National Research Foundation of Korea (NRF) grant (\mbox{RS-2024-00354947}).
The authors declare no competing interests.
\end{ack}

\fi
\ifneurips
  \bibliographystyle{plainnat}
\else\ifusenix
  \bibliographystyle{plain}
\else
  \bibliographystyle{templates/acmart/ACM-Reference-Format}
\fi\fi
\bibliography{refs,ddbench}

@misc{2026sweagent,
  title = {{{SWE-agent}}/Mini-Swe-Agent},
  year = 2026,
  month = apr,
  url = {https://github.com/SWE-agent/mini-swe-agent},
  urldate = {2026-04-25},
  copyright = {MIT},
  howpublished = {SWE-agent}
}

@inproceedings{ahmed2023recommending,
  title = {Recommending Root-Cause and Mitigation Steps for Cloud Incidents Using Large Language Models},
  booktitle = {Proceedings of the 45th {{International Conference}} on {{Software Engineering}}},
  author = {Ahmed, Toufique and Ghosh, Supriyo and Bansal, Chetan and Zimmermann, Thomas and Zhang, Xuchao and Rajmohan, Saravan},
  year = 2023,
  month = jul,
  series = {{{ICSE}} '23},
  pages = {1737--1749},
  publisher = {IEEE Press},
  address = {Melbourne, Victoria, Australia},
  doi = {10.1109/ICSE48619.2023.00149},
  url = {https://dl.acm.org/doi/10.1109/ICSE48619.2023.00149},
  urldate = {2026-04-19},
  isbn = {978-1-6654-5701-9},
  langid = {english}
}

@inproceedings{badertdinov2025swerebench,
  title = {{{SWE-rebench}}: {{An Automated Pipeline}} for {{Task Collection}} and {{Decontaminated Evaluation}} of {{Software Engineering Agents}}},
  shorttitle = {{{SWE-rebench}}},
  booktitle = {The {{Thirty-ninth Annual Conference}} on {{Neural Information Processing Systems Datasets}} and {{Benchmarks Track}}},
  author = {Badertdinov, Ibragim and Golubev, Alexander and Nekrashevich, Maksim and Shevtsov, Anton and Karasik, Simon and Andriushchenko, Andrei and Trofimova, Maria and Litvintseva, Daria and Yangel, Boris},
  year = 2025,
  month = oct,
  url = {https://openreview.net/forum?id=nMpJoVmRy1},
  urldate = {2026-04-04},
  langid = {english}
}

@inproceedings{chen2023teaching,
  title = {Teaching {{Large Language Models}} to {{Self-Debug}}},
  booktitle = {The {{Twelfth International Conference}} on {{Learning Representations}}},
  author = {Chen, Xinyun and Lin, Maxwell and Sch{\"a}rli, Nathanael and Zhou, Denny},
  year = 2023,
  month = oct,
  url = {https://openreview.net/forum?id=KuPixIqPiq},
  urldate = {2026-04-19},
  langid = {english}
}

@misc{chen2024coder,
  title = {{{CodeR}}: {{Issue Resolving}} with {{Multi-Agent}} and {{Task Graphs}}},
  shorttitle = {{{CodeR}}},
  author = {Chen, Dong and Lin, Shaoxin and Zeng, Muhan and Zan, Daoguang and Wang, Jian-Gang and Cheshkov, Anton and Sun, Jun and Yu, Hao and Dong, Guoliang and Aliev, Artem and Wang, Jie and Cheng, Xiao and Liang, Guangtai and Ma, Yuchi and Bian, Pan and Xie, Tao and Wang, Qianxiang},
  year = 2024,
  month = jun,
  number = {arXiv:2406.01304},
  eprint = {2406.01304},
  primaryclass = {cs},
  publisher = {arXiv},
  doi = {10.48550/arXiv.2406.01304},
  url = {http://arxiv.org/abs/2406.01304},
  urldate = {2026-04-19},
  archiveprefix = {arXiv}
}

@inproceedings{chen2025aiopslab,
  title = {{{AIOpsLab}}: A Holistic Framework to Evaluate {{AI}} Agents for Enabling Autonomous Clouds},
  shorttitle = {{{AIOpsLab}}},
  booktitle = {{{arXiv}}.Org},
  author = {Chen, Yinfang and Shetty, Manish and Somashekar, Gagan and Ma, Minghua and Simmhan, Yogesh and Mace, Jonathan and Bansal, Chetan and Wang, Rujia and Rajmohan, Saravan},
  year = 2025,
  month = jan,
  url = {https://arxiv.org/abs/2501.06706v1},
  urldate = {2026-04-01},
  langid = {english}
}

@misc{claude,
  title = {Claude {{Code}}},
  journal = {Claude Code Docs},
  url = {https://code.claude.com/docs/en/overview},
  urldate = {2026-04-19},
  langid = {english},
  note = {Accessed: 2026-04-19},
  year = {2026}
}

@misc{codex,
  title = {Codex {{CLI}}},
  url = {https://developers.openai.com/codex/cli},
  urldate = {2026-04-19},
  langid = {english},
  note = {Accessed: 2026-04-19},
  year = {2026}
}

@misc{deng2025swebench,
  title = {{{SWE-bench}} pro: Can {{AI}} Agents Solve Long-Horizon Software Engineering Tasks?},
  shorttitle = {{{SWE-bench}} Pro},
  author = {Deng, Xiang and Da, Jeff and Pan, Edwin and He, Yannis Yiming and Ide, Charles and Garg, Kanak and Lauffer, Niklas and Park, Andrew and Pasari, Nitin and Rane, Chetan and Sampath, Karmini and Krishnan, Maya and Kundurthy, Srivatsa and Hendryx, Sean and Wang, Zifan and Bharadwaj, Vijay and Holm, Jeff and Aluri, Raja and Zhang, Chen Bo Calvin and Jacobson, Noah and Liu, Bing and Kenstler, Brad},
  year = 2025,
  month = nov,
  number = {arXiv:2509.16941},
  eprint = {2509.16941},
  primaryclass = {cs},
  publisher = {arXiv},
  doi = {10.48550/arXiv.2509.16941},
  url = {http://arxiv.org/abs/2509.16941},
  urldate = {2026-04-04},
  archiveprefix = {arXiv},
  langid = {english}
}

@misc{fan2025sweeffi,
  title = {{{SWE-Effi}}: {{Re-Evaluating Software AI Agent System Effectiveness Under Resource Constraints}}},
  shorttitle = {{{SWE-Effi}}},
  author = {Fan, Zhiyu and Vasilevski, Kirill and Lin, Dayi and Chen, Boyuan and Chen, Yihao and Zhong, Zhiqing and Zhang, Jie M. and He, Pinjia and Hassan, Ahmed E.},
  year = 2025,
  month = sep,
  number = {arXiv:2509.09853},
  eprint = {2509.09853},
  primaryclass = {cs},
  publisher = {arXiv},
  doi = {10.48550/arXiv.2509.09853},
  url = {http://arxiv.org/abs/2509.09853},
  urldate = {2026-04-25},
  archiveprefix = {arXiv}
}

@inproceedings{gunawi2011fate,
  title = {\textbraceleft{{FATE}}\textbraceright{} and \textbraceleft{{DESTINI}}\textbraceright : A Framework for Cloud Recovery Testing},
  shorttitle = {\textbraceleft{{FATE}}\textbraceright{} and \textbraceleft{{DESTINI}}\textbraceright},
  booktitle = {8th {{USENIX Symposium}} on {{Networked Systems Design}} and {{Implementation}} ({{NSDI}} 11)},
  author = {Gunawi, Haryadi S. and Do, Thanh and Joshi, Pallavi and Alvaro, Peter and Hellerstein, Joseph M. and {Arpaci-Dusseau}, Andrea C. and {Arpaci-Dusseau}, Remzi H. and Sen, Koushik and Borthakur, Dhruba},
  year = 2011,
  url = {https://www.usenix.org/conference/nsdi11/fate-and-destini-framework-cloud-recovery-testing},
  urldate = {2026-04-19},
  langid = {english}
}

@misc{jiang2024largescale,
  title = {A Large-Scale Evaluation for Log Parsing Techniques: How Far Are We?},
  shorttitle = {A Large-Scale Evaluation for Log Parsing Techniques},
  author = {Jiang, Zhihan and Liu, Jinyang and Huang, Junjie and Li, Yichen and Huo, Yintong and Gu, Jiazhen and Chen, Zhuangbin and Zhu, Jieming and Lyu, Michael R.},
  year = 2024,
  month = mar,
  number = {arXiv:2308.10828},
  eprint = {2308.10828},
  primaryclass = {cs},
  publisher = {arXiv},
  doi = {10.48550/arXiv.2308.10828},
  url = {http://arxiv.org/abs/2308.10828},
  urldate = {2026-04-19},
  archiveprefix = {arXiv},
  langid = {english}
}

@inproceedings{jimenez2023swebench,
  title = {{{SWE-bench}}: Can Language Models Resolve Real-World Github Issues?},
  shorttitle = {{{SWE-bench}}},
  booktitle = {The {{Twelfth International Conference}} on {{Learning Representations}}},
  author = {Jimenez, Carlos E. and Yang, John and Wettig, Alexander and Yao, Shunyu and Pei, Kexin and Press, Ofir and Narasimhan, Karthik R.},
  year = 2023,
  month = oct,
  url = {https://openreview.net/forum?id=VTF8yNQM66},
  urldate = {2026-04-01},
  langid = {english}
}

@misc{jin2025llms,
  title = {From {{LLMs}} to {{LLM-based}} Agents for Software Engineering: A Survey of Current, Challenges and Future},
  shorttitle = {From {{LLMs}} to {{LLM-based}} Agents for Software Engineering},
  author = {Jin, Haolin and Huang, Linghan and Cai, Haipeng and Yan, Jun and Li, Bo and Chen, Huaming},
  year = 2025,
  month = apr,
  number = {arXiv:2408.02479},
  eprint = {2408.02479},
  primaryclass = {cs},
  publisher = {arXiv},
  doi = {10.48550/arXiv.2408.02479},
  url = {http://arxiv.org/abs/2408.02479},
  urldate = {2026-04-20},
  archiveprefix = {arXiv},
  langid = {english}
}

@article{leesatapornwongsa2016taxdc,
  title = {{{TaxDC}}: {{A Taxonomy}} of {{Non-Deterministic Concurrency Bugs}} in {{Datacenter Distributed Systems}}},
  shorttitle = {{{TaxDC}}},
  author = {Leesatapornwongsa, Tanakorn and Lukman, Jeffrey F. and Lu, Shan and Gunawi, Haryadi S.},
  year = 2016,
  month = mar,
  journal = {SIGPLAN Not.},
  volume = {51},
  number = {4},
  pages = {517--530},
  issn = {0362-1340},
  doi = {10.1145/2954679.2872374},
  url = {https://dl.acm.org/doi/10.1145/2954679.2872374},
  urldate = {2026-04-19}
}

@inproceedings{liu2025opseval,
  title = {{{OpsEval}}: A Comprehensive Benchmark Suite for Evaluating Large Language Models' Capability in {{IT}} Operations Domain},
  shorttitle = {{{OpsEval}}},
  booktitle = {Proceedings of the 33rd {{ACM International Conference}} on the {{Foundations}} of {{Software Engineering}}},
  author = {Liu, Yuhe and Pei, Changhua and Xu, Longlong and Chen, Bohan and Sun, Mingze and Zhang, Zhirui and Sun, Yongqian and Zhang, Shenglin and Wang, Kun and Zhang, Haiming and Li, Jianhui and Xie, Gaogang and Wen, Xidao and Nie, Xiaohui and Ma, Minghua and Pei, Dan},
  year = 2025,
  month = jul,
  series = {{{FSE Companion}} '25},
  pages = {503--513},
  publisher = {Association for Computing Machinery},
  address = {New York, NY, USA},
  doi = {10.1145/3696630.3728572},
  url = {https://dl.acm.org/doi/10.1145/3696630.3728572},
  urldate = {2026-04-19},
  isbn = {979-8-4007-1276-0},
  langid = {english}
}

@inproceedings{merrill2025terminalbench,
  title = {Terminal-{{Bench}}: {{Benchmarking Agents}} on {{Hard}}, {{Realistic Tasks}} in {{Command Line Interfaces}}},
  shorttitle = {Terminal-{{Bench}}},
  booktitle = {The {{Fourteenth International Conference}} on {{Learning Representations}}},
  author = {Merrill, Mike A. and Shaw, Alexander Glenn and Carlini, Nicholas and Li, Boxuan and Raj, Harsh and Bercovich, Ivan and Shi, Lin and Shin, Jeong Yeon and Walshe, Thomas and Buchanan, E. Kelly and Shen, Junhong and Ye, Guanghao and Lin, Haowei and Poulos, Jason and Wang, Maoyu and Nezhurina, Marianna and Lu, Di and Mastromichalakis, Orfeas Menis and Xu, Zhiwei and Chen, Zizhao and Liu, Yue and Zhang, Robert and Chen, Leon Liangyu and Kashyap, Anurag and Uslu, Jan-Lucas and Li, Jeffrey and Wu, Jianbo and Yan, Minghao and Bian, Song and Sharma, Vedang and Sun, Ke and Dillmann, Steven and Anand, Akshay and Lanpouthakoun, Andrew and Koopah, Bardia and Hu, Changran and Guha, Etash Kumar and Dreiman, Gabriel H. S. and Zhu, Jiacheng and Krauth, Karl and Zhong, Li and Muennighoff, Niklas and Amanfu, Robert Kwesi and Tan, Shangyin and Pimpalgaonkar, Shreyas and Aggarwal, Tushar and Lin, Xiangning and Lan, Xin and Zhao, Xuandong and Liang, Yiqing and Wang, Yuanli and Wang, Zilong and Zhou, Changzhi and Heineman, David and Liu, Hange and Trivedi, Harsh and Yang, John and Lin, Junhong and Shetty, Manish and Yang, Michael and Omi, Nabil and Raoof, Negin and Li, Shanda and Zhuo, Terry Yue and Lin, Wuwei and Dai, Yiwei and Wang, Yuxin and Chai, Wenhao and Zhou, Shang and Wahdany, Dariush and She, Ziyu and Hu, Jiaming and Dong, Zhikang and Zhu, Yuxuan and Cui, Sasha and Saiyed, Ahson and Kolbeinsson, Arinbj{\"o}rn and Rytting, Christopher Michael and Marten, Ryan and Wang, Yixin and Jitsev, Jenia and Dimakis, Alex and Konwinski, Andy and Schmidt, Ludwig},
  year = 2025,
  month = oct,
  url = {https://openreview.net/forum?id=a7Qa4CcHak&referrer=%5Bthe+profile+of+Alex+Dimakis%5D%28%2Fprofile%3Fid%3D%7EAlex_Dimakis1%29},
  urldate = {2026-04-25},
  langid = {english}
}

@inproceedings{mundler2024swtbench,
  title = {{{SWT-Bench}}: {{Testing}} and {{Validating Real-World Bug-Fixes}} with {{Code Agents}}},
  shorttitle = {{{SWT-Bench}}},
  booktitle = {The {{Thirty-eighth Annual Conference}} on {{Neural Information Processing Systems}}},
  author = {M{\"u}ndler, Niels and Mueller, Mark Niklas and He, Jingxuan and Vechev, Martin},
  year = 2024,
  month = nov,
  url = {https://openreview.net/forum?id=9Y8zUO11EQ},
  urldate = {2026-04-14},
  langid = {english}
}

@misc{sun2026sweworld,
  title = {{{SWE-world}}: Building Software Engineering Agents in Docker-Free Environments},
  shorttitle = {{{SWE-world}}},
  author = {Sun, Shuang and Song, Huatong and Huang, Lisheng and Jiang, Jinhao and Le, Ran and Lv, Zhihao and Chen, Zongchao and Hu, Yiwen and Luo, Wenyang and Zhao, Wayne Xin and Song, Yang and Xu, Hongteng and Zhang, Tao and Wen, Ji-Rong},
  year = 2026,
  month = feb,
  number = {arXiv:2602.03419},
  eprint = {2602.03419},
  primaryclass = {cs},
  publisher = {arXiv},
  doi = {10.48550/arXiv.2602.03419},
  url = {http://arxiv.org/abs/2602.03419},
  urldate = {2026-04-25},
  archiveprefix = {arXiv},
  langid = {english}
}

@misc{suri2026codescout,
  title = {{{CodeScout}}: {{Contextual Problem Statement Enhancement}} for {{Software Agents}}},
  shorttitle = {{{CodeScout}}},
  author = {Suri, Manan and Li, Xiangci and Shojaie, Mehdi and Han, Songyang and Hsu, Chao-Chun and Garg, Shweta and Deshmukh, Aniket Anand and Kumar, Varun},
  year = 2026,
  month = apr,
  number = {arXiv:2603.05744},
  eprint = {2603.05744},
  primaryclass = {cs},
  publisher = {arXiv},
  doi = {10.48550/arXiv.2603.05744},
  url = {http://arxiv.org/abs/2603.05744},
  urldate = {2026-04-30},
  archiveprefix = {arXiv}
}

@misc{thai2026sweevo,
  title = {{{SWE-EVO}}: {{Benchmarking Coding Agents}} in {{Long-Horizon Software Evolution Scenarios}}},
  shorttitle = {{{SWE-EVO}}},
  author = {Thai, Minh V. T. and Le, Tue and Manh, Dung Nguyen and Nhat, Huy Phan and Bui, Nghi D. Q.},
  year = 2026,
  month = apr,
  number = {arXiv:2512.18470},
  eprint = {2512.18470},
  primaryclass = {cs},
  publisher = {arXiv},
  doi = {10.48550/arXiv.2512.18470},
  url = {http://arxiv.org/abs/2512.18470},
  urldate = {2026-04-19},
  archiveprefix = {arXiv}
}

@inproceedings{tian2024debugbench,
  title = {{{DebugBench}}: Evaluating Debugging Capability of Large Language Models},
  shorttitle = {{{DebugBench}}},
  booktitle = {Findings of the {{Association}} for {{Computational Linguistics}}: {{ACL}} 2024},
  author = {Tian, Runchu and Ye, Yining and Qin, Yujia and Cong, Xin and Lin, Yankai and Pan, Yinxu and Wu, Yesai and Haotian, Hui and Weichuan, Liu and Liu, Zhiyuan and Sun, Maosong},
  editor = {Ku, Lun-Wei and Martins, Andre and Srikumar, Vivek},
  year = 2024,
  month = aug,
  pages = {4173--4198},
  publisher = {Association for Computational Linguistics},
  address = {Bangkok, Thailand},
  doi = {10.18653/v1/2024.findings-acl.247},
  url = {https://aclanthology.org/2024.findings-acl.247/},
  urldate = {2026-04-14},
  langid = {english}
}

@inproceedings{wang2023model,
  title = {Model {{Checking Guided Testing}} for {{Distributed Systems}}},
  booktitle = {Proceedings of the {{Eighteenth European Conference}} on {{Computer Systems}}},
  author = {Wang, Dong and Dou, Wensheng and Gao, Yu and Wu, Chenao and Wei, Jun and Huang, Tao},
  year = 2023,
  month = may,
  pages = {127--143},
  publisher = {ACM},
  address = {Rome Italy},
  doi = {10.1145/3552326.3587442},
  url = {https://dl.acm.org/doi/10.1145/3552326.3587442},
  urldate = {2025-02-24},
  isbn = {978-1-4503-9487-1},
  langid = {english}
}

@misc{wang2025openhands,
  title = {{{OpenHands}}: An Open Platform for {{AI}} Software Developers as Generalist Agents},
  shorttitle = {{{OpenHands}}},
  author = {Wang, Xingyao and Li, Boxuan and Song, Yufan and Xu, Frank F. and Tang, Xiangru and Zhuge, Mingchen and Pan, Jiayi and Song, Yueqi and Li, Bowen and Singh, Jaskirat and Tran, Hoang H. and Li, Fuqiang and Ma, Ren and Zheng, Mingzhang and Qian, Bill and Shao, Yanjun and Muennighoff, Niklas and Zhang, Yizhe and Hui, Binyuan and Lin, Junyang and Brennan, Robert and Peng, Hao and Ji, Heng and Neubig, Graham},
  year = 2025,
  month = apr,
  number = {arXiv:2407.16741},
  eprint = {2407.16741},
  primaryclass = {cs},
  publisher = {arXiv},
  doi = {10.48550/arXiv.2407.16741},
  url = {http://arxiv.org/abs/2407.16741},
  urldate = {2026-04-19},
  archiveprefix = {arXiv},
  langid = {english}
}

@misc{wang2026cloudopsbench,
  title = {Cloud-{{OpsBench}}: {{A Reproducible Benchmark}} for {{Agentic Root Cause Analysis}} in {{Cloud Systems}}},
  shorttitle = {Cloud-{{OpsBench}}},
  author = {Wang, Yilun and Yu, Guangba and Huang, Haiyu and Wang, Zirui and Huang, Yujie and Chen, Pengfei and Lyu, Michael R.},
  year = 2026,
  month = feb,
  number = {arXiv:2603.00468},
  eprint = {2603.00468},
  primaryclass = {cs},
  publisher = {arXiv},
  doi = {10.48550/arXiv.2603.00468},
  url = {http://arxiv.org/abs/2603.00468},
  urldate = {2026-04-01},
  archiveprefix = {arXiv}
}

@misc{wei2025swerl,
  title = {{{SWE-RL}}: Advancing {{LLM}} Reasoning via Reinforcement Learning on Open Software Evolution},
  shorttitle = {Swe-Rl},
  author = {Wei, Yuxiang and Duchenne, Olivier and Copet, Jade and Carbonneaux, Quentin and Zhang, Lingming and Fried, Daniel and Synnaeve, Gabriel and Singh, Rishabh and Wang, Sida I.},
  year = 2025,
  month = dec,
  number = {arXiv:2502.18449},
  eprint = {2502.18449},
  primaryclass = {cs},
  publisher = {arXiv},
  doi = {10.48550/arXiv.2502.18449},
  url = {http://arxiv.org/abs/2502.18449},
  urldate = {2026-04-19},
  archiveprefix = {arXiv},
  langid = {english}
}

@misc{xia2024agentless,
  title = {Agentless: {{Demystifying LLM-based Software Engineering Agents}}},
  shorttitle = {Agentless},
  author = {Xia, Chunqiu Steven and Deng, Yinlin and Dunn, Soren and Zhang, Lingming},
  year = 2024,
  month = oct,
  number = {arXiv:2407.01489},
  eprint = {2407.01489},
  primaryclass = {cs},
  publisher = {arXiv},
  doi = {10.48550/arXiv.2407.01489},
  url = {http://arxiv.org/abs/2407.01489},
  urldate = {2026-04-19},
  archiveprefix = {arXiv}
}

@misc{xia2025livesweagent,
  title = {Live-{{SWE-agent}}: {{Can Software Engineering Agents Self-Evolve}} on the {{Fly}}?},
  shorttitle = {Live-{{SWE-agent}}},
  author = {Xia, Chunqiu Steven and Wang, Zhe and Yang, Yan and Wei, Yuxiang and Zhang, Lingming},
  year = 2025,
  month = nov,
  number = {arXiv:2511.13646},
  eprint = {2511.13646},
  primaryclass = {cs},
  publisher = {arXiv},
  doi = {10.48550/arXiv.2511.13646},
  url = {http://arxiv.org/abs/2511.13646},
  urldate = {2026-04-25},
  archiveprefix = {arXiv}
}

@inproceedings{xie2025swefixer,
  title = {{{SWE-fixer}}: Training Open-Source {{LLMs}} for Effective and Efficient {{GitHub}} Issue Resolution},
  shorttitle = {{{SWE-fixer}}},
  booktitle = {Findings of the {{Association}} for {{Computational Linguistics}}: {{ACL}} 2025},
  author = {Xie, Chengxing and Li, Bowen and Gao, Chang and Du, He and Lam, Wai and Zou, Difan and Chen, Kai},
  editor = {Che, Wanxiang and Nabende, Joyce and Shutova, Ekaterina and Pilehvar, Mohammad Taher},
  year = 2025,
  month = jul,
  pages = {1123--1139},
  publisher = {Association for Computational Linguistics},
  address = {Vienna, Austria},
  doi = {10.18653/v1/2025.findings-acl.62},
  url = {https://aclanthology.org/2025.findings-acl.62/},
  urldate = {2026-04-19},
  isbn = {979-8-89176-256-5},
  langid = {english}
}

@inproceedings{yang2009modist,
  title = {{{MODIST}}: {{Transparent Model Checking}} of {{Unmodified Distributed Systems}}},
  author = {Yang, Junfeng and Chen, Tisheng and Wu, Ming and Xu, Zhilei and Liu, Xuezheng and Lin, Haoxiang and Yang, Mao and Long, Fan and Zhang, Lintao and Zhou, Lidong},
  year = 2009,
  pages = {213--228},
  langid = {english}
}

@inproceedings{yang2024sweagent,
  title = {{{SWE-agent}}: Agent-Computer Interfaces Enable Automated Software Engineering},
  shorttitle = {{{SWE-agent}}},
  booktitle = {The {{Thirty-eighth Annual Conference}} on {{Neural Information Processing Systems}}},
  author = {Yang, John and Jimenez, Carlos E. and Wettig, Alexander and Lieret, Kilian and Yao, Shunyu and Narasimhan, Karthik R. and Press, Ofir},
  year = 2024,
  month = nov,
  url = {https://openreview.net/forum?id=mXpq6ut8J3&referrer=%5Bthe%20profile%20of%20Shunyu%20Yao%5D(%2Fprofile%3Fid%3D~Shunyu_Yao1)},
  urldate = {2026-04-14},
  langid = {english}
}

@inproceedings{yang2025swesmith,
  title = {{{SWE-smith}}: Scaling Data for Software Engineering Agents},
  shorttitle = {{{SWE-smith}}},
  booktitle = {The {{Thirty-ninth Annual Conference}} on {{Neural Information Processing Systems Datasets}} and {{Benchmarks Track}}},
  author = {Yang, John and Lieret, Kilian and Jimenez, Carlos E. and Wettig, Alexander and Khandpur, Kabir and Zhang, Yanzhe and Hui, Binyuan and Press, Ofir and Schmidt, Ludwig and Yang, Diyi},
  year = 2025,
  month = oct,
  url = {https://openreview.net/forum?id=63iVrXc8cC&referrer=%5Bthe+profile+of+Carlos+E.+Jimenez%5D%28%2Fprofile%3Fid%3D%7ECarlos_E._Jimenez1%29},
  urldate = {2026-04-19},
  langid = {english}
}

@inproceedings{yu2023nezha,
  title = {Nezha: Interpretable Fine-Grained Root Causes Analysis for Microservices on Multi-Modal Observability Data},
  shorttitle = {Nezha},
  booktitle = {Proceedings of the 31st {{ACM Joint European Software Engineering Conference}} and {{Symposium}} on the {{Foundations}} of {{Software Engineering}}},
  author = {Yu, Guangba and Chen, Pengfei and Li, Yufeng and Chen, Hongyang and Li, Xiaoyun and Zheng, Zibin},
  year = 2023,
  month = nov,
  series = {{{ESEC}}/{{FSE}} 2023},
  pages = {553--565},
  publisher = {Association for Computing Machinery},
  address = {New York, NY, USA},
  doi = {10.1145/3611643.3616249},
  url = {https://dl.acm.org/doi/10.1145/3611643.3616249},
  urldate = {2026-04-19},
  isbn = {979-8-4007-0327-0},
  langid = {english}
}

@inproceedings{yuan2014simple,
  title = {Simple Testing Can Prevent Most Critical Failures: An Analysis of Production Failures in Distributed Data-Intensive Systems},
  shorttitle = {Simple Testing Can Prevent Most Critical Failures},
  booktitle = {11th {{USENIX Symposium}} on {{Operating Systems Design}} and {{Implementation}} ({{OSDI}} 14)},
  author = {Yuan, Ding and Luo, Yu and Zhuang, Xin and Rodrigues, Guilherme Renna and Zhao, Xu and Zhang, Yongle and Jain, Pranay U. and Stumm, Michael},
  year = 2014,
  pages = {249--265},
  url = {https://www.usenix.org/conference/osdi14/technical-sessions/presentation/yuan},
  urldate = {2026-03-21},
  isbn = {978-1-931971-16-4},
  langid = {english}
}

@inproceedings{yuan2019approach,
  title = {An Approach to Cloud Execution Failure Diagnosis Based on Exception Logs in {{OpenStack}}},
  booktitle = {2019 {{IEEE}} 12th {{International Conference}} on {{Cloud Computing}} ({{CLOUD}})},
  author = {Yuan, Yue and Shi, Wenchang and Liang, Bin and Qin, Bo},
  year = 2019,
  month = jul,
  pages = {124--131},
  issn = {2159-6190},
  doi = {10.1109/CLOUD.2019.00031},
  url = {https://ieeexplore.ieee.org/document/8814553},
  urldate = {2026-04-19},
  langid = {english}
}

@misc{yuan2026sweminisandbox,
  title = {{{SWE-MiniSandbox}}: Container-Free Reinforcement Learning for Building Software Engineering Agents},
  shorttitle = {{{SWE-MiniSandbox}}},
  author = {Yuan, Danlong and Wu, Wei and Wang, Zhengren and Zhao, Xueliang and Zhang, Huishuai and Zhao, Dongyan},
  year = 2026,
  month = mar,
  number = {arXiv:2602.11210},
  eprint = {2602.11210},
  primaryclass = {cs},
  publisher = {arXiv},
  doi = {10.48550/arXiv.2602.11210},
  url = {http://arxiv.org/abs/2602.11210},
  urldate = {2026-04-25},
  archiveprefix = {arXiv},
  langid = {english}
}

@inproceedings{zan2025multiswebench,
  title = {Multi-{{SWE-bench}}: A Multilingual Benchmark for Issue Resolving},
  shorttitle = {Multi-{{SWE-bench}}},
  booktitle = {The {{Thirty-ninth Annual Conference}} on {{Neural Information Processing Systems Datasets}} and {{Benchmarks Track}}},
  author = {Zan, Daoguang and Huang, Zhirong and Liu, Wei and Chen, Hanwu and Xin, Shulin and Zhang, Linhao and Liu, Qi and Li, Aoyan and Chen, Lu and Zhong, Xiaojian and Liu, Siyao and Xiao, Yongsheng and Chen, Liangqiang and Zhang, Yuyu and Su, Jing and Liu, Tianyu and Long, Rui and Ding, Ming and Xiang, Liang},
  year = 2025,
  month = oct,
  url = {https://openreview.net/forum?id=MhBZzkz4h9&referrer=%5Bthe+profile+of+Zhirong+Huang%5D%28%2Fprofile%3Fid%3D%7EZhirong_Huang3%29},
  urldate = {2026-04-14},
  langid = {english}
}

@inproceedings{zhang2025traceback,
  title = {Traceback of Poisoning Attacks to Retrieval-Augmented Generation},
  booktitle = {Proceedings of the {{ACM}} on {{Web Conference}} 2025},
  author = {Zhang, Baolei and Xin, Haoran and Fang, Minghong and Liu, Zhuqing and Yi, Biao and Li, Tong and Liu, Zheli},
  year = 2025,
  month = apr,
  series = {{{WWW}} '25},
  pages = {2085--2097},
  publisher = {Association for Computing Machinery},
  address = {New York, NY, USA},
  doi = {10.1145/3696410.3714756},
  url = {https://dl.acm.org/doi/10.1145/3696410.3714756},
  urldate = {2026-04-19},
  isbn = {979-8-4007-1274-6},
  langid = {english}
}

@inproceedings{zhong2024debug,
  title = {Debug like a Human: A Large Language Model Debugger via Verifying Runtime Execution Step by Step},
  shorttitle = {Debug like a Human},
  booktitle = {Findings of the {{Association}} for {{Computational Linguistics}}: {{ACL}} 2024},
  author = {Zhong, Li and Wang, Zilong and Shang, Jingbo},
  editor = {Ku, Lun-Wei and Martins, Andre and Srikumar, Vivek},
  year = 2024,
  month = aug,
  pages = {851--870},
  publisher = {Association for Computational Linguistics},
  address = {Bangkok, Thailand},
  doi = {10.18653/v1/2024.findings-acl.49},
  url = {https://aclanthology.org/2024.findings-acl.49/},
  urldate = {2026-04-19},
  langid = {english}
}

@inproceedings{zhou2021foundationdb,
  title = {{{FoundationDB}}: A Distributed Unbundled Transactional Key Value Store},
  shorttitle = {{{FoundationDB}}},
  booktitle = {Proceedings of the 2021 {{International Conference}} on {{Management}} of {{Data}}},
  author = {Zhou, Jingyu and Xu, Meng and Shraer, Alexander and Namasivayam, Bala and Miller, Alex and Tschannen, Evan and Atherton, Steve and Beamon, Andrew J. and Sears, Rusty and Leach, John and Rosenthal, Dave and Dong, Xin and Wilson, Will and Collins, Ben and Scherer, David and Grieser, Alec and Liu, Young and Moore, Alvin and Muppana, Bhaskar and Su, Xiaoge and Yadav, Vishesh},
  year = 2021,
  month = jun,
  series = {{{SIGMOD}} '21},
  pages = {2653--2666},
  publisher = {Association for Computing Machinery},
  address = {New York, NY, USA},
  doi = {10.1145/3448016.3457559},
  url = {https://dl.acm.org/doi/10.1145/3448016.3457559},
  urldate = {2026-04-19},
  isbn = {978-1-4503-8343-1},
  langid = {english}
}

@misc{zhu2023loghub,
  title = {Loghub: {{A Large Collection}} of {{System Log Datasets}} for {{AI-driven Log Analytics}}},
  shorttitle = {Loghub},
  author = {Zhu, Jieming and He, Shilin and He, Pinjia and Liu, Jinyang and Lyu, Michael R.},
  year = 2023,
  month = sep,
  number = {arXiv:2008.06448},
  eprint = {2008.06448},
  primaryclass = {cs},
  publisher = {arXiv},
  doi = {10.48550/arXiv.2008.06448},
  url = {http://arxiv.org/abs/2008.06448},
  urldate = {2026-04-19},
  archiveprefix = {arXiv}
}

@inproceedings{zhuo2024bigcodebench,
  title = {{{BigCodeBench}}: Benchmarking Code Generation with Diverse Function Calls and Complex Instructions},
  shorttitle = {{{BigCodeBench}}},
  booktitle = {The {{Thirteenth International Conference}} on {{Learning Representations}}},
  author = {Zhuo, Terry Yue and Chien, Vu Minh and Chim, Jenny and Hu, Han and Yu, Wenhao and Widyasari, Ratnadira and Yusuf, Imam Nur Bani and Zhan, Haolan and He, Junda and Paul, Indraneil and Brunner, Simon and Gong, Chen and Hoang, James and Zebaze, Armel Randy and Hong, Xiaoheng and Li, Wen-Ding and Kaddour, Jean and Xu, Ming and Zhang, Zhihan and Yadav, Prateek and Jain, Naman and Gu, Alex and Cheng, Zhoujun and Liu, Jiawei and Liu, Qian and Wang, Zijian and Hui, Binyuan and Muennighoff, Niklas and Lo, David and Fried, Daniel and Du, Xiaoning and de Vries, Harm and Werra, Leandro Von},
  year = 2024,
  month = oct,
  url = {https://openreview.net/forum?id=YrycTjllL0},
  urldate = {2026-04-04},
  langid = {english}
}

\newpage
\appendix

\section{Related Work}

\parab{Code-repair benchmarks for LLM agents.}
SWE-bench and its successors established the dominant template for evaluating LLM coding agents: mine historical issue--PR pairs from open-source repositories, package each as a faulty worktree plus a hidden test oracle, and score agents by whether their patch passes~\cite{jimenez2023swebench, zan2025multiswebench, thai2026sweevo, tian2024debugbench, deng2025swebench, wei2025swerl, badertdinov2025swerebench, yang2025swesmith, zhuo2024bigcodebench, mundler2024swtbench}.
These benchmarks have been invaluable, but they inherit two structural limits from their source material. First, they draw almost exclusively from single-process Python (and, in the multilingual extensions~\cite{zan2025multiswebench}, single-process Java/TypeScript/Go) repositories, where the reasoning challenge is primarily cross-file localization within one address space. Second, the oracle is a standard unit or integration test that can be run deterministically in a sandbox, which excludes bugs whose reproduction requires multi-node clusters, non-deterministic scheduling, or runtime-state capture.
\sysname inherits the faulty-worktree-with-hidden-oracle contract but relaxes both limits. Cases are drawn from distributed systems in five compiled and JVM languages, and the oracle is a containerized harness (oftentimes end-to-end integration tests) rather than a single test invocation, which enables us to test concurrency hazards and protocol-level bugs that unit-test oracles often structurally cannot express.

\parab{IT-operations and cloud-incident benchmarks.}
A parallel line of work evaluates LLM agents on {diagnosis} of distributed and cloud systems deployment from telemetry, utilizing logs, metrics, traces, and alerts~\cite{chen2025aiopslab, wang2026cloudopsbench, liu2025opseval, ahmed2023recommending}.
However, these benchmarks target only operational diagnosis of deployment issues rather than the repair of actual distributed systems.
These benchmarks establish that cross-service reasoning is a distinct capability axis, and many inject controlled faults (latency, CPU, packet loss, crash) into microservice deployments to generate incidents on demand.
Their scoring, however, targets root-cause identification or remediation action selection rather than producing a code patch. 
The agent's output is a natural-language attribution or a pre-defined mitigation, not a source \emph{diff} verified by a reproduce oracle.
\sysname is orthogonal and complementary. 
\sysname shares the cross-process reasoning load these benchmarks exercise but closes the loop on source-level repair, so the ground truth is an executable oracle rather than a labeled root cause.
This combination is what lets us measure the effect of debugging context on agent performance: an agent can consume runtime signals and gets scored on whether it produces a correct patch.

\parab{LLM agents for debugging and program repair.}
A growing body of work studies LLMs and agent scaffolds for bug localization, fault repair, and test-time reasoning over code~\cite{wang2025openhands, yang2024sweagent, xia2024agentless, chen2024coder, xie2025swefixer}.
Most of this work is developed and tuned against single-process repair benchmarks. 
Under this paradigm, the dominant agent patterns are shaped as read--edit--test loops with localized retrieval and repository-wide code search.
A separate strand augments LLM debugging with runtime signals. 
Namely, executing programs to collect stack traces, observing variable state, or consulting coverage and profiling tools~\cite{zhong2024debug, chen2023teaching, zhang2025traceback}.
\sysname extends this line into the distributed setting, where the relevant runtime signals (cluster logs, goroutine dumps, race-detector reports, replica state) are qualitatively different and are not consistently recoverable by re-executing a single process.
Our context-augmented condition makes this information channel a controlled input variable rather than an agent-internal implementation detail, which lets future agent and tool work be evaluated on the axis of \emph{which runtime signals it surfaces}, not only on end-to-end pass rate.

\parab{Distributed-systems bug studies and testing.}
Empirical studies of real-world distributed-systems failures have repeatedly shown that the hardest bugs arise from protocol-level invariants, cross-node concurrency, and partial failures rather than from within-process logic errors, and have motivated a long line of specialized testing and model-checking infrastructure in the distributed-system community~\cite{leesatapornwongsa2016taxdc, yuan2014simple, zhou2021foundationdb, gunawi2011fate, yang2009modist, wang2023model}.
This line of work provides both the taxonomy of failure modes we target (protocol/recovery, replication/consistency, concurrency hazards, input-triggered crashes, resource/delivery) and the evidence that such bugs cannot be reliably caught by single-process testing.
Our contribution is not a new testing technique but a benchmark that imports the bug distribution and reproducibility standards from this literature into the LLM-agent evaluation setting.
Cases are drawn from the classes these studies identify as hardest; each is paired with a hardened reproducer; and the oracle accepts any patch that makes the reproducer pass rather than requiring a textual match (our manual examinations ensure this).

\parab{Datasets for runtime signals.}
A separate line of work has released large-scale datasets of real system logs, traces, and telemetry collected from production and benchmark distributed deployments, originally built for log parsing, anomaly detection, and trace-based root-cause analysis (RCA)~\cite{zhu2023loghub, jiang2024largescale, yuan2019approach, yu2023nezha}.
These datasets are valuable runtime-signal sources, but they are not directly drop-in replacements for \sysname's evaluation harness, since they are not paired with a faulty code commit, a sanitized symptom description, or a reproduce oracle; an agent consuming them therefore cannot be scored on whether its resulting patch compiles and passes.
\sysname is designed so that such datasets can be brought in through its debug-context channel. Each case defines a bounded slot (the per-case \texttt{context} folder) that the harness mounts into the agent's container, and the content of that slot is deliberately decoupled from the oracle and the symptom file.
Future work can therefore treat these public log/trace corpora, fault-injection telemetry, and incident postmortems as alternative debug-context sources.
Researchers can select the subset whose system or failure mode matches a \sysname case, transform it into the debug-context slot's format, and re-evaluate agents under the same symptom-only vs.\ context-augmented contrast we use in \secref{sec:eval}.
This turns the existing log/trace dataset ecosystem into a first-class axis of ablation for distributed-systems agent debugging evaluation.

\section{Dataset Details}
\label{appendix:dataset-details}

\subsection{Per-Project Coverage Breakdown}

\begin{table}[h]
    \centering
    \caption{Per-project coverage of \sysname. \sysname spans \ddbenchReleaseSysSize open-source distributed systems across five languages and six system families, partitioned into three reasoning-scope tiers (tier-1, tier-2, tier-3). }
    \label{tab:coverage}
    \small
        \begin{tabular}{@{}lllrrrrr@{}}
            \toprule
            \textbf{Project} & \textbf{Language} & \textbf{System family} & \textbf{Number of Cases} & \textbf{Tier-1} & \textbf{Tier-2} & \textbf{Tier-3} \\
            \midrule
            Dragonfly  & C++    & kv-store        & 14 & 1 & 1 & 12 \\
            NATS       & Go     & messaging       &  9 & 5 & 3 & 1 \\
            Kvrocks    & C++    & kv-store        &  8 & 4 & 3 & 1 \\
            etcd       & Go     & kv-store        &  7 & 5 & 2 & 0 \\
            hashicorp/raft & Go & consensus       &  6 & 4 & 2 & 0 \\
            Consul     & Go     & service-mesh    &  3 & 2 & 1 & 0 \\
            Kafka      & Java   & streaming       &  3 & 3 & 0 & 0 \\
            ZooKeeper  & Java   & consensus       &  3 & 3 & 0 & 0 \\
            BookKeeper & Java   & consensus       &  2 & 2 & 0 & 0 \\
            raft-rs    & Rust   & consensus       &  2 & 0 & 2 & 0 \\
            Ra         & Erlang & consensus       &  1 & 1 & 0 & 0 \\
            RabbitMQ   & Erlang & messaging       &  1 & 1 & 0 & 0 \\
            RocksDB    & C++    & storage-engine  &  1 & 0 & 0 & 1 \\
            \midrule
            \textbf{Total} & 5 langs & 6 families & \textbf{60} & \textbf{31} & \textbf{15} & \textbf{14} \\
            \bottomrule
        \end{tabular}
\end{table}

\tabref{tab:coverage} reports the full per-project breakdown of \sysname's \ddbenchReleaseSize{} cases across the \ddbenchReleaseSysSize{} projects we mine from, annotated with each project's main implementation language, system family, and per-tier case count.
Three observations follow from the table.
First, case count per project is skewed. Dragonfly contributes 14 cases while Ra, RabbitMQ, and RocksDB each contribute only 1, which reflects how active each project's issue tracker is and how often its confirmed bugs satisfy our cross process reasoning criterion rather than any sampling quota on our side.
Second, tier composition varies sharply by system family. Consensus libraries (etcd, hashicorp/raft, Kafka, ZooKeeper, BookKeeper, Ra) place nearly every case in \tierone{} because correctness in these systems is defined by multi-node protocol invariants, whereas Dragonfly contributes 12 of its 14 cases to \tierthree{} because many of its confirmed bugs localize within a single process even though the deployment is distributed.
Third, programming language coverage follows the production distribution of distributed infrastructure rather than a uniform sample. Go (25 cases) and C++ (23 cases) dominate, Java contributes 8, and Erlang and Rust each contribute 2 in this dataset release.

\subsection{Licenses}
\label{appendix:license}

\begin{table}[h]
    \centering
    \caption{Per-project license and upstream repository for the \ddbenchReleaseSysSize open-source systems in \sysname.}
    \label{tab:licenses}
    \small
        \begin{tabular}{@{}lll@{}}
            \toprule
            \textbf{Project} & \textbf{License} & \textbf{Link} \\
            \midrule
            Dragonfly      & BSL 1.1               & \url{https://github.com/dragonflydb/dragonfly} \\
            NATS           & Apache-2.0            & \url{https://github.com/nats-io/nats-server} \\
            Kvrocks        & Apache-2.0            & \url{https://github.com/apache/kvrocks} \\
            etcd           & Apache-2.0            & \url{https://github.com/etcd-io/etcd} \\
            hashicorp/raft & MPL-2.0               & \url{https://github.com/hashicorp/raft} \\
            Consul         & BSL 1.1               & \url{https://github.com/hashicorp/consul} \\
            Kafka          & Apache-2.0            & \url{https://github.com/apache/kafka} \\
            ZooKeeper      & Apache-2.0            & \url{https://github.com/apache/zookeeper} \\
            BookKeeper     & Apache-2.0            & \url{https://github.com/apache/bookkeeper} \\
            raft-rs        & Apache-2.0            & \url{https://github.com/tikv/raft-rs} \\
            Ra             & Apache-2.0 / MPL-2.0  & \url{https://github.com/rabbitmq/ra} \\
            RabbitMQ       & Apache-2.0 / MPL-2.0  & \url{https://github.com/rabbitmq/rabbitmq-server} \\
            RocksDB        & Apache-2.0 / GPL-2.0  & \url{https://github.com/facebook/rocksdb} \\
            \bottomrule
        \end{tabular}
\end{table}

\tabref{tab:licenses} lists the license and upstream repository URL for every project in \sysname.
All \ddbenchReleaseSysSize{} projects are released under permissive terms (Apache 2.0, MPL 2.0, or BSL 1.1), which allows us to package each case as a pinned faulty worktree plus a containerized reproduce oracle and redistribute the bundle for benchmark use.
Two projects (Ra and RabbitMQ) are dual-licensed under Apache 2.0 and MPL 2.0, and RocksDB is dual-licensed under Apache 2.0 and GPL 2.0; in each case we rely on the Apache 2.0 grant for redistribution.
Dragonfly and Consul are released under BSL 1.1, which restricts commercial service offerings but permits research redistribution and benchmark use, the scope under which \sysname operates.
No project in \sysname is released under a license that requires us to relicense derivative artifacts, and no case in the release ships modified upstream source. 
The faulty worktree for each case is a pinned commit of the original repository with no edits applied on top.

\section{Agent-Assisted Mining Methodology}
\label{appendix:semi-agentic}

\subsection{Why Semi-Agentic?}

We elaborate on the design rationale behind the semi-agentic pipeline introduced in \secref{sec:benchmark}.

\parab{Why not purely manual curation?}
Identifying cross-process bugs reliably requires deep familiarity with each system's protocols and failure modes. 
For instance, distinguishing a genuine linearizability violation from a superficially similar client-side retry bug demands protocol-level expertise and substantial manual effort. 
Scaling such review across \ddbenchNumSystems{} systems and thousands of candidate issues demands a level of domain knowledge and expertise that is not realistic for pure manual curation.

\parab{Why not purely automated scraping?}
Faulty cases whose root-causing genuinely requires cross-process or interleaving-aware reasoning are a small fraction of any project's issue tracker, and the discriminator is rather semantic. 
Namely, it cannot be easily expressed as a regex over titles or labels, nor reliably recovered from issue metadata alone. 
Agents, in contrast, can perform semantic analysis by reading discussion threads and linked PRs and applying a tightly scoped specification of what qualifies as a cross-process bug that \sysname demands.

\parab{Why the specific division of labor.}
The pipeline assigns to agents the tasks where breadth and heavy domain-expertise dominate, \eg walking issue trackers, drafting symptom files, hardening reproducers, producing debug-context bundles.
This pipeline preserves for human reviewers the tasks where ground-truth judgement dominates, \eg final acceptance, quality gating, and sanitization of any upstream-fix leakage. 
Each agent stage operates under a narrow skill specification and is followed by either an automated verification pass (e.g., oracle verification) or a human review pass.

\subsection{Agentic Mining Protocol}
\label{appendix:mining-protocol}

We document the operational protocol that the agent follows during the mining, synthesis, and verification stages summarized in \secref{sec:benchmark}.
The protocol is codified as a set of machine-readable specifications shipped with the \sysname release, and every agent stage in our pipeline executes against a subset of these specifications.
The protocol decomposes into four phases: candidate admission, case-artifact synthesis, hermetic containerization, and pre-merge validation.

\parab{Phase 1: Candidate admission.}
Before any artifact is synthesized, a candidate bug must pass three gatekeeping tests that together enforce the cross-process reasoning contract.
The \emph{dual-reasoning test} rejects candidates for which a stack trace or an error log alone is sufficient for diagnosis, as well as candidates for which static code reading alone is sufficient.
The sweet spot is bugs where runtime signals point toward a subsystem but code reasoning is required to understand why that subsystem fails under the observed conditions.
The \emph{fix-locus test} rejects candidates whose fix is scattered across many unrelated modules with no identifiable center of gravity, since such cases test breadth of refactoring rather than depth of diagnosis.
The \emph{issue-text test} rejects candidates whose GitHub issue thread already names the faulty function or describes the fix strategy, since an agent could derive the patch from the issue alone and bypass the diagnostic process the benchmark is designed to measure.
The agent applies these tests in sequence and records, for each admitted candidate, the fix commit hash and the buggy commit.

\parab{Phase 2: Case artifact synthesis.}
An admitted candidate is expanded into eight per-case artifacts plus one isolated solution document.
The per-case artifacts are \texttt{case.yaml} (machine-readable metadata consumed by the orchestrator), \texttt{symptom.md} (observer-facing bug description), \texttt{reproduce.md} (human-readable reproduction steps), \texttt{verification.md} (documented correctness oracle with PASS/FAIL criteria), \texttt{setup.md} (human-readable build instructions), \texttt{setup.sh} (executable clone-and-build script with a Docker-first path and host fallback), \texttt{reproduce.sh} (executable trigger-and-verify script with the same dual-path structure), and \texttt{README.md} (entry point linking the other files).
The solution document lives in a separate \texttt{cases-solution/} tree to prevent answer leakage during evaluation, and it is required to include both a \emph{fix locus} (component name plus the one to three files changed) and a \emph{diagnosis chain} that documents the reasoning path from symptom to root cause, explicitly annotating where runtime signals and where code reasoning each contribute.
The solution file is what human reviewers read during acceptance and what the oracle-verification stage consults when resolving the case from the solution.

\parab{Phase 3: Hermetic containerization.}
Every case runs inside a deterministic, layered Docker image so that reproduction does not drift with the host environment.
The base layer is \texttt{ubuntu:24.04}, on top of which sit language base images (\texttt{bug-bench/base-cpp}, \texttt{bug-bench/base-go}, and analogous images for Java, Erlang, and Rust) that install the pinned toolchain, debuggers, and standard client utilities such as \texttt{redis-cli}, \texttt{curl}, and \texttt{netcat}.
Each project adds a thin image on top of its language base with only the build dependencies that project requires, and no source code is baked into any image.
Source is instead mounted at runtime through a per-case Docker volume of the form \texttt{bug-bench-case-\textless case\textgreater-src} that holds a git repository containing only the ancestry of the case's buggy commit. That repository is fetched from a per-project full-history cache (\texttt{bug-bench-\textless project\textgreater-cache}) that is mounted only during setup, so no commit after the buggy commit, including the upstream fix, is visible to the agent.
Elevated capabilities are declared per case in \texttt{case.yaml} under \texttt{docker.caps} (for instance \texttt{NET\_ADMIN} for cases that manipulate network namespaces) and \texttt{SYS\_PTRACE} is always granted to support in-container debuggers.
For environments without Docker, every \texttt{setup.sh} and \texttt{reproduce.sh} falls through to a host path that installs dependencies via \texttt{apt-get} under Ubuntu 24.04, which we retain only as a fallback since it sacrifices isolation.

\parab{Phase 4: Pre-merge validation.}
Before a case is admitted into the agent release package (before the human review), the protocol requires it to pass a fixed sequence of automated checks that jointly enforce oracle soundness, hermetic reproducibility, and absence of answer leakage.
\textit{Oracle soundness} is checked in both directions: \texttt{reproduce.sh} must report \texttt{FAIL} when run on the buggy commit and must report \texttt{PASS} when run on the commit produced by applying the solution document, so that the oracle has neither false positives (buggy code passing) nor false negatives (fixed code failing).
\textit{Hermetic reproducibility} is checked by running the full \texttt{setup} and \texttt{reproduce} pipeline end-to-end inside a fresh container, by running setup a second time to confirm idempotency (clone and build are skipped when artifacts already exist), and by running the full pipeline after a \texttt{clean} to confirm the case recovers from a clean slate.
\textit{Absence of answer leakage} is checked by cross-referencing \texttt{symptom.md} against \texttt{solution.md} for a fixed list of prohibited patterns.
These patterns are:
\begin{itemize}
    \item function or method names that appear in the fix diff;
    \item struct-field names or internal constants from the buggy code path;
    \item file paths that point to the fix location;
    \item descriptions of the fix strategy (\eg ``make X idempotent'');
    \item error strings visible only in source code rather than in user-facing output;
    \item PR or commit references that identify the fix.
\end{itemize}
A case that fails any check is either repaired and re-verified or rejected from the release.

\parab{Protocol artifacts.}
The full protocol, including the selection criteria, the case anatomy specification, the containerization rules, the authoring walkthrough, the pre-merge checklist, and the templates the agent copies from, is released with \sysname (in \texttt{mining-protocol} folder) so that future case additions and independent replications follow the same contract.

\subsection{Case Containerization Verification Prompt}
\label{appendix:prompt-verify-cases}

This prompt drives the hermetic reproducibility check of Phase 4 in the mining protocol (\secref{appendix:mining-protocol}).
The agent runs \texttt{bench.sh setup} followed by \texttt{bench.sh reproduce} across the target cases and classifies each outcome, so that the release only admits cases whose faulty worktree builds cleanly inside the pinned container and whose \texttt{reproduce.sh} correctly reports \texttt{FAIL} on the buggy commit.
The prompt additionally carries a static anti-pattern scan that flags fix-specific reproducers (for instance, tests extracted at runtime from the fix commit via \texttt{FIX\_COMMIT}, reflection-based assertions, or source-grep branches) so that fragile oracles can be flagged and remediated before evaluation.
The prompt below is a shortened version of the skill file shipped with \sysname. 
We extract only the procedural core and omit implementation boilerplate, example report templates, and project-specific remediation tables.

\begingroup
\definecolor{promptheader}{HTML}{3C6FA6}
\definecolor{promptbody}{HTML}{DCE6F3}

\begin{tcolorbox}[
  enhanced,
  breakable,
  colback=promptbody,
  colframe=promptheader,
  coltitle=white,
  colbacktitle=promptheader,
  fonttitle=\small\normalfont\bfseries,
  title={Case Containerization Verification Prompt (shortened)},
  halign title=center,
  arc=2mm,
  boxrule=0pt,
  left=3mm, right=3mm, top=2mm, bottom=2mm,
  toptitle=1.5mm, bottomtitle=1.5mm,
]
\begin{Verbatim}[breaklines=true, breakanywhere=true, fontfamily=rm, formatcom=\normalfont, fontsize=\footnotesize]
Run `bench.sh setup` followed by `bench.sh reproduce` across the specified cases and produce an aggregate containerization report.

Procedure:

1. Determine the case list (all cases under `cases/`, or the caller supplied subset). Validate that each case directory exists and contains `case.yaml`.

2. Pre flight oracle quality scan. For each `reproduce.sh`, flag (but do not block) the following anti patterns that produce fragile, fix specific oracles:
   - FIX_COMMIT dependency (`FIX_COMMIT=` variable)
   - `git show` or `git checkout` reaching into the fix commit
   - Reflection (`getDeclaredField`, `getDeclaredMethod`) used as the PASS/FAIL assertion
   - Source grep used as the PASS/FAIL branch
   - `sed -i` targeting production source paths

3. Run verification sequentially (never in parallel, to avoid Docker resource contention). For each case:
     SETUP_EXIT=$(./bench.sh setup <case-name>; echo $?)
     if [ $SETUP_EXIT -eq 0 ]; then
         REPRO_EXIT=$(./bench.sh reproduce <case-name>; echo $?)
     fi
   Enforce a 30 minute setup timeout and a 5 minute reproduce timeout. Capture the last 20 lines of output for the report.

4. Classify each case into one of:
   - VERIFIED: setup OK and reproduce exited 1 with "FAIL:" in output (bug present on the buggy commit, as expected).
   - FIXED: setup OK and reproduce exited 0 with "PASS:" in output (unexpected; the buggy_commit is likely wrong).
   - SETUP_FAIL: setup exited non zero.
   - REPRO_ERROR: reproduce exited non zero without "FAIL:" in the output (script error, not bug detection).
   - TIMEOUT: setup or reproduce exceeded the timeout.

5. Produce a markdown report with a per case status table, a summary row count per status, and, for every non VERIFIED case, the last 20 lines of output plus a suggested next step (check build dependencies, verify buggy_commit resolves, increase timeout).

6. After the main run, list cases that passed verification but still tripped a pre flight anti pattern under a "Fragile Oracles" section, with the recommended remediation (inline tests as heredocs, behavioral Mockito verify, ship regression tests as static inject_test/ files).

Notes:
 - A VERIFIED result only confirms the oracle fires on the buggy commit; soundness in the other direction is covered by the Oracle Verification prompt.
 - Repeated setup failures on cases from the same project typically indicate a stale project Dockerfile rather than per case bugs.
\end{Verbatim}

\end{tcolorbox}
\endgroup

\subsection{Oracle Verification}
\label{appendix:prompt-verify-oracle}

This subsection details the oracle-verification step from Phase 4 of the mining protocol (\secref{appendix:mining-protocol}): the two-direction soundness check, the iterative hardening loop applied when a case fails first verification, the design choice not to verify against the upstream patched commit, the role of the per-case reference solution, the manual acceptance pass, and the verification prompt the agent runs.

\parab{Two-direction soundness check.}
For each case, the agent first runs the oracle on the buggy commit and confirms \texttt{FAIL}, then applies the fix documented in \texttt{solution.md} inside the container, rebuilds, and reruns the oracle to confirm \texttt{PASS}.
A case is admitted only if the oracle is sound in both directions, carrying no false positive (buggy code passing) and no false negative (fixed code failing).
A fix-specificity scan additionally distinguishes \emph{sound} oracles from \emph{sound-but-fragile} oracles, since a fragile oracle can pass the two-direction check while still rejecting valid alternative fixes at evaluation time (\eg runtime-extracted tests, reflection-based assertions, source-grep branches).

\parab{Iterative hardening loop.}
A reproducer that passes the two-direction check on the first attempt is rare; the typical path is iterative hardening.
Each iteration tightens one of three knobs: setup determinism (pinning seeds, fixed cluster sizes, blocking until quorum is reached), timing (sleep windows replaced with explicit wait-for-condition probes, stress phases lengthened to expose interleavings), and repetition (the trigger phase is run $k$ times and the case is admitted only if it fires every time).
We cap the loop at five iterations: if a candidate's oracle has not stabilized by then, the case is dropped from the release rather than patched around.
This preserves the invariant that every released oracle is reliable on its own terms, at the cost of admitting fewer cases.

\parab{Why not verify against the upstream patched commit.}
A natural alternative would be to define the oracle as ``\texttt{reproduce.sh} fails on the buggy commit and passes on the upstream fix commit.''
We rejected this because distributed-system fixes often land months or years after the bug was filed, and by the time the upstream patch is merged the surrounding code has frequently drifted (API renames, dependency bumps, build-system changes).
An oracle pinned to the upstream patched commit then conflates an underlying-bug regression with API drift, which would inflate false negatives and reduce benchmark separability.
We therefore verify against a per-case reference solution (below) applied directly on top of the buggy commit, where the patch context is stable.

\parab{Reference solution as Markdown, not gold patch.}
The reference solution shipped in \texttt{cases-solution/} is a human-readable Markdown document, not a checked-in code diff.
Two reasons motivate this choice.
First, a Markdown solution forces the curator to articulate the diagnosis chain --- the artifact human reviewers read during acceptance and the artifact a future researcher needs to audit a contested case.
Second, since the released oracle accepts \emph{any} patch restoring the violated invariant rather than matching a textual fix, a single canonical diff would mislead readers into believing a textual match is required.
The Markdown solution is consumed only by the oracle-verification stage (which applies the fix described in the document and reruns the oracle) and by human reviewers; it is never exposed at evaluation time.

\parab{Manual acceptance.}
After the automated checks above, every surviving case is reviewed by the five distributed-systems researchers who authored \sysname.
Reviewers re-read the agent-produced \texttt{symptom.md}, \texttt{context/}, and \texttt{solution.md} against the original GitHub issue and PR, and perform a final round of symptom and debug-context sanitization on top of the automated leakage scan.
A case is admitted only when reviewers sign off; cases that fail review are repaired and re-verified, or dropped from the release.

\parab{Verification prompt.}
The prompt below is a shortened version of the skill file shipped with \sysname.
We extract the procedural core and omit the report templates and the full remediation guidance.

\begingroup
\definecolor{promptheader}{HTML}{3C6FA6}
\definecolor{promptbody}{HTML}{DCE6F3}

\begin{tcolorbox}[
  enhanced,
  breakable,
  colback=promptbody,
  colframe=promptheader,
  coltitle=white,
  colbacktitle=promptheader,
  fonttitle=\small\normalfont\bfseries,
  title={Oracle Verification Prompt (shortened)},
  halign title=center,
  arc=2mm,
  boxrule=0pt,
  left=3mm, right=3mm, top=2mm, bottom=2mm,
  toptitle=1.5mm, bottomtitle=1.5mm,
]
\begin{Verbatim}[breaklines=true, breakanywhere=true, fontfamily=rm, formatcom=\normalfont, fontsize=\footnotesize]
Verify that a case's reproduce script is a sound correctness oracle: it must FAIL on the buggy commit AND PASS after applying the documented fix.

Procedure:

1. Read case metadata. For each case, load `cases/<case-name>/case.yaml` and `cases-solution/<case-name>/solution.md`. Extract the buggy commit, the fix commit hash (from solution.md "Commit" field), and the build command. If no fix commit is recorded, abort with an explicit error, since soundness cannot be verified in both directions.

2. Direction 1 (buggy commit -> expect FAIL).
     ./bench.sh setup <case-name>
     ./bench.sh reproduce <case-name>
   Expected: exit code 1 with "FAIL:" in output. Record the exit code, the last 10 lines of output, and the verdict FAIL (correct) or PASS (unsound, a false positive).

3. Fix specificity scan. Before running Direction 2, scan `reproduce.sh` for anti patterns that pass the soundness check but reject valid alternative fixes:
   - FIX_COMMIT dependency (HIGH): test pulled from the fix commit.
   - `git show` or `git checkout` from the fix commit (HIGH): files pulled from the fix tree at run time.
   - Reflection used as the PASS/FAIL assertion (HIGH): checks code structure rather than behavior.
   - Source grep branching (MEDIUM): detects fix text instead of fix effect.
   - `sed -i` on production source (MEDIUM): injects accessors to read internal state.
   If a HIGH severity pattern is found, the oracle is flagged SOUND-BUT-FRAGILE in the final report even if both directions pass, together with a remediation pointer (inline test as a heredoc, replace reflection with a behavioral mock, ship regression tests as static inject_test/ files).

4. Direction 2 (fix commit -> expect PASS). Apply the fix inside the container and rerun the oracle:
     ./bench.sh shell <case-name> --cmd "
       cd /workspace/worktrees/<case-name> &&
       git checkout <fix-commit> &&
       [git submodule update --init --recursive] &&
       <rebuild-command> &&
       bash /bench/reproduce.sh
     "
   Expected: exit code 0 with "PASS:" in output. Record the exit code, the last 10 lines of output, and the verdict PASS (correct) or FAIL (unsound, a false negative). For race or flaky oracles, repeat at least three times.

5. Reset the worktree back to the buggy commit (`./bench.sh reset <case-name>`) so the case is clean for downstream use.

6. Emit a per case report that states both directions, their verdicts, and the overall oracle verdict:
   - SOUND: both directions correct and no fix specific anti pattern found.
   - SOUND-BUT-FRAGILE: both directions correct but at least one HIGH severity anti pattern; the oracle works as a regression test but will reject valid alternative fixes.
   - UNSOUND: either direction fails. Classify as false positive (buggy code passes) or false negative (fixed code fails), and point at the likely cause (timing sensitivity, insufficient iterations, wrong condition, wrong fix commit).
\end{Verbatim}

\end{tcolorbox}
\endgroup

\subsection{Debug-Context Generation Prompt}
\label{appendix:prompt-context}

This prompt drives debug-context synthesis for the context-augmented condition introduced in \secref{sec:benchmark}.
The agent acts as a developer who has read only \texttt{symptom.md} and has entered the container at the buggy commit, gathers debugging clues using standard tools (stack traces, race-detector reports, goroutine or thread dumps, anomalous logs, test failures, and short code-reading notes), and writes them into a per-case \texttt{context/} folder.
Every debug-context bundle in this release was produced by running this prompt in Claude Code with Claude Opus~4.6, followed by the human sanitization review described in~\secref{sec:benchmark}.
The prompt enforces two invariants critical to benchmark validity: (a) every piece of debug context must be obtainable with only the artifacts visible to the evaluation agent (no \texttt{reproduce.sh}, no injected regression tests, no access to \texttt{solution.md} during investigation), and (b) the debug context must remain one step removed from the root cause (it may point at an area worth investigating, but it must not name the defect, contrast buggy with correct code, or trace the full causal chain).
The prompt below is a shortened version of the skill file shipped with \sysname. 
We extract the procedural core and the anti-leak checklist, and omit the per-language tool examples, placeholder templates, and file-naming table.

\begingroup
\definecolor{promptheader}{HTML}{3C6FA6}
\definecolor{promptbody}{HTML}{DCE6F3}

\begin{tcolorbox}[
  enhanced,
  breakable,
  colback=promptbody,
  colframe=promptheader,
  coltitle=white,
  colbacktitle=promptheader,
  fonttitle=\small\normalfont\bfseries,
  title={Debug-Context Generation Prompt (shortened)},
  halign title=center,
  arc=2mm,
  boxrule=0pt,
  left=3mm, right=3mm, top=2mm, bottom=2mm,
  toptitle=1.5mm, bottomtitle=1.5mm,
]
\begin{Verbatim}[breaklines=true, breakanywhere=true, fontfamily=rm, formatcom=\normalfont, fontsize=\footnotesize]
You are a developer investigating a bug from inside the Docker container, starting from `symptom.md` and working only with artifacts visible to the evaluation agent at the buggy commit. Your job is to gather debugging clues (crash backtraces, anomalous logs, race detector reports, test failures, server behavior, code reading notes) and write them into `cases/<case-name>/context/` without revealing the fix.

Core principle. The container holds the full built source and standard tools (GDB, Go toolchain, Maven, cargo, rebar3, redis-cli, etcdctl, jstack, and similar). It does NOT hold `reproduce.sh`, `bench.sh`, injected test files, the `cases/` tree, or the `cases-solution/` tree. Every piece of debug context must be obtainable from that strictly limited view.

Two debug-context shapes, a case may have both:
 - Runtime context: raw output from a tool, with a single provenance line naming the command. No interpretation, no annotations, no decoded constants. Only the tool's own output.
 - Code investigation context: a short "What was searched", a block of real grep/sed/cat output from inside the container (not hand transcribed), and a one to two sentence developer hypothesis. The hypothesis must stop at "this area is worth investigating further" and must NOT name the exact defect, contrast buggy with correct code, or trace the full causal chain.

Procedure:

1. Read only `symptom.md`, `reproduce.md`, `verification.md`, and `case.yaml`. Do NOT read `solution.md` yet. Reading the solution before investigating contaminates the debug context by letting the agent reverse engineer clues from the known answer.

2. Read `reproduce.sh` as a guardrail only, to detect whether it injects test files. Any test file it injects is off limits for debug context, since the evaluation agent does not have that file.

3. Enter the container at the clean buggy commit via `bench.sh setup` then `bench.sh shell` (never via `bench.sh reproduce`). Reset first if the workspace was dirtied.

4. Investigate. Pick tools appropriate to the bug category: GDB or ASAN for C++ crashes, `go test -race` or goroutine dumps for Go, `jstack` or existing JUnit tests for Java, `rebar3 ct` for Erlang, `cargo test` for Rust. Capture all output to `/tmp/context-raw-<case-name>.txt` with `tee` so every written line has a traceable source.

5. Write debug-context files under `cases/<case-name>/context/`. Runtime files carry only a provenance line and raw output. Code investigation files carry "What was searched", actual grep/sed output with surrounding context (20 to 40 lines, not a three line snippet), and a one to two sentence hypothesis. The provenance line must reference a command a developer runs inside the container (`go test`, `gdb`, `mvn test`, `grep -n`), never `bench.sh` or host paths.

6. Anti leak checks. Only now, read `solution.md` and run these checks against the drafted debug context. Delete or rewrite any file that fails:
   - Agent reproducibility: could an agent inside the container actually run this exact command on the buggy commit and get this output?
   - Injected test: is this output from a test that `reproduce.sh` would inject? If yes, delete.
   - Native output: every line in the runtime `## Output` block is raw tool output, with no added commentary.
   - Fix from context: could a reader identify the exact function and change from this file alone? If yes, the file is a diagnosis, not a clue.
   - Curation: for runtime output, include realistic surrounding context (other goroutines in a dump, surrounding log lines) so the smoking gun is not artificially isolated.
   - Term check: the debug context does not carry variable names from the fix diff or internal constants decoded from source.
   - Solution comparison: the debug context does not restate the solution's root cause in different words.

7. Report the investigation approach, the files written, and confirm that the leak checks and the agent reproducibility check both passed.
\end{Verbatim}

\end{tcolorbox}
\endgroup

\subsection{Sanitization Prompt}
\label{appendix:prompt-sanitize}

This prompt implements the automated leakage pass described in Phase 4 of the mining protocol (\secref{appendix:mining-protocol}) and complements the human sanitization step noted in \secref{sec:benchmark}.
The agent cross-references \texttt{symptom.md} against \texttt{solution.md} along both a lexical axis (prohibited identifiers, paths, commit hashes, prior issue references) and a semantic axis (internal-mechanism disclosure, regression provenance, subsystem or code-path narrowing, inverted fix hints, and diagnostic equivalence to the full solution).
Each finding is triaged into one of three \emph{information tiers}, which are an internal sanitization construct and should not be confused with the reasoning-scope tiers (\tierone{}/\tiertwo{}/\tierthree{}) used elsewhere in the paper to partition cases.
The information tiers classify individual facts about a bug: \emph{observable} facts remain in \texttt{symptom.md}; \emph{runtime-obtainable} facts are extracted into the per-case \texttt{context/} slot; \emph{source-derived} facts are redacted entirely.
This tiering is the mechanism that keeps the symptom-only and context-augmented conditions cleanly separated across the benchmark.
The prompt below is a shortened version of the skill file shipped with \sysname.
We extract the tier model and the scan procedure, and omit the debug-context-file format templates, the per-oracle-type recommendation table, and the example violation reports.

\begingroup
\definecolor{promptheader}{HTML}{3C6FA6}
\definecolor{promptbody}{HTML}{DCE6F3}

\begin{tcolorbox}[
  enhanced,
  breakable,
  colback=promptbody,
  colframe=promptheader,
  coltitle=white,
  colbacktitle=promptheader,
  fonttitle=\small\normalfont\bfseries,
  title={Sanitization Prompt (shortened)},
  halign title=center,
  arc=2mm,
  boxrule=0pt,
  left=3mm, right=3mm, top=2mm, bottom=2mm,
  toptitle=1.5mm, bottomtitle=1.5mm,
]
\begin{Verbatim}[breaklines=true, breakanywhere=true, fontfamily=rm, formatcom=\normalfont, fontsize=\footnotesize]
Scan `cases/<case-name>/symptom.md` for information leakage (both lexical and semantic), triage each finding into one of three tiers, extract tier 2 content into `context/`, and redact tier 3 content entirely.

Three tier information model:
 - Tier 1 (keep in symptom.md): what an operator can observe by running the software. Crash messages, wrong CLI output, missing data, timeouts, user facing errors.
 - Tier 2 (extract to context/): what a developer can gather with standard runtime tools (logs, stack traces, goroutine dumps, metrics, config state, protocol level captures), provided it does not reveal the fix.
 - Tier 3 (redact entirely): anything that requires reading the project's source code to know. Root cause, fix location, internal mechanism, regression provenance. Goes nowhere.

Fresh developer test. Every sentence in symptom.md must be writeable by someone who has only run the software and observed the failure, never read the source.

Procedure:

1. Build the reference set. Load `cases-solution/<case-name>/solution.md` and extract the prohibited terms: function and method names cited in "Root cause" / "Fix description" / "Diagnosis chain"; variable, struct field, and constant names; file paths and component name from "Fix locus"; PR numbers and commit hashes; prior issue or PR IDs cited as regression sources.

2. Lexical scan of `symptom.md`. Grep for each prohibited term, for file path fragments, internal constants and magic numbers, and for PR, commit, or prior issue references.

3. Semantic scan of `symptom.md`. For each sentence, check these leakage shapes:
   - Internal mechanism disclosure: sentences that describe how the bug works internally rather than what the user sees.
   - Regression provenance: references to a specific prior change, issue, PR, or commit that introduced the bug.
   - Subsystem or code path narrowing: naming internal protocol phases, subsystems, or code paths not visible in user facing output.
   - Fix direction hints: sentences whose inverse is effectively the fix ("does not populate X", "does not call Y after Z").
   - Diagnostic equivalence: the symptom file, taken as a whole, conveys the same information as the solution's "Root cause" or "Diagnosis chain" even in different words.

4. Triage each finding:
   - Externally observable without tools -> tier 1, keep in symptom.md, reword if the phrasing leaks.
   - Obtainable with runtime tools AND does not reveal the fix location -> tier 2, extract into `cases/<case-name>/context/`.
   - Requires source knowledge OR reveals fix location or strategy -> tier 3, redact entirely.

5. Extract tier 2 findings into debug-context files (e.g., `stack-trace.md`, `logs.md`, `metrics.md`, `config-state.md`, `runtime-behavior.md`, `test-output.md`, `error-output.md`), each with a "How this was obtained" section (standard runtime tool, not source reading), an "Evidence" section (raw tool output), and a "What this tells a developer" section (one to two sentences of direction, without naming the fix or the buggy function). Sanitize stack traces by preserving full shape (call depth, thread or goroutine count) while redacting frames whose name alone would identify the buggy function.

6. Rewrite `symptom.md` to contain tier 1 only. Replace internal data structure names with observable consequences, protocol phase names with user visible events, and regression references with version ranges. Append a single line pointer to the new context directory if debug-context files were created.

7. Re run both the lexical and the semantic scans against the rewritten `symptom.md` and every debug-context file. The symptom file must have zero violations. Debug-context files must have zero tier 3 violations. Apply the litmus tests: grep test, diagnosis test, fresh developer test, narrowing test, regression diff test; plus, for debug context, the fix from context, curation, and interpretation tests.

8. Report: a per case table of findings (status, type, tier, action taken), the list of debug-context files created, a summary count per tier, and the remaining violation count (must be zero).
\end{Verbatim}

\end{tcolorbox}
\endgroup

\section{Benchmark Features}
\label{appendix:bench-features}

\subsection{Difficulty: Reasoning, Not Patch Size}

\begin{figure}[h]
    \centering
    \includegraphics[width=\textwidth]{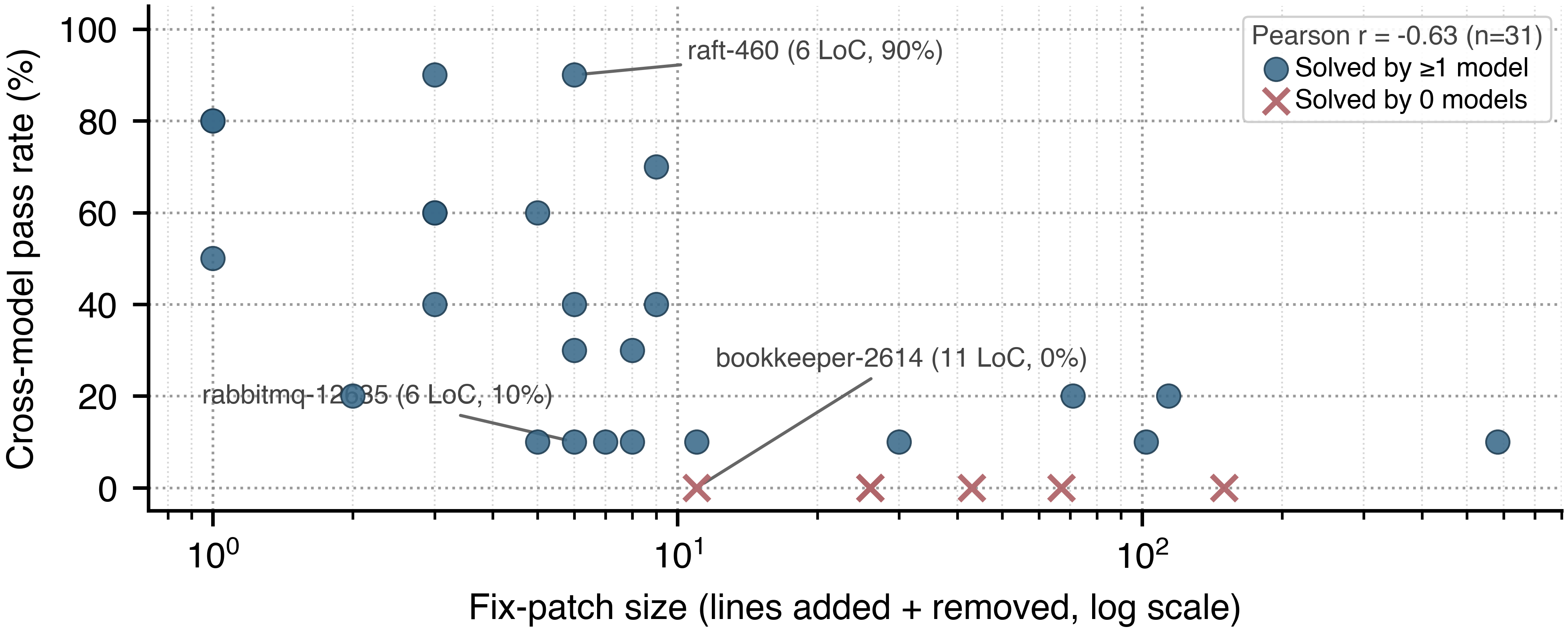}
    \caption{Fix-patch size vs.\ cross-model pass rate on \tierone{} ($n\!=\!31$). Each marker is a case; LoC is the upstream production-file diff size (lines added\,$+$\,removed, tests/mocks excluded), sourced from \texttt{cases-solution/*/solution.md} (packaged with \sysname) or from the upstream GitHub PR diff when not annotated directly in \sysname, with the smallest passing agent patch as a fallback (the executable oracle accepts only correct fixes, so passing-patch size faithfully bounds upstream scope). Red $\times$~markers denote cases unsolved by every model. Patch size correlates only moderately with difficulty ($r\!=\!-0.63$); within the $6$-LoC bucket alone, cross-model pass rate ranges from $10\%$ to $90\%$, confirming that reasoning load, not the surface size of the fix, drives difficulty.}
    \label{fig:patch-loc}
\end{figure}

The difficulty of \sysname does not come from the size of the fix.
We analyzed the size of the upstream fix patches for \sysname's tier-1 case-set and found most upstream patches are surgical: median 8 lines added plus removed across \tierone{}.
A manual review of these patches reveals that the underlying issues are typically a misplaced ordering, a missing fence, or a violated invariant rather than large-scale, multi-file code errors.
The difficulty comes from the path required to \emph{reach} the fix: the agent must reconstruct cross-process causality from incomplete logs, reason about non-deterministic interleavings, and validate hypotheses against protocol-level invariants that source-code reasoning alone cannot easily capture.
\figref{fig:patch-loc} quantifies this observation by plotting cross-model pass rate against fix-patch size.
An 11-line BookKeeper recovery-fencing fix is solved by zero of ten models, while a 6-line Raft log-cache ordering fix is solved by nine; within that same $6$-LoC bucket, a RabbitMQ quorum-queue snapshot crash is solved by only one, showing an $80$~pp difference on patches of identical size.
This is the reasoning mode that drives our headline result, producing a wide cross-model spread despite surgical patches. Single-process repair benchmarks structurally cannot exercise this reasoning mode on LLMs.

\subsection{Bounded Debug-Context Channel}
\label{appendix:context-bounds}

\begin{figure}[h]
    \centering
    \includegraphics[width=0.92\linewidth]{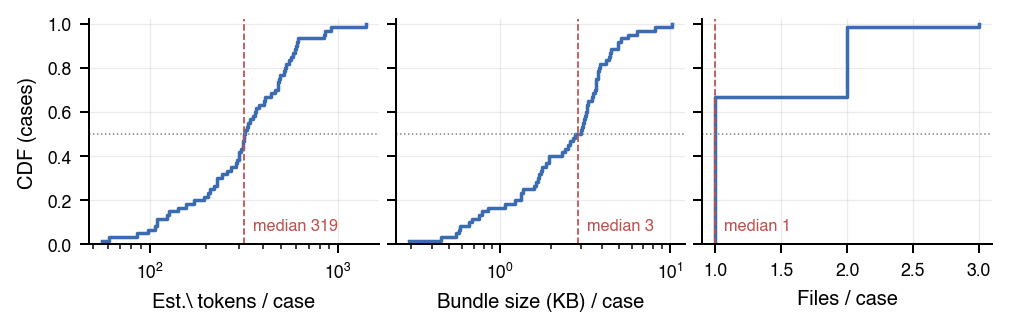}
    \caption{Per-case debug-context bundle size across all \ddbenchReleaseSize{} cases. Red dashed lines mark the median; the horizontal grey line is the 50th percentile. The channel is bounded across every case: median 321 estimated tokens, max 1{,}403 tokens, $\leq$3 markdown files per case.}
    \label{fig:context-cdf}
\end{figure}

\figref{fig:context-cdf} plots the per-case empirical CDF of debug-context-bundle size across all \ddbenchReleaseSize{} cases along three axes: estimated tokens (using a $1.3\times$ word-count proxy that approximates common tokenizers), bundle size in kilobytes, and file count.
The distribution is heavy-tailed but tightly bounded. 
Specifically, the median is 321 tokens with a maximum of 1{,}403, median 3.0\,KB with a maximum of 10.4\,KB, and at most three markdown files per case. 
Even the largest bundle occupies a small fraction of every evaluated model's context window, so model differences in the context-augmented condition cannot be attributed to differential context-budget pressure.

\begin{wraptable}[10]{r}{0.42\linewidth}
    \vspace{-1.5em}
    \small
    \centering
    \caption{Debug-context-bundle file-type composition (\ddbenchReleaseSize{} cases). Categories are not mutually exclusive, so shares do not sum to 100\%.}
    \label{tab:context-composition}
    \begin{tabular}{@{}lrr@{}}
        \toprule
        \textbf{Category} & \textbf{n} & \textbf{Share} \\ \midrule
        Code investigation       & 33 & 55.0\% \\
        Test output              & 25 & 41.7\% \\
        Crash output             & 12 & 20.0\% \\
        Runtime state            & 3  & 5.0\%  \\
        Race/sanitizer report  & 3  & 5.0\%  \\
        Debugger trace (GDB)     & 2  & 3.3\%  \\
        Goroutine dump           & 1  & 1.7\%  \\
        \bottomrule
    \end{tabular}
\end{wraptable}

\tabref{tab:context-composition} reports the file-type composition: code investigation appears in 55.0\% of cases, test output in 41.7\%, and crash output in 20.0\%, with the remainder split among structured runtime artifacts (goroutine dumps, race-detector and sanitizer reports, GDB backtraces, and runtime state captures). 
Most cases combine one or two of these categories, mirroring what a developer would realistically capture during a short triage session.

\subsection{\sysname Position Relative to Existing Benchmarks.}

\begin{table}[h]
    \centering
    \caption{\sysname versus existing code-repair and IT-operations benchmarks. Existing code-repair benchmarks supply an executable oracle but restrict to single-process workloads; IT-operations benchmarks exercise cross-service reasoning but ask agents to diagnose from telemetry rather than repair source. \sysname is the only benchmark that combines an executable repair oracle, cross-process bugs, and a controllable debug-context channel.}
    \label{tab:bench-comparison}
    \resizebox{\textwidth}{!}{%
        \begin{tabular}{@{}lccccc@{}}
            \toprule
            \textbf{Benchmark} & \textbf{Cross-process} & \textbf{Code fix} & \textbf{Executable oracle} & \textbf{Debug-context channel} & \textbf{Languages}  \\
            \midrule
            SWE-bench Verified~\citep{jimenez2023swebench}   & \xmark & \cmark & \cmark (unit tests)         & \xmark & 1 (Python)  \\
            Multi-SWE-Bench~\citep{zan2025multiswebench}     & \xmark & \cmark & \cmark (unit tests)       & \xmark & 8             \\
            DebugBench~\citep{tian2024debugbench}            & \xmark & \cmark & \cmark (unit tests)     & \xmark & 3              \\
            AIOpsLab~\citep{chen2025aiopslab}                & \cmark & \xmark & \cmark (telemetry)      & \cmark (telemetry) & ---         \\
            CloudOpsBench~\citep{wang2026cloudopsbench}      & \cmark & \xmark & \cmark (telemetry)      & \cmark (telemetry) & ---       \\
            \midrule
            \textbf{\sysname} (ours)                          & \cmark & \cmark & \cmark (E2E \& unit tests)  & \cmark (controlled) & 5       \\
            \bottomrule
        \end{tabular}%
    }
\end{table}

\tabref{tab:bench-comparison} situates \sysname against the closest existing benchmarks along four axes: whether the workload requires cross-process reasoning, whether the agent is asked to produce a code fix (as opposed to only a diagnosis), whether the harness exposes a controllable debug-context channel, and the language coverage.
Code-repair benchmarks, such as SWE-bench, Multi-SWE-Bench, and DebugBench, provide executable oracles but restrict to single-process workloads.
IT-operations benchmarks, such as AIOpsLab and CloudOpsBench, exercise cross-service reasoning but ask agents to identify operational problems from telemetry rather than repair source code.
\sysname is the only benchmark that combines an executable repair oracle, cross-process bugs, and a controlled debug-context channel, making the effect of debugging context on agent performance directly measurable.

\section{Benchmark Infrastructure}
\label{appendix:infrastructure}

This section walks through the end-to-end flow of running a single \sysname case against an agent under evaluation.
The flow is driven by a single orchestrator script, \texttt{bench.sh}, that exposes six verbs (\texttt{setup}, \texttt{reproduce}, \texttt{shell}, \texttt{agent}, \texttt{reset}, \texttt{clean}) and resolves each verb against a declarative \texttt{case.yaml} plus the layered Docker images described in \secref{appendix:mining-protocol}.
From the caller's perspective, evaluating one case is a three-step sequence: bring up a hermetic container with the faulty worktree built inside it (\texttt{setup}), launch the agent against that container under either of the two information conditions (\texttt{agent}), and score the resulting patch by running the hidden oracle (\texttt{reproduce}).
We describe each step in turn, then the reset and teardown path used between conditions and models.

\parab{Step 1: Container bring-up.}
\texttt{bench.sh setup <case>} reads \texttt{case.yaml} for the target case, resolves the project name, the repository URL, the buggy commit, the submodule flag, and the build command, and then performs four actions.
First, it builds the project's Docker image by chaining \texttt{ubuntu:24.04}, the language base image (\texttt{bug-bench/base-cpp}, \texttt{bug-bench/base-go}, and analogous images for Java, Erlang, and Rust), and the thin per-project image that adds only build dependencies.
Second, it creates or reuses a per-project cache volume (\texttt{bug-bench-\textless project\textgreater-cache}) holding a bare clone of the upstream repository, so that the clone cost is paid once per project; this cache is mounted only during setup, never into agent or scoring containers.
Third, it builds the case's own repository in a per-case volume (\texttt{bug-bench-case-\textless case\textgreater-src}) under \texttt{/workspace/worktrees/\textless case\textgreater}, fetching only the ancestry of the buggy commit from the cache and checking it out as a detached \texttt{HEAD}.
It then deletes every ref and remote, expires the reflog, and prunes unreachable objects, and it fails unless the repository has no refs and its commit objects are exactly the buggy commit's ancestry.
Commits after the buggy commit, including the upstream fix, are therefore absent from the repository rather than merely hidden, and giving each case its own volume keeps a sibling case at a later commit from exposing this case's fix.
Fourth, it runs the build command inside a throwaway container whose working directory is the worktree, writes the sanitized \texttt{SYMPTOM.md} into the worktree root, and generates a \texttt{REBUILD.sh} script tailored to the detected build system (\texttt{ninja}, \texttt{make}, \texttt{go build}, \texttt{cargo}, \texttt{mvn}, or \texttt{gradle}) so that the agent has a single, stable incremental-rebuild entry point regardless of the underlying toolchain.
After \texttt{setup}, the worktree holds the faulty source, the built binaries, \texttt{SYMPTOM.md}, and \texttt{REBUILD.sh}.

\parab{Step 2: Agent launch under a chosen information condition.}
\texttt{bench.sh agent <case> --addon <agent-name> [-{}-context]} starts a long-running container named after the case, installs the requested agent (Claude Code, Aider, Codex, or mini-swe-agent) via a runtime \emph{addon} script that caches the agent's binaries and language runtimes in a shared Docker volume (\texttt{bug-bench-tools} mounted at \texttt{/opt/tools}), and then execs the agent inside the container with its working directory pinned to the case worktree.
Addons decouple agent installation from the project Docker images, so adding a new agent requires a single shell script under \texttt{addons/} rather than a rebuild of each project image.
The two evaluation conditions differ in whether the per-case \texttt{context/} folder is mounted into the worktree as a read-only directory, and in the prompt, which directs the agent to that folder (Appendix~\secref{appendix:prompt}).
Under the symptom-only condition, the agent sees only \texttt{SYMPTOM.md}, the full source tree at the buggy commit, the built binaries, and \texttt{REBUILD.sh}.
Under the context-augmented condition, \texttt{bench.sh agent} additionally mounts \texttt{cases/\textless case\textgreater/context/} at \texttt{\textless worktree\textgreater/context/} so that the agent can list and read the debug-context files generated by the procedure in \secref{appendix:prompt-context}.
API keys (\texttt{ANTHROPIC\_API\_KEY}, \texttt{OPENAI\_API\_KEY}, and similar) are forwarded from the host environment, and agent configuration directories (\eg \texttt{\textasciitilde/.claude/}) are bind-mounted read-only.
The agent is invoked with the instance template documented in Appendix~\secref{appendix:prompt}. 
In summary, it reads \texttt{SYMPTOM.md}, inspects \texttt{context/}, browses and edits source files, iteratively runs \texttt{REBUILD.sh} to keep the worktree compiling, and signals task completion by printing a sentinel string.
A hard wall-clock budget bounds the trajectory, and trajectory logs are written to a mounted output directory on the host for later analysis.

\parab{Step 3: Oracle-based scoring.}
Once the agent terminates, \texttt{bench.sh reproduce <case>} is run in a fresh container against the same worktree volume.
\texttt{reproduce} first re-runs the incremental build so that any source edits made by the agent are compiled, then executes \texttt{/bench/reproduce.sh} inside the container with the case scripts bind-mounted at \texttt{/bench/} (read-only) and the built worktree at \texttt{/workspace/worktrees/\textless case\textgreater}.
The reproduce script implements the per-case correctness oracle described in Appendix~\secref{appendix:mining-protocol}.
It starts the system (for server-based cases), triggers the bug with a scripted client workload or an injected regression test, checks the outcome, and prints exactly one terminal line of the form \texttt{PASS: ...} or \texttt{FAIL: ...} with a matching exit code (0 for PASS, 1 for FAIL).
The oracle is hermetic in two respects.
First, the agent never sees \texttt{reproduce.sh} or anything under \texttt{/bench/} during its trajectory (the mount is added only at scoring time), so the agent cannot game the oracle by reading it.
Second, the oracle was validated in both directions by the procedure in Appendix~\secref{appendix:prompt-verify-oracle}, so a correct patch on the buggy commit should give a \texttt{PASS}.
Textual match against the upstream fix is \emph{not} required.
\sysname's headline metric is the fraction of cases on which \texttt{reproduce} returns \texttt{PASS} after the agent's trajectory. 
All other metrics (tokens consumed, number of actions, wall-clock latency) are derived from the trajectory logs captured during Step 2.

\parab{Step 4: Reset and cleanup between runs.}
Evaluating multiple models or both information conditions on the same case requires returning the worktree to its pristine buggy state between runs, without repaying the clone or build cost.
For this purpose, the command \texttt{bench.sh reset <case>} is provided to revert the worktree. 
It preserves \texttt{SYMPTOM.md} and \texttt{REBUILD.sh}, discards every source modification with \texttt{git checkout -- .} and \texttt{git clean -fd}, moves \texttt{HEAD} back to the recorded buggy commit if the agent moved it, restores the two preserved files, and re-initializes submodules (if applicable).
The agent evaluation container and the debug-context mount are stopped and discarded, but the per-case source volume and the addon tool cache are kept so that the next \texttt{agent} invocation pays only an incremental-rebuild cost.
\texttt{bench.sh clean <case>} removes a single case's volume (leaving the per-project cache and other cases untouched), and \texttt{bench.sh clean-all} removes every \sysname volume, image, and addon cache for a full reset.
Across the full benchmark, this design means that a same-project family of cases (for instance, the 14 Dragonfly cases) shares a single clone cache and a single project image, while every case still evaluates against its own isolated repository pinned to its own buggy commit.

\section{\sysname Prompt}
\label{appendix:prompt}

The templates below are the versions shipped with the \sysname harness. The runs reported in this paper used earlier revisions that differ in three respects. First, the debug-context folder was named \texttt{evidence/} and the corresponding block \texttt{<evidence>}, with ``debugging evidence'' in place of ``debug context''. Second, most runs used templates that did not yet include the constraint on ancestor commits (the line beginning ``You may look into''). Third, runs that drive the agent through mini-swe-agent's text-based interface rather than native tool calls also receive that interface's command-format instructions (one bash command per response, in a fenced block).

\subsection{System Prompt}

\begingroup
\definecolor{promptheader}{HTML}{3C6FA6}
\definecolor{promptbody}{HTML}{DCE6F3}

\begin{tcolorbox}[
  enhanced,
  breakable,
  colback=promptbody,
  colframe=promptheader,
  coltitle=white,
  colbacktitle=promptheader,
  fonttitle=\small\normalfont\bfseries,
  title={System Prompt},
  halign title=center,
  arc=2mm,
  boxrule=0pt,
  left=3mm, right=3mm, top=2mm, bottom=2mm,
  toptitle=1.5mm, bottomtitle=1.5mm,
]
\begin{Verbatim}[breaklines=true, breakanywhere=true, fontfamily=rm, formatcom=\normalfont, fontsize=\footnotesize]
You are a software debugging agent working on a distributed systems codebase.

You interact with the environment by calling the bash tool.
Think carefully before each command. Be systematic and precise.

Include a THOUGHT section before your command to explain your reasoning.
Failure to follow this format will cause your response to be rejected.
\end{Verbatim}

\end{tcolorbox}

\subsection{Symptom-Only Condition Instance Template}

\begingroup
\definecolor{promptheader}{HTML}{3C6FA6}
\definecolor{promptbody}{HTML}{DCE6F3}

\begin{tcolorbox}[
  enhanced,
  breakable,
  colback=promptbody,
  colframe=promptheader,
  coltitle=white,
  colbacktitle=promptheader,
  fonttitle=\small\normalfont\bfseries,
  title={Instance Template},
  halign title=center,
  arc=2mm,
  boxrule=0pt,
  left=3mm, right=3mm, top=2mm, bottom=2mm,
  toptitle=1.5mm, bottomtitle=1.5mm,
  fontupper=\small
]
\begin{Verbatim}[breaklines=true, breakanywhere=true, fontfamily=rm, formatcom=\normalfont, fontsize=\footnotesize]
<symptom_file>
A file called SYMPTOM.md in the current directory describes the bug you need to fix.
</symptom_file>

<instructions>
## Task

You are debugging a distributed systems codebase. The bug is described in SYMPTOM.md.
Your goal is to identify and fix the root cause by modifying source files in place.

{{task}}

## Constraints

- Read SYMPTOM.md first to understand the observable failure.
- You may freely browse the codebase, build, and run tests.
- You may access the internet for API docs, language references, and general knowledge.
- You MUST NOT search for the specific GitHub issue, pull request, or known fix.
  Do not search GitHub, Google, or any other source for the specific issue number or fix.
- You may look into ancenster commits, but not future commits in order to cheat.
- Make your fix by editing source files directly.
- You MUST ensure ./REBUILD.sh completes successfully after your changes.
  A failing REBUILD.sh means your fix is incomplete.
- Your fix should address the root cause, not just mask symptoms.
- Your should never early declare your finish unless you have very strong evidence that the bug is fixed.

## Workflow

1. Read SYMPTOM.md to understand the failure.
2. Explore the relevant source code to understand the component architecture.
3. Form a hypothesis about the root cause.
4. Implement a targeted fix (prefer minimal, correct changes).
5. If the fix doesn't solve the bug, go back to step 2. If you are confident that you have fixed the bug, go to step 6.
6. Before submitting the final output, you MUST run `./REBUILD.sh` and confirm it exits successfully. 
  Fix any compile errors before proceeding. (this is not related to your bug, it is just a sanity check)
7. When you are absolutely confident that the bug is fixed, submit the job completion by running: `echo COMPLETE_TASK_AND_SUBMIT_FINAL_OUTPUT`
  Do not combine it with any other command.
  <important>After this command, you cannot continue working on this task.</important>

## Command Execution Rules

Each response should include:

1. A **THOUGHT** section explaining your reasoning and plan
2. Other things.

**CRITICAL REQUIREMENTS:**

- Your response SHOULD include a THOUGHT section explaining your reasoning
- Directory or environment variable changes are not persistent across commands
- You can prefix any action with `cd /path/to/dir && ...` or write/load env vars from files

## Editing Files

### Create a new file:

```bash
cat <<'EOF' > newfile.go
package main
import "fmt"
func main() { fmt.Println("hello") }
EOF
```

### Edit files with sed:

```bash
# Replace all occurrences
sed -i 's/old_string/new_string/g' filename.go

# Replace only in specific line range
sed -i '10,20s/old_string/new_string/g' filename.go
```

### View file content:

```bash
nl -ba filename.go | sed -n '10,20p'
```
</instructions>
\end{Verbatim}

\end{tcolorbox}
\endgroup

\subsection{Context-Augmented Condition Instance Template}

\begingroup
\definecolor{promptheader}{HTML}{3C6FA6}
\definecolor{promptbody}{HTML}{DCE6F3}

\begin{tcolorbox}[
  enhanced,
  breakable,
  colback=promptbody,
  colframe=promptheader,
  coltitle=white,
  colbacktitle=promptheader,
  fonttitle=\small\normalfont\bfseries,
  title={Instance Template},
  halign title=center,
  arc=2mm,
  boxrule=0pt,
  left=3mm, right=3mm, top=2mm, bottom=2mm,
  toptitle=1.5mm, bottomtitle=1.5mm,
  fontupper=\small
]
\begin{Verbatim}[breaklines=true, breakanywhere=true, fontfamily=rm, formatcom=\normalfont, fontsize=\footnotesize]
<symptom_file>
A file called SYMPTOM.md in the current directory describes the bug you need to fix.
</symptom_file>

<debug_context>
The `context/` folder in the current directory contains debug context that a developer has gathered during their debugging process. This debug context may provide additional insights into the root cause of the bug.

**You MUST review the debug context early in your investigation:**
1. Run `ls context/` to see what files are available.
2. Read each file with `cat context/<filename>`.

Debug-context files may contain test outputs, log snippets, stack traces, crash reports, or code-level observations that can help you narrow down the root cause more quickly.
However, treat this debug context critically: it reflects one developer's investigation and may be incomplete.
</debug_context>

<instructions>
## Task

You are debugging a distributed systems codebase. The bug is described in SYMPTOM.md.
Your goal is to identify and fix the root cause by modifying source files in place.

{{task}}

## Constraints

- Read SYMPTOM.md first to understand the observable failure.
- Review the context/ folder to see what debug context is available.
- You may freely browse the codebase, build, and run tests.
- You may access the internet for API docs, language references, and general knowledge.
- You MUST NOT search for the specific GitHub issue, pull request, or known fix.
  Do not search GitHub, Google, or any other source for the specific issue number or fix.
- You may look into ancenster commits, but not future commits in order to cheat.
- Make your fix by editing source files directly.
- You MUST ensure ./REBUILD.sh completes successfully after your changes.
  A failing REBUILD.sh means your fix is incomplete.
- Your fix should address the root cause, not just mask symptoms.
- Your should never early declare your finish unless you have very strong evidence that the bug is fixed.

## Workflow

1. Read SYMPTOM.md to understand the failure.
2. Run `ls context/` and `cat` each file to review the debugging clues the developer gathered.
3. Explore the relevant source code to understand the component architecture.
4. Form a hypothesis about the root cause, informed by the symptom and any debug context.
5. Implement a targeted fix (prefer minimal, correct changes).
6. If the fix doesn't solve the bug, go back to step 3. If you are confident that you have fixed the bug, go to step 7.
7. Before submitting the final output, you MUST run `./REBUILD.sh` and confirm it exits successfully.
  Fix any compile errors before proceeding. (this is not related to your bug, it is just a sanity check)
8. When you are absolutely confident that the bug is fixed, submit the job completion by running: `echo COMPLETE_TASK_AND_SUBMIT_FINAL_OUTPUT`
  Do not combine it with any other command.
  <important>After this command, you cannot continue working on this task.</important>

## Command Execution Rules

Each response should include:

1. A **THOUGHT** section explaining your reasoning and plan
2. Other things.

**CRITICAL REQUIREMENTS:**

- Your response SHOULD include a THOUGHT section explaining your reasoning
- Directory or environment variable changes are not persistent across commands
- You can prefix any action with `cd /path/to/dir && ...` or write/load env vars from files
## Editing Files

### Create a new file:

```bash
cat <<'EOF' > newfile.go
package main
import "fmt"
func main() { fmt.Println("hello") }
EOF
```

### Edit files with sed:

```bash
# Replace all occurrences
sed -i 's/old_string/new_string/g' filename.go

# Replace only in specific line range
sed -i '10,20s/old_string/new_string/g' filename.go
```

### View file content:

```bash
nl -ba filename.go | sed -n '10,20p'
```
</instructions>
\end{Verbatim}

\end{tcolorbox}
\endgroup

\section{Benchmark Results}

\subsection{Pairwise Model Differentiation on Distributed System Debugging}
\label{appendix:pairwise-models}

\begin{figure}[h]
    \centering
    \includegraphics[width=\linewidth]{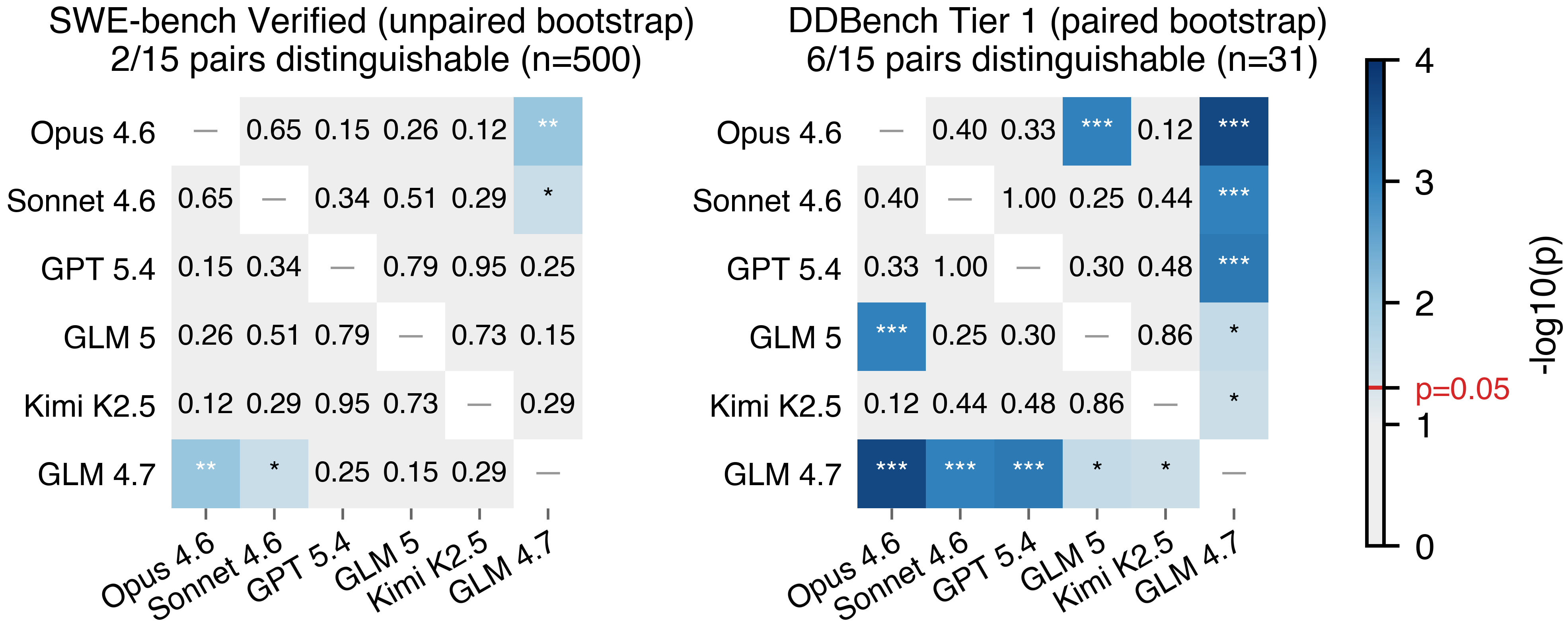}
    \caption{Pairwise bootstrap tests on the six models with published SWE-bench Verified scores (10{,}000 resamples each). \sysname{} Tier-1 uses a paired bootstrap over \ddbenchTierOneSize{} per-case outcomes; SWE-bench Verified uses an unpaired bootstrap over 500 per-case outcomes per model whose pass count equals the vendor-reported rate (as per-case data not publicly released). Cells show the two-sided $p$ per pair; color encodes $-\log_{10}(p)$ clipped to $[0,4]$, with a red mark at $p=0.05$ ($*\!<\!0.05$, $*\!*\!<\!0.01$, $*\!*\!*\!<\!0.001$). On the same six models, \sysname Tier-1 separates 6 of 15 pairs at $p<0.05$, compared with 2 on SWE-bench Verified. The contrast is consistent with \sysname exercising a reasoning dimension that single-process code repair does not.}
    \label{fig:differentiability-grid}
\end{figure}

We present the full pairwise bootstrap matrix in \figref{fig:differentiability-grid}, comparing the six models with published SWE-bench Verified scores across both benchmarks (evaluated on \sysname tier-1 case set).
SWE-bench Verified dinstinguishes only 2 of 15 pairs at $p<0.05$, while \sysname distinguishes 6 of 15 pairs.
For instance, SWE-bench Verified does not clearly separate GPT~5.4 from GLM~4.7 ($p=0.25$), while \sysname does with $p < 0.001$.

\subsection{Contamination Check}
\label{appendix:contamination}

Every \sysname case is a confirmed, fixed, public issue, so the upstream fix commit and its issue discussion may appear in a model's pretraining data.
We test for memorization along two axes: how closely passing agent patches match the upstream fix, and whether pass rate tracks bug recency or repository popularity.

\parab{Patch similarity to the upstream fix.}
We compared all 204 passing agent patches with the production-code changes in the recorded upstream fix commits.
Of these, 44 patches share no line with the upstream changes, and the remaining 160 share at least one line.
We further analyzed the 160 overlapping patches with Tree-sitter; 5 could not be fully parsed and are excluded.
Of the remaining 155 patches, 38 (24.5\%) make logically equivalent but not identical edits, 10 (6.5\%) are subsets of the upstream fix, 5 (3.2\%) contain the upstream fix as a subset, 90 (58.1\%) overlap with it only partially, and 12 (7.7\%) have no structural overlap.
The 44 patches with no shared line and the 102 with only partial or no structural overlap also show that the oracles accept repairs that differ from the upstream fix.

\parab{Audit of near-matches.}
We manually audited the trajectories of all 53 patches strongly similar to the upstream fix (the 38 equivalent, 10 subset, and 5 superset patches).
Forty-seven belong to six cases whose fixing style is inherently limited by the nature of the bug, so independent repairs converge on similar edits; the other six involve broader repairs than the upstream fix.

\parab{Recency and popularity.}
Under the symptom-only condition, the Spearman correlation between fix-commit date and pass rate is $-0.109$, and between GitHub star count and pass rate is $+0.014$.
Neither correlation indicates that pass rate depends on how recent a bug is or how popular its repository is.
Taken together, these checks show no strong evidence of memorization, although patch-level overlap cannot rule it out entirely.

\begin{table}[h]
    \caption{Per-model pass rates (\%) on \tiertwo (\ddbenchTierTwoSize cases) under symptom-only and context-augmented conditions. $\Delta$ denotes the percentage-point improvement. Eight of ten models improved; two tied (Claude Sonnet~4.6, GPT OSS~120B); none regressed.}
    \label{tab:headline-tier2}
    \centering
    \resizebox{\textwidth}{!}{%
        \begin{tabular}{@{}lll|rrr|rrr@{}}
            \toprule
            \multirow{2}{*}{\textbf{Model}}               &
            \multirow{2}{*}{\textbf{Thinking}}            &
            \multirow{2}{*}{\textbf{Form}}                &
            \multicolumn{3}{c|}{\textbf{Total Cost (\$)}} &
            \multicolumn{3}{c}{\textbf{Pass Rate \%}}                                                                                         \\
                                                          &
                                                          &
                                                          &
            Base                                          &
            +Context                     &
            \multicolumn{1}{l|}{$\Delta$ (\%)}            &
            Base                                          &
            +Context                     &
            $\Delta$                                                                                                                          \\ \midrule
            Claude Opus 4.6                               & Adaptive High & API             & 32.40 & 25.29 & -\,21.9\% & 66.7 & 73.3 & +\,6.7   \\
            Claude Sonnet 4.6                             & Adaptive High & API             & 29.34 & 18.53 & -\,36.8\% & 73.3 & 73.3 & ---      \\
            GPT 5.4                                       & High          & API             & 10.75 & 10.35 & -\,3.7\%  & 60.0 & 86.7 & +\,26.7  \\
            GPT 5.4 mini                                  & High          & API             & 2.70  & 2.25  & -\,16.7\% & 66.7 & 73.3 & +\,6.7   \\
            GLM 5.1 (FP8)                                 & Enabled       & Open, Self-Host & 7.12  & 5.33  & -\,25.1\% & 40.0 & 46.7 & +\,6.7   \\
            GLM 5                                         & Enabled       & Open, API       & 36.68 & 22.01 & -\,40.0\% & 53.3 & 66.7 & +\,13.3  \\
            GLM 4.7                                       & Enabled       & Open, API       & 19.88 & 11.30 & -\,43.2\% & 33.3 & 66.7 & +\,33.3  \\
            Gemma 4 (31B)                                 & Enabled       & Open, Self-Host & 0.94  & 0.77  & -\,18.1\% & 26.7 & 53.3 & +\,26.7  \\
            Kimi K2.5                                     & N/A           & Open, API       & 8.30  & 2.39  & -\,71.2\% & 33.3 & 60.0 & +\,26.7  \\
            GPT OSS (120B BF16)                           & Enabled       & Open, Self-Host & 1.95  & 1.65  & -\,15.4\% & 26.7 & 26.7 & ---      \\ \bottomrule
        \end{tabular}%
    }
\end{table}

\begin{table}[h]
    \caption{Per-model pass rates (\%) on \tierthree (\ddbenchTierThreeSize cases) under symptom-only and context-augmented conditions. $\Delta$ denotes the percentage-point improvement. Six of ten models improved; two tied (GLM 4.7, GLM 5); two regressed.}
    \label{tab:headline-tier3}
    \centering
    \resizebox{\textwidth}{!}{%
        \begin{tabular}{@{}lll|rrr|rrr@{}}
            \toprule
            \multirow{2}{*}{\textbf{Model}}               &
            \multirow{2}{*}{\textbf{Thinking}}            &
            \multirow{2}{*}{\textbf{Form}}                &
            \multicolumn{3}{c|}{\textbf{Total Cost (\$)}} &
            \multicolumn{3}{c}{\textbf{Pass Rate \%}}                                                                                         \\
                                                          &
                                                          &
                                                          &
            Base                                          &
            +Context                     &
            \multicolumn{1}{l|}{$\Delta$ (\%)}            &
            Base                                          &
            +Context                     &
            $\Delta$                                                                                                                          \\ \midrule
            Claude Opus 4.6                               & Adaptive High & API             & 38.05 & 14.61 & -\,61.6\% & 78.57 & 92.86 & +\,14.29  \\
            Claude Sonnet 4.6                             & Adaptive High & API             & 33.82 & 17.55 & -\,48.1\% & 71.43 & 92.86 & +\,21.43  \\
            GPT 5.4                                       & High          & API             & 13.17 & 9.50  & -\,27.9\% & 78.57 & 92.86 & +\,14.29 \\
            GPT 5.4 mini                                  & High          & API             & 2.37  & 1.75  & -\,26.2\% & 85.71 & 92.86 & +\,7.14  \\
            GLM 5.1 (FP8)                                 & Enabled       & Open, Self-Host & 9.47  & 10.41 & +\,9.9\%  & 57.14 & 50.00 & -\,7.14  \\
            GLM 5                                         & Enabled       & Open, API       & 32.78 & 23.70 & -\,27.7\% & 85.71 & 85.71 & ---      \\
            GLM 4.7                                       & Enabled       & Open, API       & 20.67 & 15.35 & -\,25.7\% & 85.71 & 85.71 & ---      \\
            Gemma 4 (31B)                                 & Enabled       & Open, Self-Host & 1.55  & 2.75  & +\,77.4\% & 57.14 & 50.00 & -\,7.14  \\
            Kimi K2.5                                     & N/A           & Open, API       & 12.01 & 3.87  & -\,67.8\% & 42.86 & 85.71 & +\,42.86 \\
            GPT OSS (120B BF16)                           & Enabled       & Open, Self-Host & 1.98  & 1.83  & -\,7.6\%  & 35.71 & 42.86 & +\,7.14  \\ \bottomrule
        \end{tabular}%
    }
\end{table}

\subsection{Pass Rate on Tier-2 and Tier-3}

We report the per-model pass rates and total cost on \tiertwo{} (\ddbenchTierTwoSize{} cases, \tabref{tab:headline-tier2}) and \tierthree{} (\ddbenchTierThreeSize{} cases, \tabref{tab:headline-tier3}) under both the symptom-only and context-augmented conditions, complementing the aggregate tier comparison in \secref{sec:eval:tiers}.

\parab{Results on \tiertwo{}.}
On \tiertwo{}, 8 of \ddbenchNumModels{} models improved under the context-augmented condition, 2 tied, and none regressed.
The two ties are Claude Sonnet 4.6 at 73.3\% and GPT OSS (120B BF16) at 26.7\% in both conditions.
GLM 4.7 posted the largest gain at $+$33.3~pp, doubling its base pass rate from 33.3\% to 66.7\%.
Three further models gained $+$26.7~pp: GPT 5.4 rose from 60.0\% to 86.7\%, Gemma 4 (31B) from 26.7\% to 53.3\%, and Kimi K2.5 from 33.3\% to 60.0\%.
GPT 5.4's 86.7\% is the highest pass rate on \tiertwo{} under either condition, consistent with the pattern on \tierone{} where frontier capability combined with debug context produces the strongest joint lift.
Total cost fell for every model on \tiertwo{}, ranging from $-$3.7\% on GPT 5.4 to $-$71.2\% on Kimi K2.5.
GLM 5, GLM 4.7, and Kimi K2.5 each realized cost reductions above 40\%, the strongest compression of the exploration phase across the cohort.

\parab{Results on \tierthree{}.}
On \tierthree{}, the picture shifts toward saturation. 6 of \ddbenchNumModels{} models improved, 2 tied at 85.71\%, and 2 regressed by $-$7.14~pp each.
The two ties are GLM 5 and GLM 4.7.
Each regression (GLM 5.1 and Gemma 4 (31B)) is a net loss of a single case at $n=14$ within binomial noise.
Under the context-augmented condition, four models converge at 92.86\% pass rate, Claude Opus 4.6, Claude Sonnet 4.6, GPT 5.4, and GPT 5.4 mini, reflecting the ceiling effect expected on this tier since \tierthree{} cases are predominantly single-process, input-triggered crashes whose root cause often localizes within one file.
The largest gain comes from the weaker end of the cohort.
Kimi K2.5 climbed $+$42.86~pp from 42.86\% to 85.71\%, reaching the open-weight cluster anchored by the GLM 5 and GLM 4.7 tie point.
Cost trajectories are more mixed than on \tiertwo{}.
8 of \ddbenchNumModels{} models reduced total cost, with Kimi K2.5 dropping $-$67.8\% and the two Claude models dropping 61.6\% (Opus) and 48.1\% (Sonnet), while Gemma 4 (31B) nearly doubled its cost ($+$77.4\%) without a net pass-rate gain (2 cases gained, 3 lost).
GLM 5.1 and GPT OSS (120B BF16) saw small cost shifts of $+$9.9\% and $-$7.6\% respectively.

\parab{Takeaway.}
The per-model tables reinforce the aggregate trend in \secref{sec:eval:tiers}.
Debug context lifts pass rate broadly but with a narrower margin as the tier's reasoning scope narrows, and the cost benefit is not uniform. Models already near the ceiling realize most of the benefit as cost reduction, while weaker models realize it as pass-rate gains.

\begin{figure}[h]
    \centering
    \includegraphics[width=\linewidth]{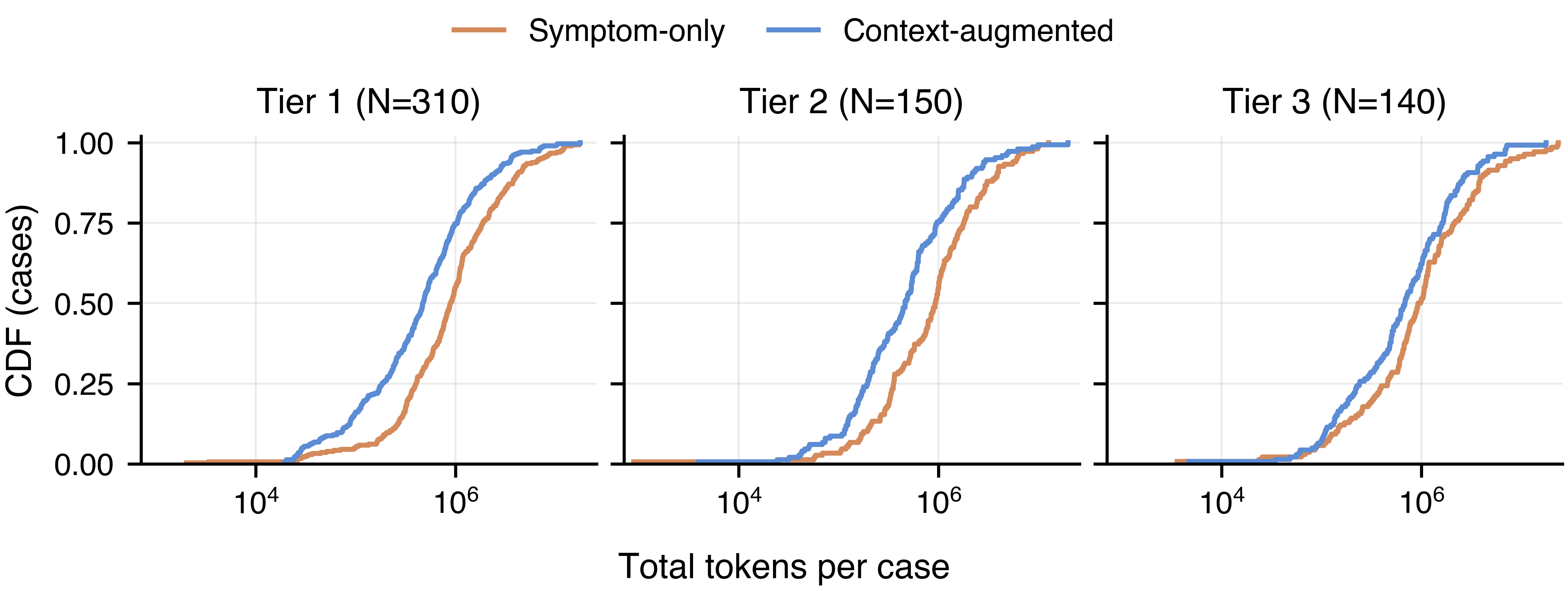}
    \caption{Per-case total-token CDFs on \tierone--\tierthree. Each panel overlays the symptom-only (orange) and context-augmented (blue) distributions, pooling all evaluated models. Medians shift from 0.89M to 0.47M on \tierone ($-$47\%), 0.93M to 0.46M on \tiertwo ($-$50\%), and 0.95M to 0.68M on \tierthree ($-$28\%). The context-augmented CDF lies to the left of the symptom-only CDF at every percentile on \tierone{} and \tiertwo; on \tierthree the two curves converge in the upper tail.}
    \label{fig:efficiency-tokens-cdf}
\end{figure}

\begin{figure}[h]
    \centering
    \includegraphics[width=\linewidth]{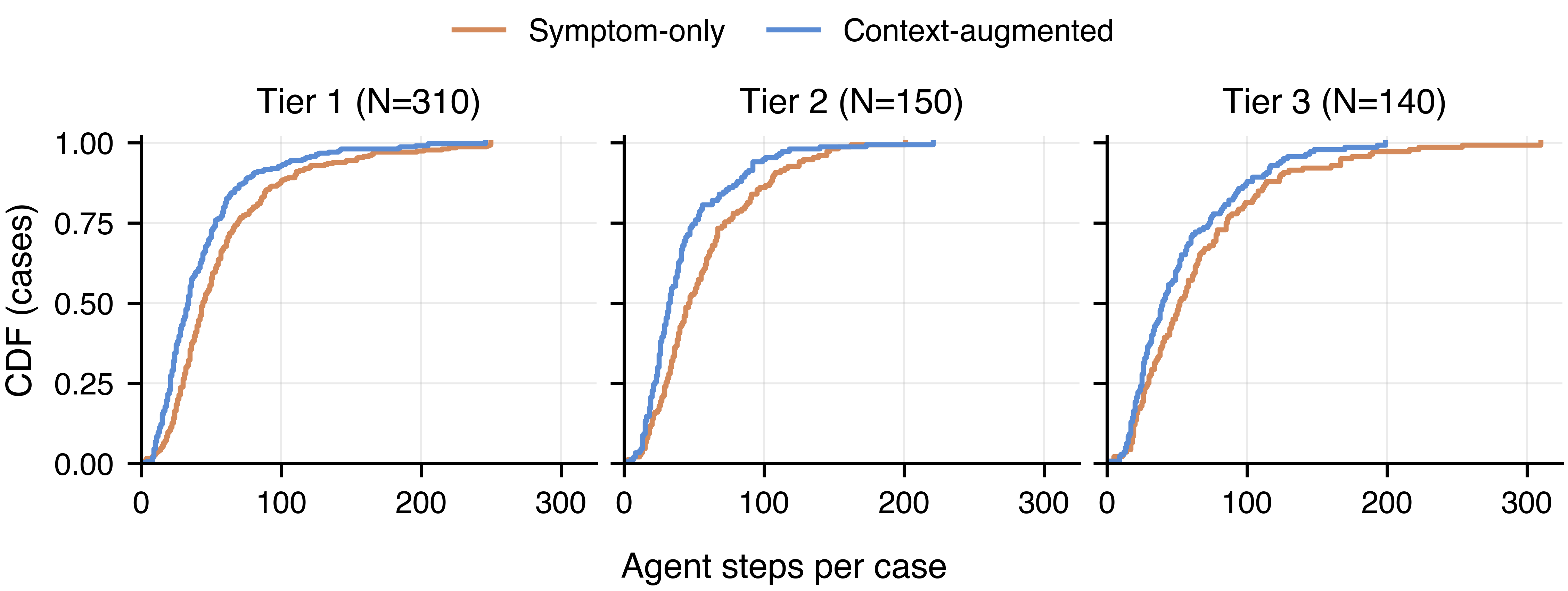}
    \caption{Per-case agent-step CDFs on \tierone--\tierthree. Medians drop from 44 to 34 on \tierone ($-$23\%), 46 to 32 on \tiertwo ($-$30\%), and 52 to 40 on \tierthree ($-$23\%). Unlike the token distributions, the step CDFs separate roughly uniformly across the support, so the savings are not driven by a few long-running outliers.}
    \label{fig:efficiency-steps-cdf}
\end{figure}

\subsection{Efficiency Gain from Debug Context}
\label{appendix:efficiency-cdf}

The trajectory analysis (\secref{sec:eval:trajectory}) reported mean steps and tokens on \tierone{}. \figref{fig:efficiency-tokens-cdf} and \figref{fig:efficiency-steps-cdf} extend the result along two axes.
We plot the full per-case distribution rather than the mean, and we repeat the measurement on \tiertwo{} and \tierthree{} to validate that the effect is not tier-specific.

\parab{Tokens.} Debug context cut the median per-case token budget by nearly half on \tierone{} (0.89M\,$\to$\,0.47M) and \tiertwo{} (0.93M\,$\to$\,0.46M), and by 28\% on \tierthree{} (0.95M\,$\to$\,0.68M).
The context-augmented CDF dominates the symptom-only CDF on \tierone{} and \tiertwo{}: at every quantile a context-augmented run consumed fewer tokens than the matched symptom-only run.
On \tierthree{}, the two curves cross in the upper tail. Specifically, a small number of cases ran to the step budget in both conditions, indicating that when the abstraction distance between the supplied symptom and the fault is large (\secref{sec:eval:case-studies}), debug context does not shorten the exploration loop.

\parab{Agent steps.} The step distributions shifted more uniformly than tokens: medians dropped 23--30\% across tiers, and the two CDFs sit roughly parallel.
The per-tier median reductions (\tierone: $-$10 steps; \tiertwo: $-$14; \tierthree: $-$12) are consistent with the previously reported mean of $-$14.4 steps on \tierone{} (\tabref{tab:trajectory}).
Debug context therefore removes a near-constant number of exploratory steps per case, rather than only trimming the long tail.

\parab{Why the CDF matters.}
The mean-based summary in \tabref{tab:trajectory} leaves open whether the efficiency gain is driven by a handful of pathological runs or by a broad shift. 
\figref{fig:efficiency-tokens-cdf} and \figref{fig:efficiency-steps-cdf} resolve this by showing that the shift is distributional and persists across tiers. 
Combined with the paired-outcome analysis (\secref{appendix:outcome-matrix}), which shows efficiency gains on concordant-PASS, concordant-FAIL, and FAIL$\to$PASS cases, these CDFs support the conclusion that debug context acts as a search-space reducer for agent diagnosis, narrows the set of hypotheses the agent must explore, and is independent of whether the final patch changes.

\subsection{Per-Model Step-Count Distributions}
\label{appendix:steps-per-model}

\begin{figure}[!h]
    \centering
    \includegraphics[width=\linewidth]{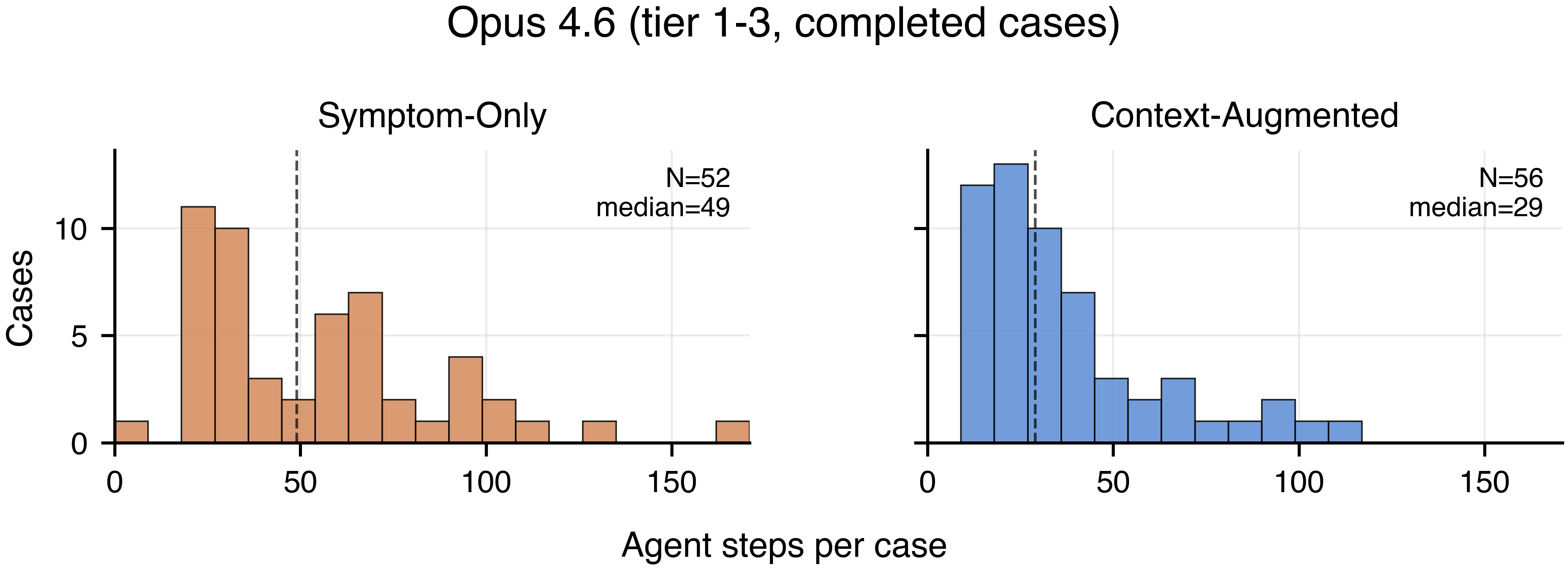}
    \caption{Per-case agent-step distribution for Claude Opus 4.6 across \tierone--\tierthree, restricted to runs that terminated under \texttt{Submitted}. The median drops from 49 to 29 steps under context augmentation ($\approx$41\% reduction). The right tail reaches 167 steps under symptom-only and 114 steps under context augmentation,indicating that even the strongest model in our setup encounters cases where the cross-process search dominates the trajectory regardless of debug context.}
    \label{fig:steps-distribution-opus}
\end{figure}

\begin{figure}[!h]
    \centering
    \includegraphics[width=\linewidth]{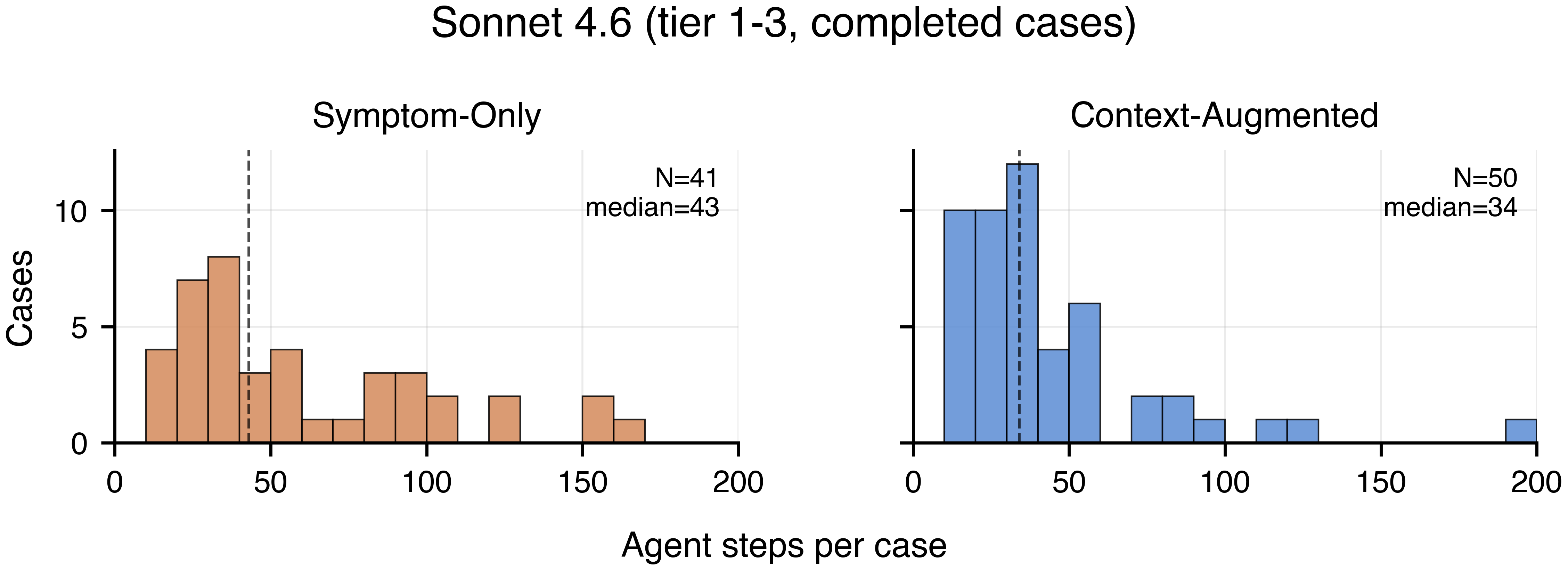}
    \caption{Per-case agent-step distribution for Claude Sonnet 4.6 across \tierone--\tierthree. The median drops from 43 to 34 steps. The base count ($N\!=\!41$) is lower than the context-augmented count ($N\!=\!50$): nine additional cases reached \texttt{Submitted} under context augmentation. The recovered cases are dominated by runs that produced no exit status under the symptom-only condition (most plausibly wall-clock timeouts or framework aborts) rather than by step-cap exceptions; under either reading, the asymmetry is consistent with debug context shortening the path to a candidate patch and is not visible in the median alone.}
    \label{fig:steps-distribution-sonnet}
\end{figure}

\begin{figure}[!h]
    \centering
    \includegraphics[width=\linewidth]{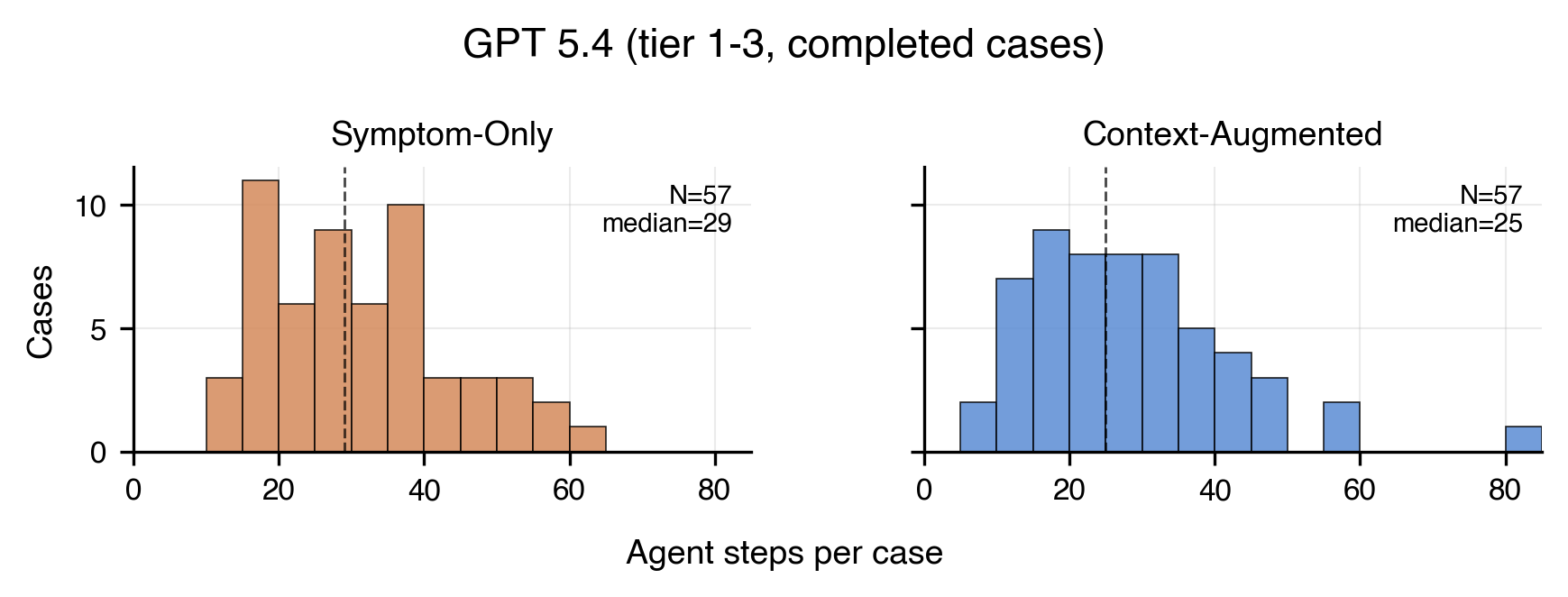}
    \caption{Per-case agent-step distribution for GPT 5.4 across \tierone--\tierthree. The median drops from 29 to 25 steps. Both distributions concentrate tightly below 85 steps with negligible right-tail mass, indicating that GPT 5.4 either converges on a candidate patch quickly or abandons the case, rather than continuing to explore.}
    \label{fig:steps-distribution-gpt54}
\end{figure}

\begin{figure}[!h]
    \centering
    \includegraphics[width=\linewidth]{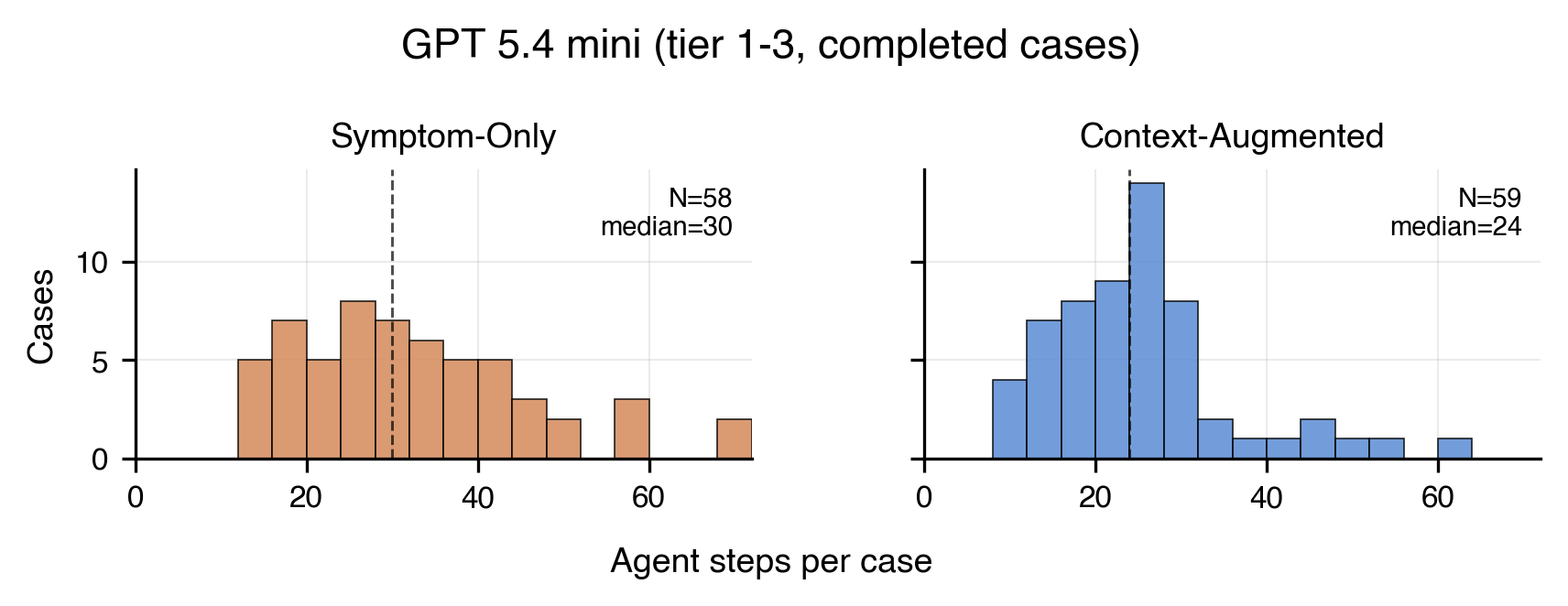}
    \caption{Per-case agent-step distribution for GPT 5.4 mini across \tierone--\tierthree. The median drops from 30 to 24 steps. The shape mirrors GPT 5.4: short-tailed and tightly concentrated. The smaller absolute reduction ($-$6 steps) is consistent with the already-compact base distribution---there are fewer wasted exploratory steps for debug context to remove.}
    \label{fig:steps-distribution-gpt54-mini}
\end{figure}

\begin{figure}[!h]
    \centering
    \includegraphics[width=\linewidth]{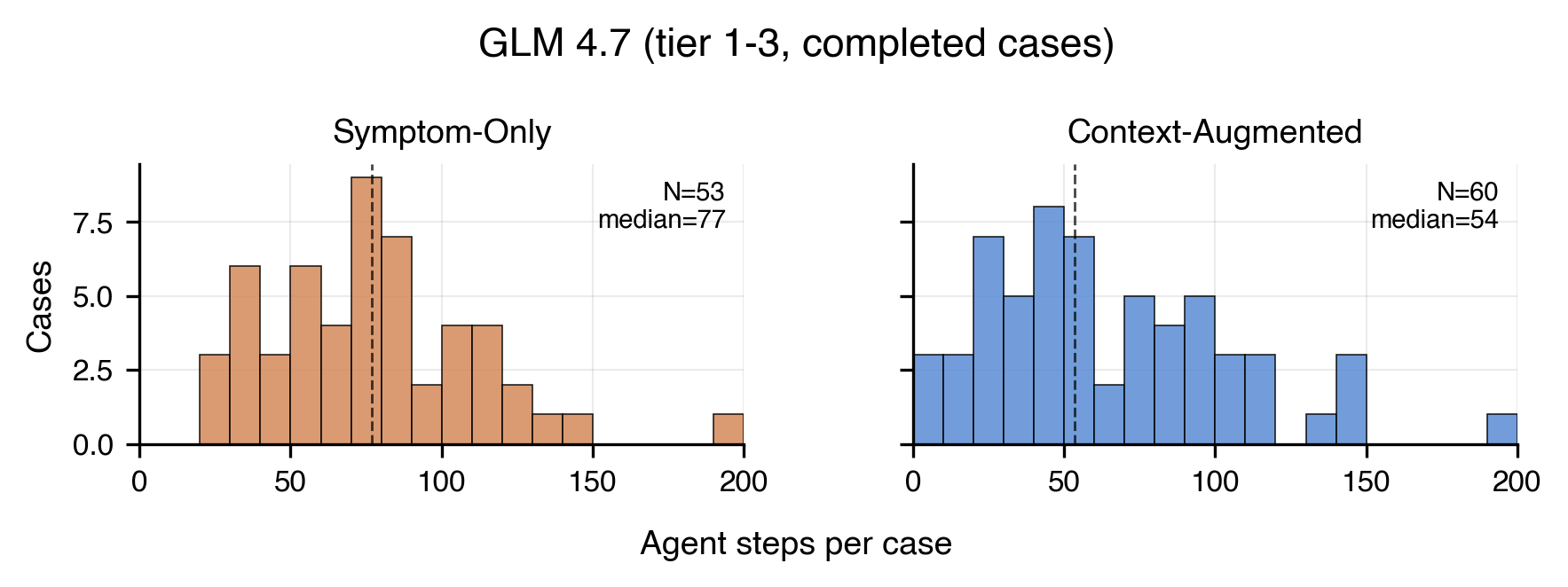}
    \caption{Per-case agent-step distribution for GLM 4.7 across \tierone--\tierthree. The median drops from 77 to 54 steps---the highest medians in this subset under either condition. The right tail extends past 190 steps in both conditions, reflecting GLM 4.7's tendency to run a long exploratory phase on hard cases. The base count ($N\!=\!53$) is lower than the context-augmented count ($N\!=\!60$): all seven cases that did not terminate under \texttt{Submitted} in the symptom-only condition (one step-cap exception and six runs with no recorded exit status) reached \texttt{Submitted} when given debug context, so every GLM 4.7 case terminated normally under context augmentation.}
    \label{fig:steps-distribution-glm47}
\end{figure}

\figref{fig:efficiency-steps-cdf} pools per-case step counts across the full cohort, which leaves open whether the median reduction reflects a uniform per-model shift or is driven by a subset of the cohort.
\figref{fig:steps-distribution-opus}--\figref{fig:steps-distribution-glm47} decompose this view by plotting the per-case step histogram for each of five models that span the cohort's capability range, pooling \tierone--\tierthree{} cases per model so that each panel carries enough mass to estimate the distribution shape.
We restrict each panel to runs that terminated with the \texttt{Submitted} exit status so that the histogram reflects deliberation length rather than abnormal termination (step-cap or context-window exceptions, or wall-clock timeouts).
The two side-by-side panels in each figure share a step axis and overlay the per-condition median as a dashed reference inside each panel.

\parab{Cross-model synthesis.}
Three patterns emerge from the per-model panels.
First, the median reduction is universal in this subset: every model shows a leftward shift under context augmentation, ranging from $-$4 steps for GPT 5.4 to $-$23 steps for GLM 4.7 ($-$20 for Claude Opus 4.6).
This rules out the hypothesis that the cohort-level effect in \figref{fig:efficiency-steps-cdf} is carried by a few frontier models; lighter models show the same direction even when the magnitude is small.
Second, the absolute reduction tracks the base-condition median.
Models that already settle quickly under the symptom-only condition (the GPT 5.4 family) have less slack to compress, while models that explore extensively under symptom-only conditions (Claude Opus 4.6, GLM 4.7) realize the largest absolute step savings.
This is consistent with the search-space-reduction reading from \secref{appendix:efficiency-cdf}: debug context removes exploratory steps, and a model that takes few exploratory steps in the first place gains less.
Third, the heavy right tails are concentrated in the two Claude models and GLM 4.7, all of which run long deliberation loops on hard cases rather than abandoning early; the GPT 5.4 family in contrast caps trajectories well below 85 steps in both conditions, suggesting an earlier give-up policy.
These two failure modes---long-tail persistence vs.\ early abandonment---are visible at the distribution level even when the medians are similar, and they suggest that scaffold-design work aiming to extract more from a fixed step budget should treat the two regimes separately.

\subsection{Per-case Paired Outcome in Tier-1 under Debug-Context Augmentation}
\label{appendix:outcome-matrix}

We present the full per-case paired outcomes on \tierone{} under debug-context augmentation in \figref{fig:context-flip-matrix}.
Weak models (bottom rows) are dominated by green gains, where GLM~4.7 flips 13 of 31 cases.
Strong models (top rows) are dominated by blue tie-passes. 
Losses are rare (13 of 310 paired runs) and have mixed causes. For example, \texttt{kvrocks-3356} (the last column of Resource \& delivery), lost by Claude~Opus~4.6 and GLM~5, is the highlighted example analyzed in \secref{sec:eval:case-studies}.

\begin{figure}[h]
    \centering
    \includegraphics[width=\linewidth]{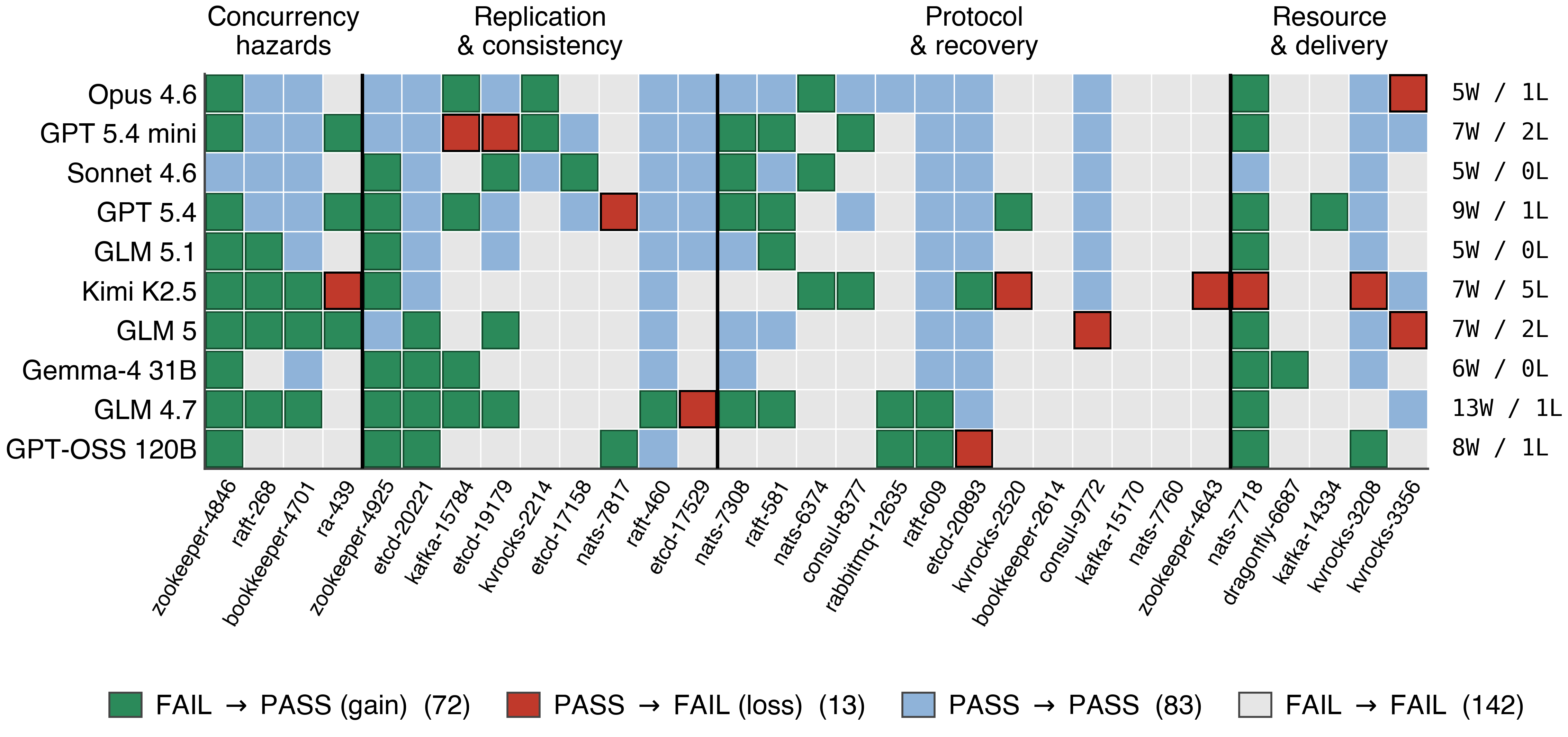}
    \caption{Per-case paired outcomes across all \ddbenchNumModels models on \tierone (rows: models, ordered by symptom-only pass rate; columns: the 31 cases, grouped by bug category and ordered within each group by cross-model win count). Green cells are gains (FAIL$\to$PASS), red cells are regressions (PASS$\to$FAIL). Weak models (bottom rows) are dominated by green---GLM~4.7 flips 13 of 31 cases---while strong models (top rows) are dominated by blue tie-passes. Losses are rare and have mixed causes. For instance, \texttt{kvrocks-3356} (last column of Resource \& delivery) is one of the highlighted examples in Appendix~\secref{app:case-studies}.}
    \label{fig:context-flip-matrix}
\end{figure}

\subsection{Case Studies: When Debug Context Helps, and When It Misleads}
\label{app:case-studies}

This section expands the qualitative analysis summarized in \secref{sec:eval:case-studies}.
We examine three \tierone cases that together span the qualitative regimes of the debug-context effect: one in which debug context supplies a signal unreachable from code alone (ZooKeeper-4846), one in which debug context compresses a tractable but expensive search (Raft-268), and one in which debug context anchors the agent on the wrong layer across an abstraction gap (Kvrocks-3356).
The first two cases were selected by cross-model consensus, so that a majority of the 10~models show the same directional effect and the pattern reflects task structure rather than noise from any single model; the third is analyzed at the level of individual trajectories.

\parab{ZooKeeper-4846: debug context supplies a signal that no code reading can recover (9/10 FAIL$\to$PASS).}
The bug corrupts access-control state after a crash-recovery path, so that child nodes are re-linked to stale permission entries whenever the on-disk snapshot and the in-memory state disagree.
Under the symptom-only condition, 9~of~10 models failed, and not because they did not try: agents explored the snapshot and restore code paths broadly, yet the fault only manifests as a specific post-recovery runtime state that static reading of the codebase cannot reconstruct.
When we provided the post-recovery permission table as debug context, all 10 models succeeded and converged on a precise fix in \texttt{createNode()}.
The behavioral shift is consistent with an information-gap hypothesis.
Kimi~K2.5's exploration actions dropped from 9 to 2, its read actions from 26 to 8, and its first-edit step from 46 to 17.
Gemma~4, which had given up after 4 steps without ever editing, now navigated directly to the relevant file and patched the bug in 12 steps.
The near-unanimous flip across a heterogeneous model set indicates that debug context here acts as a capability unlock: when the root cause is a runtime-only invariant, the missing ingredient is signal rather than reasoning ability.

\parab{Raft-268: debug context compresses a tractable but expensive search (4/10 FAIL$\to$PASS, 0~regressions).}
The bug is a shutdown-path deadlock, in which one goroutine blocks on a channel that a second, already-stopped goroutine was supposed to service.
The root cause is in principle derivable from the source, but doing so requires reasoning over the full space of goroutine interleavings, a search that scales badly with context length.
The debug context we provide is a goroutine stack dump that points directly at the blocked operation, which collapses this combinatorial reasoning problem into a localized one.
With debug context, 4 additional models flip to PASS and no model regresses.
The more revealing effect, however, is on trajectory rather than outcome.
Kimi~K2.5 went from 225~steps (first edit at step~174) to 36~steps (first edit at step~9); its symptom-only patch was a surface-level workaround in the wrong file, whereas its context-augmented patch correctly restructured the blocking \texttt{select}.
Even Claude~Opus, which passed under both conditions, halved its step count (58 to 30) and its first-edit step (51 to 21).
Debug context here functions as a search-space reducer, and we argue this is where most of the aggregate trajectory effect in \secref{sec:eval:trajectory} originates, since the efficiency benefit is visible even when the final outcome does not change.

\parab{Kvrocks-3356: faithful but distant debug context anchors the agent on the wrong layer (0~wins, 2~losses).}
The root cause of this bug is in the master's replication feed thread, which blocks on a socket write to a slow replica; while it is blocked, the replica falls behind and the master keeps retaining its write-ahead log (WAL), so disk usage grows without bound. The debug context is the output of the case's reproducer on the faulty build, reporting that the master's data directory grew by 541{,}596\,KB, above the test's 512\,MB threshold. This observation is faithful but sits several causal steps from the fault (blocked feed thread, replica lag, WAL retention, disk usage), and its threshold matches Kvrocks' default 512\,MB limit on total WAL size (\texttt{max\_total\_wal\_size}). Claude~Opus~4.6 read the faulty feed-thread loop early (step~15), yet 52 of its 205 reasoning steps return to the 512\,MB figure or the WAL size limit, and 45 steps read the source of the underlying storage engine (RocksDB). At step~128 it identified the actual mechanism (the blocking socket write stalls the feed thread), but at step~158 it concluded that ``the WAL growth is purely due to the RocksDB configuration,'' and its final patch only lowered WAL-retention defaults before its 30-minute budget ran out. Without debug context, the same model patched the feed-thread loop in 49 steps; GLM~5 likewise patched the feed thread in 35 steps without debug context but timed out without a patch when given it. Here, debug context inherits the abstraction level of the observable symptom, and when that symptom sits several causal steps above the fault, it can act as a confident but wrong prior rather than a neutral input. We present this case as evidence of anchoring in the agent's reasoning rather than as a measured effect on pass rate: the symptom-only patches that pass do not add the upstream fix's send timeout and replica disconnect, and they clear the oracle's 512\,MB threshold by only about 6\%.

\ifneurips
  \newpage
  \section*{NeurIPS Paper Checklist}
\begin{enumerate}
\item {\bf Claims}
    \item[] Question: Do the main claims made in the abstract and introduction accurately reflect the paper's contributions and scope?
    \item[] Answer: \answerYes{} 
    \item[] Justification:
    We provide a dataset, benchmark infrastructure, and a thorough evaluation on them that collectively reflect the technical, conceptual, and major contributions of this paper.
    The claims and findings mentioned in abstract and introduction are thoroughly supported and discussed in the later sections (\secref{sec:benchmark} and \secref{sec:eval}).
    The benchmark shape is fully described in \secref{sec:benchmark}, and the evaluation results are fully described in \secref{sec:eval}.
    The claims on the model differentiations on distributed debugging and controlled study of symptom-only and context-augmented are fully discussed in \secref{sec:eval}, which further reveals several findings that is backed by the evaluation results.
    All number that is mentioned in the abstract and introduction are accurate and also supported by the evaluation results in \secref{sec:eval}.

\item {\bf Limitations}
    \item[] Question: Does the paper discuss the limitations of the work performed by the authors?
    \item[] Answer: \answerYes{} 
    \item[] Justification:
    We fully discussed the limitations of our work in \secref{sec:limitations}, including the limitations of the evaluation (single agent scaffold, no content-level ablation) and the single curated release.
    We also fully discussed the potential future work to address these limitations, including multi-scaffold evaluation, content-level ablation, and scale-up across additional systems and bug families.

\item {\bf Theory assumptions and proofs}
    \item[] Question: For each theoretical result, does the paper provide the full set of assumptions and a complete (and correct) proof?
    \item[] Answer: \answerNA{} 
    \item[] Justification: 
    This work does not include theoretical results and we do not have any theorems to prove, so this question is not applicable.

    \item {\bf Experimental result reproducibility}
    \item[] Question: Does the paper fully disclose all the information needed to reproduce the main experimental results of the paper to the extent that it affects the main claims and/or conclusions of the paper (regardless of whether the code and data are provided or not)?
    \item[] Answer: \answerYes{} 
    \item[] Justification:
    We fully disclosed all the information needed to reproduce the main experimental results in \secref{sec:eval}, including the details of the models evaluated, the agent scaffold used, and the evaluation protocol.
    We also provided the details of the benchmark features and data curation process in \secref{sec:benchmark}. Appendix~\secref{appendix:infrastructure} also includes the details of data mining and prompts we used for the semi-agentic mining process.
    We released the codebase, dataset, and benchmark infrastructure (including the semi-agentic mining tools) to facilitate reproducibility.

\item {\bf Open access to data and code}
    \item[] Question: Does the paper provide open access to the data and code, with sufficient instructions to faithfully reproduce the main experimental results, as described in supplemental material?
    \item[] Answer: \answerYes{} 
    \item[] Justification: 
    We released the codebase, dataset, and benchmark infrastructure (including the semi-agentic mining tools) to facilitate reproducibility.
    The dataset and codebase are also submitted alongside the paper and will be made publicly available upon publication.
    We also provided detailed instructions for how to use the codebase and reproduce the main experimental results in the README and documentation.

\item {\bf Experimental setting/details}
    \item[] Question: Does the paper specify all the training and test details (e.g., data splits, hyperparameters, how they were chosen, type of optimizer) necessary to understand the results?
    \item[] Answer: \answerYes{} 
    \item[] Justification: 
    This paper does not include training details as we do not train any models, but we provide all the details of the experimental setting and evaluation protocol in \secref{sec:eval}, including the models evaluated, the agent scaffold used, and the evaluation protocol.
    We had dedicated subsections in evaluation section describes the model we used, the conditions (symptom-only vs. context-augmented) we evaluated, and the emprical bounds we set for evaluting agent performance (bounded max steps, time limits, and token cost).

\item {\bf Experiment statistical significance}
    \item[] Question: Does the paper report error bars suitably and correctly defined or other appropriate information about the statistical significance of the experiments?
    \item[] Answer: \answerYes{} 
    \item[] Justification:
    Every claim that supports the headline contributions is accompanied by an explicit uncertainty estimate, computed by a method documented in \secref{sec:eval:setup}.
    Pairwise model-differentiation results (\figref{fig:differentiability-grid}) are obtained by a paired case-level bootstrap over the \ddbenchTierOneSize{} \tierone{} cases (10{,}000 resamples) and reported as a two-sided $p$-value per model pair with 95\% CIs on the pass-rate difference; the SWE-bench Verified comparison panel uses the analogous unpaired bootstrap over per-model outcome vectors whose pass counts equal the vendor-reported rates since per-case outcomes are not publicly released.
    Per-model debug-context effects use the same 10{,}000-resample paired bootstrap (\secref{sec:eval}, ``Statistical significance''), and the aggregate debug-context effect is estimated by a mixed-effects logistic regression with random intercepts for model and case over all 620 paired outcomes, reported with point estimate, standard error, 95\% CI, odds ratio, and Wald $z$-statistic.
    The factors of variability captured by these intervals are the random sampling of cases (bootstrap) and the model$\times$case grid the regression is fit on.
    All intervals are two-sided 95\% unless otherwise stated.

\item {\bf Experiments compute resources}
    \item[] Question: For each experiment, does the paper provide sufficient information on the computer resources (type of compute workers, memory, time of execution) needed to reproduce the experiments?
    \item[] Answer: \answerYes{} 
    \item[] Justification:
    We fully described the containerized evaluation infrastructure which implicitly specifies the compute resources required. 
    As we did not train any models in this paper, so we do not have any strong requirement on the type of compute workers.
    Any commodity CPU server with sufficient memory and storage should be able to run this benchmark.
    
\item {\bf Code of ethics}
    \item[] Question: Does the research conducted in the paper conform, in every respect, with the NeurIPS Code of Ethics \url{https://neurips.cc/public/EthicsGuidelines}?
    \item[] Answer: \answerYes{} 
    \item[] Justification:
        Yes, the entire research process, methodology and results fully conform to the NeurIPS code of ethics. We have carefully reviewed the code of ethics and ensured that our research adheres to all the guidelines and principles outlined in it. 
        We have taken necessary precautions to ensure that our research does not cause harm, respects privacy, and is conducted with integrity and transparency.
        We did not conduct any experiments involving human subjects or crowdsourcing, so we did not need to obtain IRB approval or equivalent.

\item {\bf Broader impacts}
    \item[] Question: Does the paper discuss both potential positive societal impacts and negative societal impacts of the work performed?
    \item[] Answer: \answerYes{} 
    \item[] Justification: 
        We do not anticipate any direct negative societal impacts of our work, but we do discuss the potential positive impacts and insights throughout the evaluation section \secref{sec:eval}.
        We discussed the potential future work on enhancing agent engineering and model capabilities for distributed debugging, which could have positive impacts on software engineering practices and productivity in general. 
    
\item {\bf Safeguards}
    \item[] Question: Does the paper describe safeguards that have been put in place for responsible release of data or models that have a high risk for misuse (e.g., pre-trained language models, image generators, or scraped datasets)?
    \item[] Answer: \answerNA{} 
    \item[] Justification: 
    We do not anticipate any risk of misuse of our dataset, benchmark, or evaluation results, so we do not have any specific safeguards in place for responsible release.
    However, the case used in this benchmark are all sanitized and runs inside an isolated container by default.

\item {\bf Licenses for existing assets}
    \item[] Question: Are the creators or original owners of assets (e.g., code, data, models), used in the paper, properly credited and are the license and terms of use explicitly mentioned and properly respected?
    \item[] Answer: \answerYes{} 
    \item[] Justification: 
        We properly credited the creators and original owners of all assets used in the paper, including the code repository, datasets, and literature.
        We also included a dedicated section in Appendix~\secref{appendix:license} that lists all licenses of code repositories that is included in this benchmark release.

\item {\bf New assets}
    \item[] Question: Are new assets introduced in the paper well documented and is the documentation provided alongside the assets?
    \item[] Answer: \answerYes{} 
    \item[] Justification: 
        The codebase for benchmark infrastructure is well documented with a README and detailed instructions for how to use the codebase and reproduce the main experimental results.
        The cases curated in this release are also all well documented with detailed descriptions of each case in both human-readable and machine-readable formats.

\item {\bf Crowdsourcing and research with human subjects}
    \item[] Question: For crowdsourcing experiments and research with human subjects, does the paper include the full text of instructions given to participants and screenshots, if applicable, as well as details about compensation (if any)? 
    \item[] Answer: \answerNA{} 
    \item[] Justification: 
        We did not conduct any crowdsourcing experiments or research with human subjects, so this question is not applicable.

\item {\bf Institutional review board (IRB) approvals or equivalent for research with human subjects}
    \item[] Question: Does the paper describe potential risks incurred by study participants, whether such risks were disclosed to the subjects, and whether Institutional Review Board (IRB) approvals (or an equivalent approval/review based on the requirements of your country or institution) were obtained?
    \item[] Answer: \answerNA{} 
    \item[] Justification: 
        We did not conduct any research with human subjects, so this question is not applicable.

\item {\bf Declaration of LLM usage}
    \item[] Question: Does the paper describe the usage of LLMs if it is an important, original, or non-standard component of the core methods in this research? Note that if the LLM is used only for writing, editing, or formatting purposes and does \emph{not} impact the core methodology, scientific rigor, or originality of the research, declaration is not required.
    \item[] Answer: \answerYes{} 
    \item[] Justification: 
        We used semi-agentic mining protocol to mine the cases in this benchmark, which involves LLMs and agents to semantically parse the information from GitHub issues and PRs and synthesize the artifacts used later in the evaluation. 
        The full details are described in \secref{sec:benchmark} and Appendix~\secref{appendix:semi-agentic}.
        Our usage of LLMs does not impact the core methodology, scientific rigor, or originality of the research, but it is an important component of the data curation process, so we declare it here for transparency.

\end{enumerate}
\fi

\end{document}